\documentclass[twocolumn]{emulateapj}
\pdfoutput=1
\usepackage{graphicx}
\usepackage{natbib}
\usepackage{multirow}
\usepackage{amsmath}
\usepackage{amssymb}
\usepackage{rotating}
\usepackage{hyperref}
\newcommand{\hii}{\ion{H}{2}}
\newcommand{\am}{NH$_{3}$}
\newcommand{\deut}{NH$_{2}$D}
\newcommand{\cyano}{HC$_{3}$N}
\newcommand{\amiso}{$^{15}$NH$_{3}$}
\newcommand{\ammain}{$^{14}$NH$_{3}$}
\newcommand{\meth}{CH$_3$OH}
\newcommand{\methmain}{$^{12}$CH$_3$OH}
\newcommand{\methiso}{$^{13}$CH$_3$OH}
\newcommand{\cm}{cm$^{-3}$}

\newcommand{\kms}{km s$^{-1}$}

\begin{document}
\title{A Centimeter-wave Study of Methanol and Ammonia Isotopologues in Sgr B2(N): Physical and chemical differentiation between two hot cores}

\author{E.A.C. Mills}
\affil{Physics Department, Brandeis University, 415 South Street, Waltham, MA 02453}
\email{elisabeth.ac.mills@gmail.com}

\author{J. Corby}
\affil{Department of Physics, University of South Florida, 4202 East Fowler Ave, Tampa, FL 33605, USA}

\author{A.R. Clements}
\affil{Department of Chemistry, University of Virginia, Charlottesville, VA, 22904}

\author{N. Butterfield}
\affil{Green Bank Observatory, 155 Observatory Rd, PO Box 2, Green Bank, WV 24944, USA  }

\author{P. Jones}
\affil{School of Physics, University of New South Wales, NSW 2052, Australia}

\author{M. Cunningham}
\affil{School of Physics, University of New South Wales, NSW 2052, Australia}

\author{J. Ott}
\affil{National Radio Astronomy Observatory 1003 Lopezville Rd Socorro, NM 87801}

\begin{abstract}
We present new radio-frequency interferometric maps of emission from the \ammain, \amiso, and \deut\, isotopologues of ammonia, and the \methmain\, and \methiso\, isotopologues of methanol toward Sgr B2(N). With a resolution of $\sim3''$ (0.1 pc), we are able to spatially resolve emission from two hot cores in this source and separate it from absorption against the compact \hii\, regions in this area. The first (N1) is the well-known v = 64 \kms\, core, and the second (N2) is a core 6$''$ to the north at v = 73 \kms. Using emission from \amiso\, and hyperfine satellites of \ammain\, metastable transitions we estimate the \ammain\, column densities of these sources and compare them to those of \deut. We find that the ammonia deuteration fraction of N2 is roughly 10-20 times higher than in N1. We also measure an [\amiso/\ammain] abundance ratio that is apparently 2-3 times higher in N2 than N1, which could indicate a correspondingly higher degree of nitrogen fractionation in N2. In addition, we find that N2 has a factor of 7 higher methanol abundance than N1. Together, these abundance signatures suggest that N2 is a younger source, for which species characteristic of grain chemistry at low temperatures are currently being actively liberated from ice mantles, and have not yet reached chemical equilibrium in the warm gas phase. The high D abundance and possible high $^{15}$N abundance in \am\, found in N2 are interesting for studying the potential interstellar origin of abundances in primitive solar system material. 

\end{abstract}

\section{Introduction}
\label{int}

The Sagittarius B2 (Sgr B2) molecular cloud, with a mass of 3$\times10^6$ M$_{\odot}$, is the largest molecular cloud in the central 500 pc of the Galaxy, and one of the largest in the Galaxy as a whole \citep[e.g.,][]{Scoville1975}. The Milky Way's central black hole lies at a distance of $\sim$ 8.1 kpc \citep{Boehle16,Gravity18}, with Sgr B2 believed to be $\sim$130 pc from the center of the Milky Way and located slightly closer to the Sun due to the inclination of the central bar \citep{Reid09}. It is an active site of star formation, containing numerous signposts of this activity from masers \citep[e.g.,][]{Mehringer94,Reid88} and more than 50 compact and ultra-compact \hii\, regions \citep{Gaume95,dePree98, Zhao11}, most of which are clustered in two massive concentrations of star formation in the northern and middle part of this elongated cloud, which are commonly referred to as Sgr B2(N) and (M). In addition to \hii\, regions, each of these concentrations hosts hot molecular cores having warm temperatures \citep[100$\leq$ T $\leq$ 300 K;][]{Nummelin00,Pei00}, high densities \citep[$10^6 < $n$ <10^8$ \cm;][]{Lis90,Qin11}, and high gas phase abundances of complex organic molecules \citep{Miao95,Belloche08,Belloche13}. Sgr B2 is perhaps best known for its complex chemistry: of the more than 180 known circumstellar and interstellar molecules, half were first detected in Sgr B2 \citep{Snyder94}, and most of these were detected in line surveys toward the ``Large Molecule Heimat'' (LMH; where heimat means home in german) hot core in Sgr B2(N).

Because of its enormous column density \citep[N$_{\mathrm{H}2}$=$10^{24}-10^{25}$ cm$^{-2}$;][]{Qin11,Schmiedeke16} and rich chemistry, the LMH is one of the most well-known hot cores in the Galaxy. Indeed, this core exhibits the most spectral line-dense spectrum of any position in Sgr B2 \citep{Corby15}. While interferometric observations prior to the Atacama Large Millimeter Array (ALMA) typically only detected the LMH source, centered on Right Ascension = 17$^{\mathrm{h}}$47$^{\mathrm{m}}$19.9$^{\mathrm{s}}$, Declination = -28$\degr$22$'$19$''$ at v=63.5 \kms\, \citep{Vogel87,deVicente00}, a second line-rich, high-excitation (T$_{ex} > $80 K) hot core lies $\sim$6$''$ to the north, centered on Right Ascension =17$^{\mathrm{h}}$47$^{\mathrm{m}}$19.9$^{\mathrm{s}}$, Declination = -28$\degr$22$'$13$''$ at a velocity of 73 \kms\, \citep{Belloche13,Corby15,Halfen17}. This second core was first noticed in emission from several molecules, including ethyl cyanide \citep[CH$_3$CH$_2$CN;][]{Hollis03}, and the 44 GHz $7_0\rightarrow6_1 A^+$ transition of \meth, in which quasithermal emission is seen in both this source and the LMH \citep[sources h and i respectively;][]{MM97}. Following \cite{Belloche16}, we designate the LMH and the northern hot core as N1 and N2, respectively. Three other, less prominent hot cores in this region (N3, N4, and N5) have also been recently identified in ALMA line observations \citep{Bonfand17}.

At sub-millimeter wavelengths, the emission from Sgr B2(N) becomes dominated by thermal dust emission from N1 and N2 and their surroundings  \citep{Qin11,SM17}. However, at radio to millimeter wavelengths the continuum emission from Sgr B2(N) is dominated by seven compact, ultracompact, and expanding shell-shaped \hii\, regions \citep[K1-K7;][]{dePree98,dePree15}. Both N1 and N2 appear to have an embedded ultra- or hyper-compact \hii\, region \citep[K2 and K7;][]{Gaume95,dePree15}, although \cite{Bonfand17} suggest that the hyper-compact \hii\, region k7 may be separate from the N2 hot core.

From the sub-millimeter and molecular line emission, N1 and N2 have been measured to have masses ranging from 500 to 5000 M$_{\odot}$ \citep{Vogel87,deVicente00,Qin11}, sizes of 0.04 - 0.07 pc \citep[with no apparent fragmentation into smaller structures;][]{Qin11}, and densities greater than 10$^7$ \cm\, \citep{Lis90,MM97,Qin11}. Rotation temperatures from 70 to 500 K have been inferred from observations of complex molecules toward N1 \citep{Nummelin00,Pei00}. At the high densities inferred for these sources, most molecules should be thermalized and representative of the gas kinetic temperature. Molecular line observations toward Sgr B2 (N) also probe absorption at 64 and 82 \kms\, from material in the foreground of the \hii\, regions \citep[e.g.,][]{Huttem93a}. Unlike the cores, the excitation temperatures of the foreground material are low \citep[$\sim10 K$;][]{Zaleski13,Loomis13,Corby15}, indicating subthermal excitation in relatively diffuse gas. However, metastable ammonia transitions from high energy states \citep[up to the (14,14) transition;][]{Huttem95}, are also seen in the absorbing gas toward Sgr B2 (N), indicating that it may be hot: having temperatures from 150 to 600 K \citep{Flower95,Ceccarelli02,Huttem95,Wilson06}. Alternatively, it has been suggested that this apparently high temperature might be nonthermal in origin, similar to what is observed with H$_3$O$^+$. For this molecule, the apparently high temperature component is believed to be the result of formation pumping, resulting from the downward decay of a population of molecules that formed in highly excited states, and have not yet achieved thermal equilibrium \citep{Lis14}. 

Centimeter-wave observations are particularly important for characterizing sources with rich chemistry such as Sgr B2(N), as line blending and confusion become significant at millimeter and submillimeter wavelengths. This makes separating individual spectral features for analyzing kinematics and making spatially-resolved measurements of physical conditions difficult. Two important centimeter-wave probes are ammonia (\am), a symmetric top that is one of the more reliable probes of gas kinetic temperature, and methanol (\meth), a tracer of conditions in shocked gas and hot cores. Sgr B2 was one of the first interstellar sources observed in \am, and the cloud on large scales is an extremely strong source of emission \citep{Kaifu75,Morris83}. Interferometric observations have detected \am\, in absorption toward Sgr B2(N), however these lines of \am\, have not yet yielded robust temperatures in the cores, as they have only been mapped up to the (4,4) transition, and only N1 has been detected in emission in these lines \citep{Vogel87}. Nonmetastable \am\, lines up to (11,10) have been mapped in emission toward N1 by \cite{Huttem93a}, but at low resolution with a single dish (40$''$) that does not clearly localize this emission. Unlike the metastable lines, these lines have short decay times and so are only detected in the presence of a strong far-infrared radiation field or high densities sufficient to maintain the population of these levels \citep[][]{Sweitzer79}. \deut\, ($1_{11}-1_{01}$) has been previously detected in Sgr B2(N) by \cite{Peng93}, but the resolution of this observation  ($8''\times5''$) was insufficient to determine whether the emission originated from two separate sources. While N2 has not previously been detected in \am, both N2 and N1 have been seen in \meth. N1 has also been observed to be a strong source of emission in the $K = 2-1$ centimeter lines of \meth, first observed toward Sgr B2 by \cite{Menten86}. High-J \meth\, lines in this series have also been mapped in N1, in transitions up to $J$ = 20 \citep{Pei00}, yielding a measured rotation temperature of 170 K for this source. ALMA observations of isotopologues of \meth\, in N2 yield comparable rotation temperatures for N2 of 140-160 K \citep{Muller16}. 

We present new centimeter-wave interferometric ($\sim3''$) observations of Sgr B2(N) that spatially resolve emission from more than 30 transitions of \am\, and \meth\, isotopologues toward both N1 and N2, including \ammain\, (both metastable and nonmetastable transitions), \amiso, and \deut\, as well as \meth\, and \methiso. Maps and spectra of these species are presented in Section \ref{res}, from which we measure column densities and present the first characterization of the rotational temperatures of \ammain, \amiso, \deut, \methmain, and \methiso\, in both N1 and N2. We use these to make new estimates of abundances, including the deuteration fraction of \am\, in both cores. Finally, in Section \ref{dis}, we analyze several mechanisms for deuteration of species in Sgr B2, and contrast the physical and chemical properties of the spatially-resolved substructures in Sgr B2(N).

\section{Observations and Data Calibration}
\label{obs}

The observations presented in this paper include centimeter-wave spectral line data collected with the Karl G. Jansky Very Large Array (VLA), a facility of the National Radio Astronomy Observatory\footnotemark[1]\footnotetext[1]{The National Radio Astronomy Observatory is a facility of the National Science Foundation operated under cooperative agreement by AUI.}, the Robert C. Byrd Green Bank Observatory (GBT), and the Australian Telescope Compact Array (ATCA). All of these observations utilize the new-generation bandwidth capabilities of these instruments to achieve broad spectral line coverage. The observing setups and strategies are described below.

\subsection{VLA observations}

The primary observations presented in this paper were made with the VLA on January 13 and 14, 2012 as part part of a larger survey that covers the majority of the Sgr B2 cloud as well as several other Galactic center clouds, and is described further in separate papers \citep{Mills14,Mills15,Ludovici16,Butterfield18}. The calibration and imaging of these data are the same as described in \cite{Mills15}, and additional details of these procedures can be found there. 

The VLA observations were made with two different receivers (K and Ka bands), taking advantage of the large bandwidths afforded by the WIDAR correlator. Ka-band observations were made with a single correlator setup covering 27-28 and 36-37 GHz, and K-band observations were made with a single correlator setup covering 23.5-24.5 and 25-26 GHz. The spectral resolution was 250 kHz (2.7-3.1 \kms) for all lines but the \ammain\, (1,1) and (2,2) lines, for which the spectral resolution was 125 kHz (1.6 \kms) in order to resolve their hyperfine structure. These observations cover 16 lines of isotopologues of \am: the (6,6), (7,7), and (8,8) metastable transitions of \amiso, the (4$_{14}$-4$_{04}$) line of \deut, and the (1,1) through (7,7) and (9,9) metastable lines and the (10,9), (12,11), (14,13) and (19,18) nonmetastable lines of \ammain. Included as well are 16 lines of two \meth\, isotopologues: the J$_2$-J$_1$ transitions of $^{12}$\meth\, for J=6 through J=10, J=13, J=25 and J=26, and the J$_2$-J$_1$ transitions of \methiso\, for J=3 to J=10. Properties of all observed transitions are given in Table \ref{data}. We estimate that the flux calibration of these data is accurate to 10\%.

In this paper we focus only on the N1 and N2 cores of Sgr B2(N), both of which are entirely covered in a single pointing in both K and Ka bands. The Ka-band pointing was centered on (RA = $17^\mathrm{h}47^\mathrm{m}19^\mathrm{s}.00$, Dec = $-28\degr22'08''.18$), 15$''.$4 (13$''.$2) from N1(N2), with a primary beam size of 1$'.$25. The K-band pointing was centered on (RA=$17^{\mathrm{h}}47^{\mathrm{m}}18.87^{\mathrm{s}} $, Dec=$-28\degr22'28''.43$), 16$''$.5 (20$''$.8) from N1(N2), with a primary beam size of 2$'$. The total integration time at both frequencies was $\sim$25 minutes, yielding a typical per-channel RMS noises of 0.25-0.5 K. The data were taken in the hybrid DnC array configuration of the VLA, a configuration that yields a nearly circular beam shape given the low elevation of the Galactic center from the VLA site. The resulting angular resolution of the VLA images is $\sim2''.7$, which corresponds to a spatial resolution of 0.1 pc at the assumed Galactocentric distance of 8.1 kpc.

\subsection{ATCA observations} 

The VLA data were supplemented by ATCA observations of the 2$_{1 2}$-2$_{0 2}$ and 3$_{1 3}$-3$_{0 3}$ transitions of \deut. The data consist of a single pointing toward Sgr B2(N) and are part of a full 7 mm band survey from 30-50 GHz. The data utilized the broad band mode of the Compact Array Broadband Backend \citep[CABB;][]{Wilson11}, providing 1 MHz channel widths, equivalent to 6-10 \kms\, for this frequency range. Full details on the observing and data reduction strategies are provided in \cite{Corby15}. The 2$_{1 2}$-2$_{0 2}$ transition at 49.96284 GHz was observed on 21 Oct 2011 in the H75 array configuration, providing 11$''.2\times8''.5$ resolution; the 3$_{1 3}$-3$_{0 3}$ transition at 43.04243 GHz was observed on 4 Apr 2013 in the H214 array configuration, providing $4''.4\times3''.4$ resolution. Additional properties of these lines are given in Table \ref{data}. The flux calibration of these data is estimated to be accurate to 30\%.

\subsection{GBT observations}

Additional observations of \amiso\, were obtained from the publicly-available PRebiotic Interstellar MOlecular Survey (PRIMOS)\footnotemark[1]\footnotetext[1]{\href{http://www.cv.nrao.edu/~aremijan/PRIMOS/}{http://www.cv.nrao.edu/$\sim$aremijan/PRIMOS/}} survey, a legacy program of the GBT.
PRIMOS provides a nearly frequency-complete survey of Sgr B2(N) from 1-50 GHz towards a single pointing centered on N1. These data were mostly collected in the Spring of 2007, and details on the observing setup and data calibration are given in \cite{Neill12}. For our analysis we use the (1,1) through (6,6) lines of \amiso; the (6,6) line is then covered in both the GBT and the VLA observations, and can be used to determine a relative calibration offset between these observations. Further properties of these lines are given in Table \ref{data}. The selected transitions span the frequency range of 22.6-29.9 GHz, with corresponding beam sizes ranging from 31$''$.6 to 33$''$.4. As a result, this beam size covers core N2 as well, placing it at approximately the 93\% power point of the GBT beam. 
We estimate that the flux calibration of these data is accurate to 10\%. 

\section{Results}	
\label{res}

\subsection{Maps}

Using the VLA, we detect 33 lines of isotopologues of \meth\, and \am\, toward N1, and 26 lines toward source N2 (several lines of \ammain, \amiso, \methmain, and \methiso\, are not detected in N2, likely due to the lower column density in this source compared to N1). All detected transitions are listed in Table \ref{data}. In addition, toward both sources we detect 6 lines of \amiso\, (one of which is also detected with the VLA) and 2 lines of \deut\, in supplementary data from the GBT and ATCA. Below, we present further analysis of the spatial distribution of each isotopologue. We follow this with an analysis of the properties of N1 and N2 that can be derived from their source-averaged spectra. 

To examine the spatial distribution of the observed molecules, we have made channel maps showing both the emission and absorption that is present in this region (Figure \ref{channelmap}) as well as peak intensity maps (Figure \ref{ammomap} and \ref{methmap}) which emphasize the emission. We show these rather than maps of the integrated intensity that would be dominated by the absorption present in species like \am.

\subsubsection{Ammonia Isotopologues}
Emission from \ammain\, is extended over the Sgr B2(N) region in the (1,1) through (9,9) lines. The strongest emission is centered on N1. Absorption (due to the widespread distribution of \ammain\, and the low-energy of the metastable lines) is also seen against the compact and extended \hii\, regions in Sgr B2(N). Figure \ref{channelmap} shows the intermixing of emission and absorption in channel maps of the \ammain\, (7,7) line toward Sgr B2(N). Prominent absorption against the radio continuum sources including the compact \hii\, regions K1 and K3 is seen at velocities from 50 to 90 \kms. The ultracompact \hii\, regions K2 (at the center of N1) and K7 \citep[newly discovered to lie at the center of N2;][]{dePree15} are not resolved by these data and do not appear as absorption sources. N1 (with a central velocity of 64 \kms)is seen in nearly continuous emission at velocities from 22-104 \kms, however the emission in these channel maps at large velocities (v $<$ 50 \kms\, and v $>$ 90 \kms) actually originates from the hyperfine satellite lines of the (7,7) line. In contrast to the emission from N1, emission from the N2 core is only detected in the hyperfine satellite lines, as at the central velocity of N2 (73 \kms) the emission from \ammain\, is entirely overwhelmed by absorption.

\begin{figure*}
\includegraphics[scale=0.5]{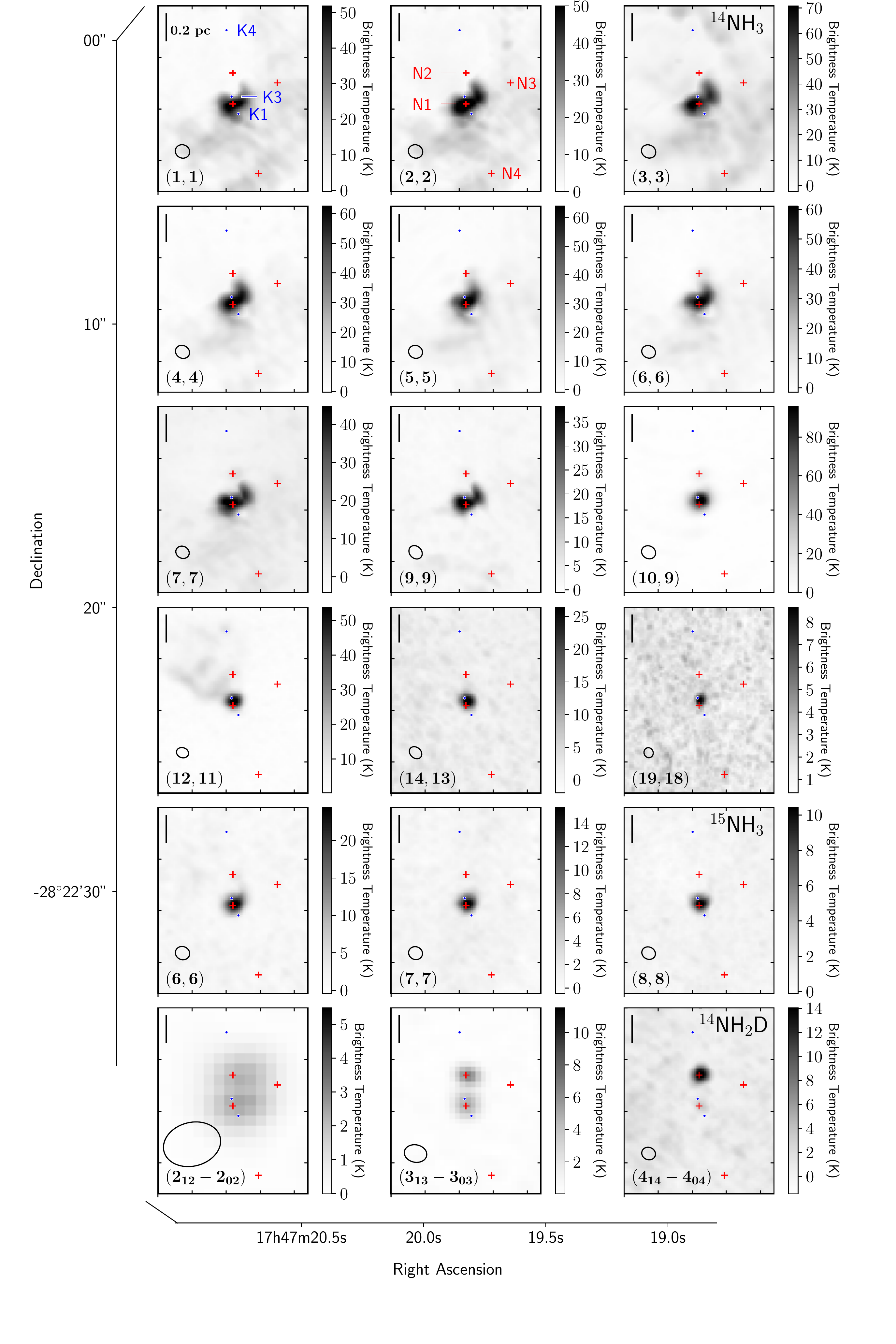}
\caption{Maps of the peak emission over all velocities from observed transitions of the \am\, isotopologues \ammain, \amiso, and \deut\, toward Sgr B2(N). The positions of the hot cores N1, N2, N3 and N4 are shown as red plus marks. The positions of the compact \hii\, regions K1, K3, and K4 are shown as blue points. All images shown in this figure are from the VLA observations except for the $2_{12}-2_{02}$ and $3_{13}-3_{03}$ lines of \deut, which are from observations with the ATCA. The beam size of each image is plotted in the bottom left corner, and exact spatial resolutions are given in Table \ref{data}.  A 0.2 pc scale bar is plotted in the top left of each panel.}
\label{ammomap}
\end{figure*}

Maps of the peak emission in each line are shown in Figure \ref{ammomap}. While these maps largely suppress the absorption features, areas of strong absorption can still be seen as regions where emission is absent at every velocity. The most prominent feature in these maps is emission from the N1 hot core. Emission from N1 and its environs is extended in all of the metastable \ammain\, lines, with a size of $\sim$0.3 parsecs. The emission is elongated from southeast to northwest, which is especially apparent in the (1,1) and (4,4) lines. In several lines, including (3,3), (7,7), and (9,9), the peak emission from N1 is offset to the east of the peak of submillimeter emission in the core, from \cite{Qin11}. The elongated appearance of this central emission is at least partly a result of absorption in this core against the compact \hii\, regions K3 (evident as an indentation in the emission on the northeast edge of N1 in all of these lines) and K1 (the silhouette of which is apparent on the southwest edge in the J $>$4 lines). Additional structure is also visible in the more extended \ammain\, emission surrounding N1: there is a bright ridge of emission to the northwest of N1 that extends diagonally up to the position of N2, which is especially noticeable in the (3,3), (7,7), and (9,9) lines. A southeast extension along the edge of the K1 absorption is also prominent in the (1,1) and (5,5) lines. In the J$<$7 metastable lines, very extended emission from the larger Sgr B2 cloud can also be seen primarily to the southwest of the hot cores. The extended emission is especially apparent in the (3,3) line, the brightest metastable line. In the peak emission map of this line, absorption can also be seen (though it is actually present in all of the metastable \ammain\, lines) toward the large, shell-shaped \hii\, region K5, lying to the northeast of N1. Although N2 is largely hidden by strong absorption from this superposed K5 shell, \ammain\, emission from N2 is detected to the north of N1 in the (7,7) and (9,9) lines, the first reported detection of N2 in \ammain. While N1 appears extended, N2 is largely unresolved by the VLA observations (having a size $<$0.1 pc). 

Unlike the metastable \ammain\, lines, the nonmetastable lines of \ammain, also shown in Figure \ref{ammomap} are largely confined to the N1 and N2 cores. Neither the northwest ridge seen in the metastable lines nor the extended emission from the Sgr B2 cloud are detected in this line. The nonmetastable \ammain\, emission from N1 is also more symmetrically oriented around the centroid of the submillimeter emission from \cite{Qin11}. Compared to the metastable lines, emission from N2 is much more prominent in the nonmetastable lines, which appear largely unaffected by the absorption seen in the metastable lines. Note that the (12,11) nonmetastable line is severely confused with H63$\alpha$ line emission from the \hii\, regions. 

Emission from \amiso\ is also detected in our VLA data toward both N1 and N2. As seen in Figure \ref{ammomap}, N2 is only clearly visible in the lowest excitation line that is observed, J,K= (6,6), while N1 is seen in emission in (6,6), (7,7) and (8,8). Faint emission from the northwestern ridge connecting emission from N2 and N1 can also be seen in the (6,6) line. Significant absorption is not seen in the \amiso\, line. In general, emission from N1 is much more compact and circularly symmetric in \amiso\, than in \ammain, and like the nonmetastable \ammain\, emission is better centered on the location of the submillimeter core. We note that the (3,3) line \amiso\, has previously been mapped with VLA toward Sgr B2(N), however it did not separately resolve the N1 and N2 cores \citep{Peng93}.

Finally, we have also mapped emission from 3 lines of \deut, shown in the bottom row of Figure \ref{ammomap}. Even more than \amiso, the  \deut\, emission is compact and centered on the two cores N1 and N2. Notably, in the 4$_{14}$ $\rightarrow$ 4$_{04}$ VLA data the N1 and N2 cores are clearly separated, and the line emission from source N2 is significantly stronger than that from N1. We supplement this VLA detection with observations of two lower-excitation lines of \deut\, from ATCA: 2$_{1 2}$-2$_{0 2}$ and 3$_{1 3}$-3$_{0 3}$. In these lower-resolution \deut\, observations, N1 and N2 are again the only sources of emission toward Sgr B2(N), and are not fully resolved. Despite their overlap, N1 and N2 appear to be comparably bright in these lower-excitation lines. The second source to the west of Sgr B2(N) seen in the lower resolution ($8''\times5''$) \citealt{Peng93} observations of the 1$_{1 0}$-1$_{0 1}$) line of \deut\, is not detected in any of these data, suggesting it may be spurious.

\subsubsection{Methanol Isotopologues}

\begin{figure*}
\hspace{0.6cm}
\includegraphics[scale=0.5]{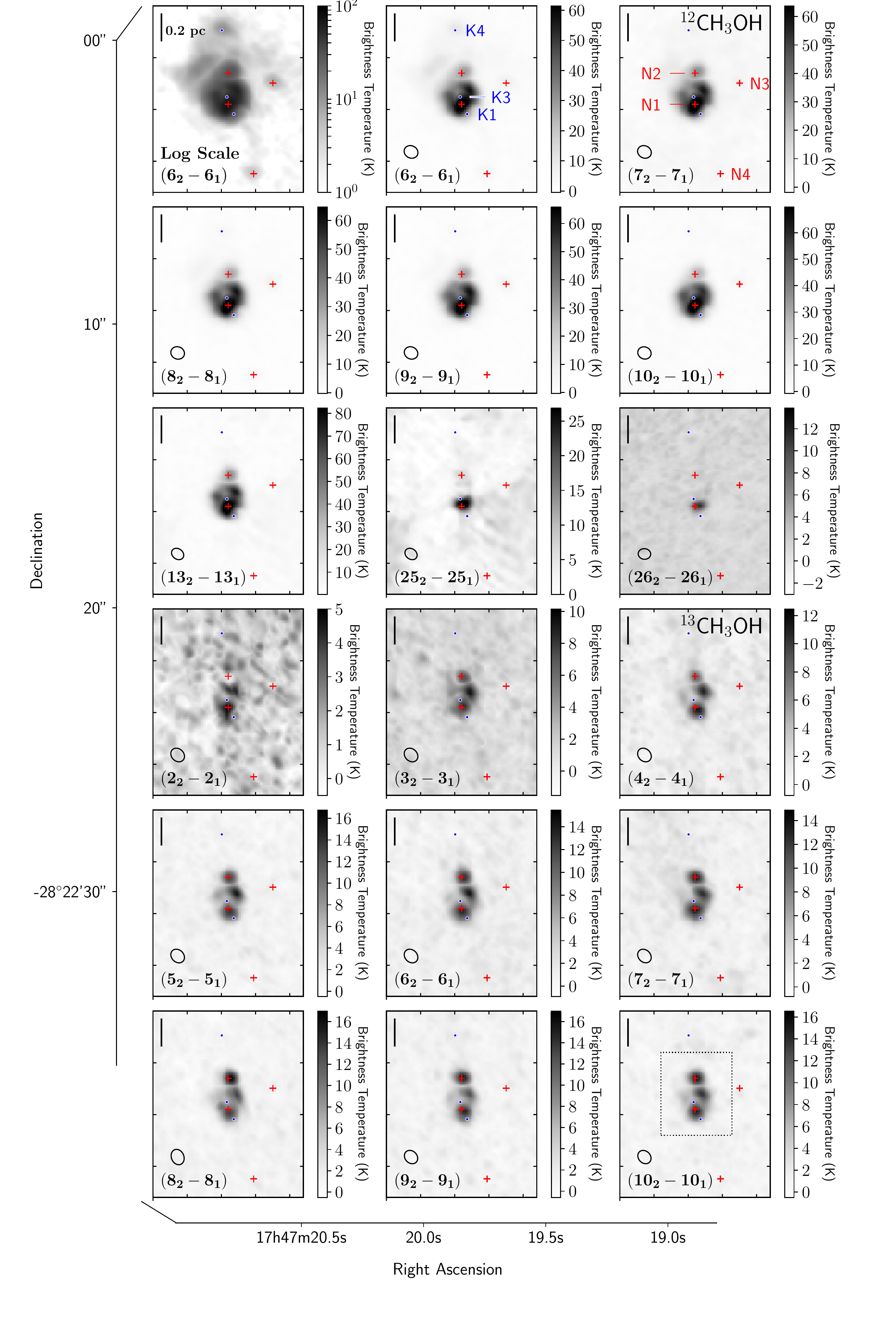}
\caption{ Maps of peak emission over all velocities from the observed transitions of the \meth\, isotopologues \meth\, (top) and \methiso\, (bottom) toward Sgr B2(N). The positions of the hot cores N1, N2, N3 and N4 are shown as red plus marks. The positions of the compact \hii\, regions K1, K3, and K4 are shown as blue points. The \meth\, ($6_2-6_1$) transition is plotted twice, first with a logarithmic scaling to highlight the fainter structure detected outside of the N1 and N2 cores. All images shown in this figure are from the VLA observations. The beam size of each image is plotted in the bottom left corner, and exact spatial resolutions are given in Table \ref{data}.   A 0.2 pc scale bar is plotted in the top left of each panel. The rectangular region marked with dotted lines in the last panel indicates the area shown in Figures \ref{schematic} and \ref{ratiomap}}
\label{methmap}
\end{figure*}

In addition to these three isotopologues of \am, we observe the J$_2$-J$_1$ ($\Delta$ K=1) transitions of two isotopologues of \meth: \methmain\, and \methiso. Emission from these species is shown in Figure \ref{methmap}. Compared to the observed \ammain\, transitions, the \methmain\, transitions observed in the Sgr B2(N) core appear to be significantly less affected by absorption from the K5 shell. The northeast edge of N1 is not seen in \ammain\, emission, but in \methmain\, it forms a nearly complete ring to the northeast of N1, surrounding the \hii\, region K3. Source N2 is less clearly separated from N1, and is much more prominent in \methmain\, than in \ammain, as it is not obscured by absorption. Further, unlike the maps of \ammain, which largely lack clumpy emission outside of N1 and N2, there is additional structure outside of the two primary hot cores apparent in the \methmain\, maps. This is best seen in a logarithmic stretch of the \methmain\, ($6_2-6_1$) line, shown in the first panel of Figure \ref{methmap}. There is a weak ridge of emission to the northeast of N1 and N2 (roughly aligned with the southwest edge of the K6 shell-shaped \hii\, region) that can be seen in the J=6,7,8, and 9 lines. There are also three additional cores. The first is $\sim8''$ north of N2, and is cospatial with the K4 compact \hii\, region. The second is a brighter core $\sim5''$ to the west of N2 and N1, which is identified as a hot core and designated N3 by \citep{Bonfand17} and which is seen in all lines of \methmain, but has no radio continuum counterpart. Like N2, this core is also associated with a rare formaldehyde maser \citep{Mehringer94}. Finally, we also detect weak \methmain\, emission from a core $\sim13''$ to the south of N1, which is also identified as a hot core by \citep{Bonfand17} and is designated as N4. 

In \methiso, Sgr B2(N) appears to fragment into three rather than two prominent structures. Sources N2 and N1 are still present, with N2 now comparably bright with N1, however the peak emission from N1 is slightly offset to the east of the submillimeter core centroid in the J$>$ 5 transitions. In addition, the northwest ridge structure that is seen in \ammain\, and which forms part of the ring of emission around K3 in \methmain\, appears in \methiso\, as a separate compact core in between N2 and N1 at a velocity of 64 \kms, which is roughly coincident with several dust continuum sources seen in the high-resolution ALMA continuum maps of \cite{SM17}. Weak \methiso\, emission is also seen to the northeast of K3, contributing to the ring of emission surrounding this absorption source which is also seen in \methmain. 

Comparing these observations with prior observations of this region we present an updated schematic of the structure of the Sgr B2(N) region in Figure \ref{schematic}. The positions of the compact \hii\, regions K1, K2, K3, and K7 are shown along with the positions of multiple dust continuum cores seen in high resolution Band 6 ALMA observations by \cite{SM17}, and an outflow detected in ALMA SiO data by \cite{Higuchi15}. \methiso\, emission is shown in the background of this figure, and an absence of emission at the positions of K1 and K3 can be seen. Emission from \methiso\, and other molecules in source N2 is relatively well-aligned with the dust continuum emission, though the poorer resolution of the VLA data does not allow us to definitively associate the line emission with a specific continuum peak (e.g., `a2' or  `a3') from \cite{SM17} , or to assess the claim of \cite{Bonfand17} that the N2 core may be separate from the K7 \hii\, region. However, the \methiso\, emission in N1 more clearly does not peak at the location of K2 / `a1', and instead peaks several arcseconds to the south (additional emission to the northwest corresponds with the continuum peaks, `a4' and `a5' identified by \citealt{SM17}). It is possible that this offset is due to the averaging of the emission and the unresolved absorption against K2. Alternatively, this offset could be real, and due to the association of the strongest \meth\, emission with the inner regions of the outflow coming from K2 / SMA 1 \citep{Lis93} rather than the actual hot core. 

\subsection{Spectra}

For the remainder of the analysis, we focus on the emission from the N1 and N2 hot cores, and the physical and chemical properties that can be derived for these sources. We first smooth all of the maps to a common resolution of $4''.4\times3''.38$, set by the ATCA \deut\, 3$_{13}$ $\rightarrow$ 3$_{03}$ observations (The ATCA 2$_{12}$ $\rightarrow$ 2$_{02}$ line observations have a much coarser spatial resolution, and so instead of degrading all of the data, we exclude this line from further analysis). We then extract spectra from $4''.4$ circular apertures centered on N1 and N2. In Figure \ref{spectra} we show the spatially-averaged spectra of all of the lines detected toward each source. In addition to these spectra from the VLA and ATCA maps shown in Figures \ref{ammomap} and \ref{methmap}, we also include \amiso\, spectra from the GBT (beam size $\sim 32''$) for a pointing centered on N1. As both N1 and N2 are observed in the same pointing, the spectra appear identical, apart from a slight scaling factor (1.075) applied to the spectrum of N2 to correct for its location in the sensitivity pattern of the GBT primary beam. The GBT spectra are subject to a larger amount of beam dilution than the VLA spectra, which we address when computing the column densities. We perform Gaussian fits to all the observed spectra, which are described in the Appendix.

\subsubsection{Ammonia Isotopologues}
\label{spec_amiso}

\begin{figure*}
\hspace{-1cm}
\includegraphics[scale=0.65]{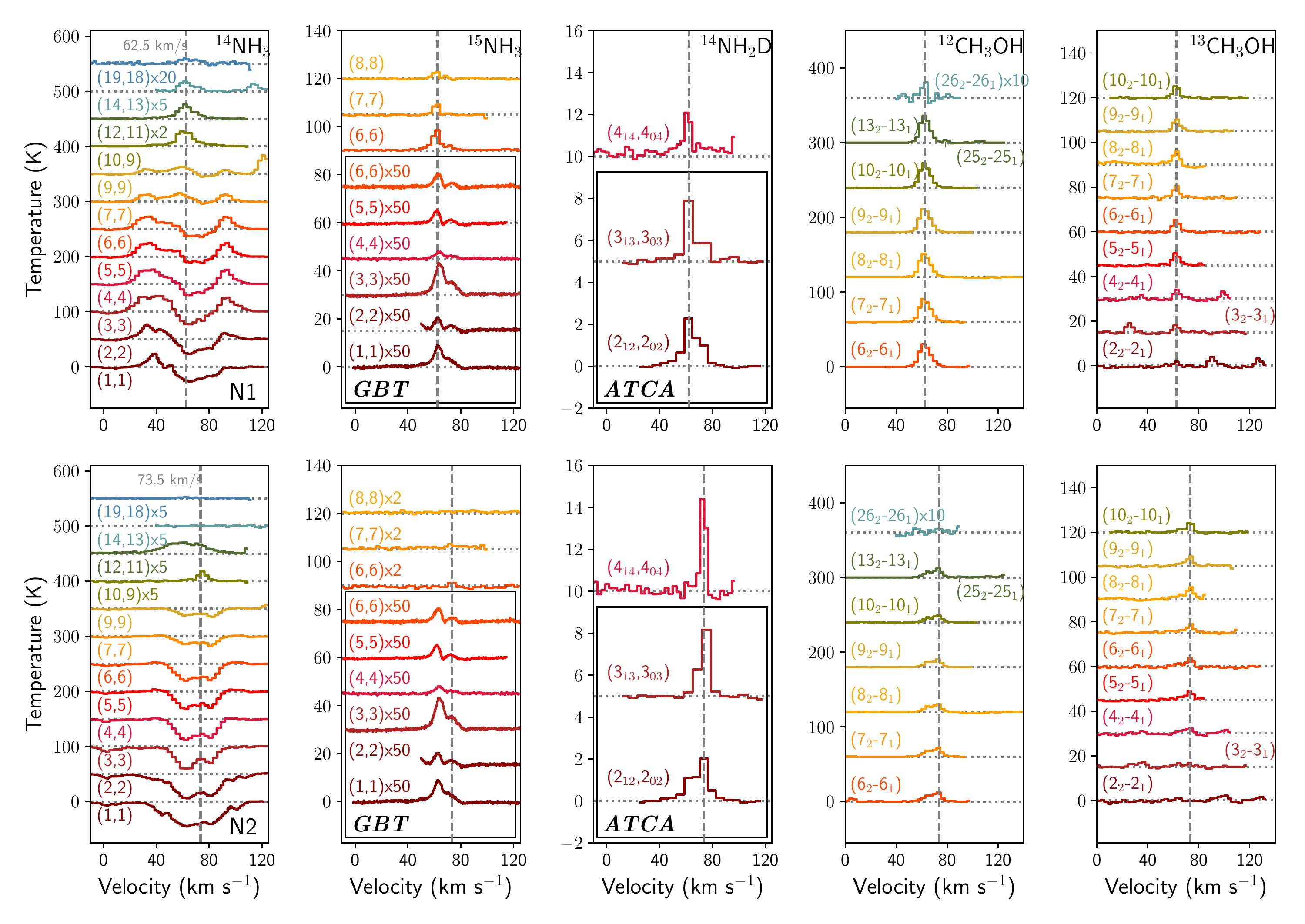}
\caption{Spectra of the observed transitions of \am, \amiso, \deut, \meth, and \methiso\, toward the two hot cores in Sgr B2(N): N1 ({\bf Top}) and N2 ({\bf Bottom}). From red to blue, colors correspond to transitions from low-$J$ to high-$J$ values. Spectra are from VLA observations unless otherwise indicated by a box showing spectra obtained from lower spatial-resolution observations with ATCA and the GBT. The GBT spectra for N1 and N2 are the same, as the observations are from a single pointing containing both sources that is centered on N1, with N2 at the 93\% power position. The spatial and spectral resolutions of all of the data are given in Table \ref{data}.}
\label{spectra}
\end{figure*}

Consistent with previous observations of \cite{Vogel87}, the metastable \ammain\, lines in N1 seen in Figure \ref{spectra} are overwhelmed by foreground absorption (likely local to the Sgr B2 cloud, given its similar velocity), meaning that for J$<$7, only the hyperfine satellites are seen in emission for these transitions. We also see that the emission from the hyperfine satellites becomes more symmetric in velocity with increasing J: In N1, the blue-shifted hyperfine satellites are significantly stronger than the redshifted satellites in the (1,1) and (2,2) lines, but are nearly symmetric for J$>$3. However, the absorption affecting the central (main) hyperfine line component remains asymmetric: the absorption appears shifted to more positive velocities with higher J, such that the lower-velocity side of the main line begins to appear in emission for J$>$3. 
Amazingly the hyperfine satellites are still detected in the \ammain\, (7,7) and (9,9) lines, despite the fact that the (9,9) satellite lines should be $>$250 times fainter than the main line. In N1, the absorption has diminished sufficiently in these transitions for the main, central component of these lines to be detected in emission, though this component is only slightly stronger than the hyperfine satellite lines.

\begin{figure*}
\includegraphics[scale=0.55]{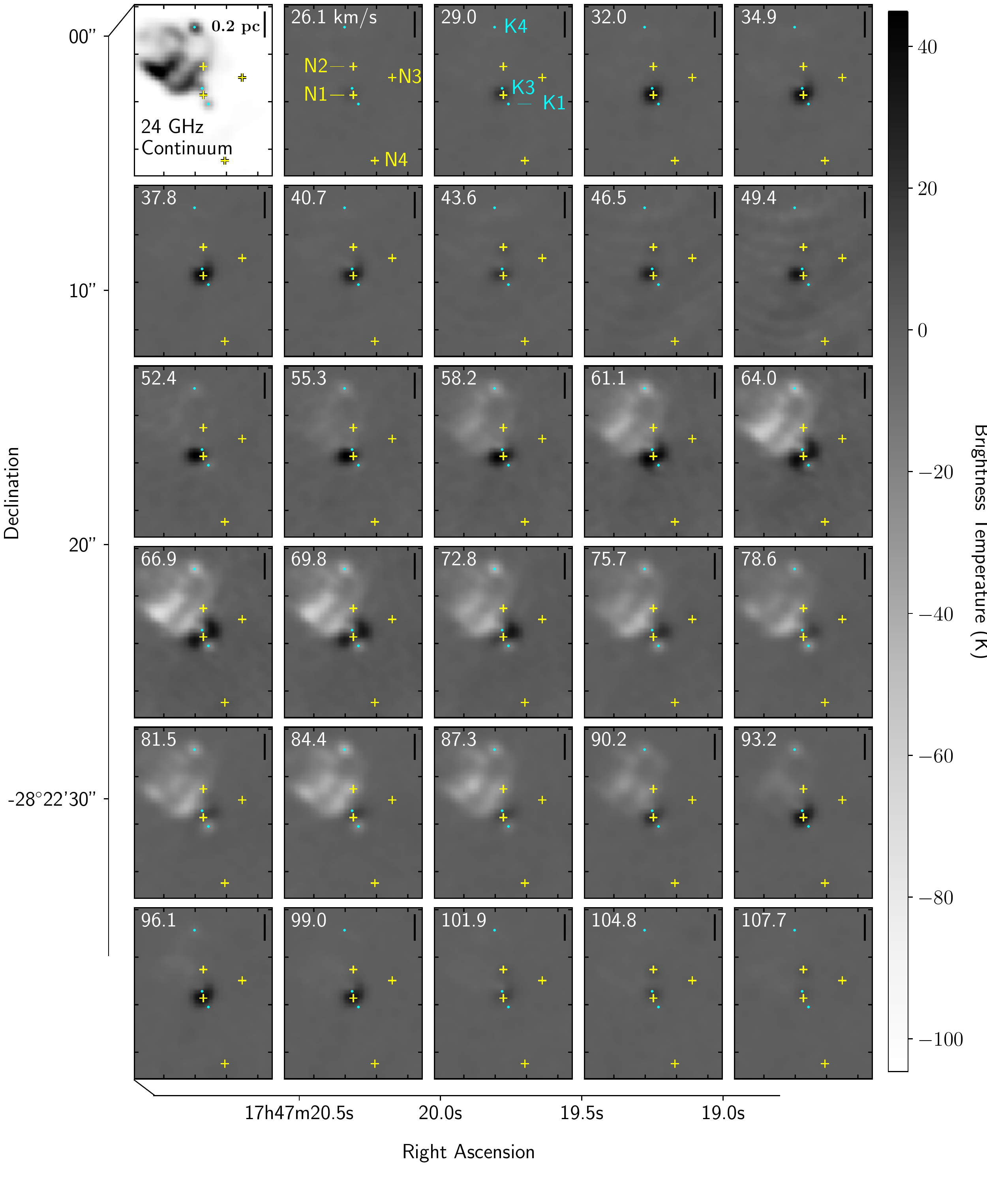}\\
\caption{Channel maps of emission from \ammain\, (7,7) toward Sgr B2(N). The top left frame shows 24 GHz continuum emission, to illustrate the structure of the ionized gas in the region. The positions of the hot cores N1, N2, N3 and N4 are shown as yellow plus marks. The positions of the compact \hii\, regions K1, K3, and K4 are shown as cyan points. A 0.2 pc scale bar is plotted in the top right of each panel. In the \ammain\, line the emission from N1 and N2 do not peak at their central velocities ( $\sim$63 and 73 \kms, respectively), but instead peak at velocities offset by $\pm$30\kms\, from the central velocity, at the location of the optically-thinner hyperfine satellite lines. Widespread absorption is seen against continuum emission from the `K' complex of compact and ultracompact \hii\, regions at velocities of $\sim$55-85 \kms. Semi-resolved absorption against the two compact \hii\, regions K1 and K3 is seen toward N1 at most of the velocities in this series of channel maps. The spatial resolution of these data is not sufficient to resolve the embedded hypercompact \hii\, regions K2 in N1 or K7 in N2, and thus no absorption against these features is detected.}
\label{channelmap}
\end{figure*}

In N2, the spectrum of the metastable \ammain\, lines is dominated by absorption in the beam toward the large, shell-shaped \hii\, region K5 (Figure \ref{channelmap}). No emission is detected at the central velocity ($\sim$74 \kms); any apparent `peak' at this velocity seen in the absorption spectra is likely because the central velocity of N2 is midway between the two absorbing components in the cloud at 64 and 82 \kms. However, emission in source N2 from the hyperfine satellites can be seen in the (5,5) through (7,7) lines.

Given the large column densities in Sgr B2(N), we suggest that even some of the hyperfine satellites of \ammain\, are likely to be optically thick. Toward N1, the hyperfine satellites in the (2,2)-(5,5) lines are observed to have roughly consistent brightness temperatures of $\sim$30 K ($\sim$65 K in unconvolved images with resolutions of $\sim2''.8\times2''.5$). However, the strength of these lines relative to the central peak should decrease significantly (from being 13\% of the central peak brightness in the 2,2 line to 1.6\% in the 5,5 line, assuming that the satellite lines are blended together), and for any thermalized, finite temperature, non-masing gas, the intensity of the central component should generally be decreasing as well (though the larger statistical weights of the $K=3n$ ortho-\am\, lines make these somewhat brighter). A large optical depth may also be responsible for broadening the satellite lines \citep[e.g., as discussed for CO;][]{Phillips79,Hacar16}: the blueshifted velocity extent of the spectra of Sgr B2(N1) is greater for the (2,2) through (6,6) lines than for the (7,7) and (9,9) lines, even though the velocity offset of the satellite lines from the line center should be (slightly) larger for these higher-J lines. If the satellite lines are indeed optically thick in N1, and \ammain\, is thermalized, the brightness temperature of the spectra from the unconvolved images would indicate that the gas kinetic temperature is greater than 65 K, with the value depending upon the filling factor of the emission (for example, if much of the emission originated in a region the size of the SMA source ($1''.72\times1''.28$), the kinetic temperature would be $\sim$200 K). Due to the substantial absorption seen in the \ammain\, spectra, we perform Gaussian fits to the hyperfine satellites in order to constrain the properties of the \ammain\, emission. The details of this fitting are given in the Appendix. 

In contrast to the \ammain\, observations, the spectra of \amiso\, are dominated by emission in both N1 and N2. In addition to VLA observations of the (6,6), (7,7), and (8,8) lines of \amiso, we include in our analysis observations of the (1,1) to (6,6) \amiso\, lines from the GBT PRIMOS survey. Unlike the VLA data, the GBT data are pointed spectra from a single position centered on N1 with a beam size of $\sim32''$. The large aperture of these observations means that emission from both cores is present in the blended spectra shown in Figure \ref{spectra}. Prior observations of the \amiso\, (3,3) line in Sgr B2(N) indicated that this line was marginally optically thick ($\tau = 1.2$), based on its measured absorption against one of the compact \hii\, regions \citep{Peng93}. Using detected absorption against K3 in our VLA map of the (6,6) \amiso\, line in N1, we measure $\tau<0.1$, indicating that this transition is optically thin. Ultimately, we do not attempt to correct any of the lines of \amiso\, for their optical depth, but we note that in subsequent derivations of column density and rotational temperature, the column densities of the J$\le3$ lines may be slightly underestimated.The GBT spectra exhibit significant variations in the shapes of the line profiles, with the contributions from N1 and N2 being less clearly separated in the (1,1) and (3,3) lines, possibly due to slight broadening due to increased optical depth in these lines toward N1. A more pronounced absorption dip is also seen in the (5,5) and (6,6) lines, compared to the lower-$J$ lines. 

Like \amiso\, the spectra of \deut\, are also dominated by emission in N1 and N2. The VLA observations of the $4_{14}-4_{04}$ line of \deut\, are supplemented with lower-resolution ATCA data of the 2$_{1 2}$-2$_{0 2}$ and 3$_{1 3}$-3$_{0 3}$ lines, the spectra of which also show blended emission from both N2 and N1. As previously mentioned, the $4_{14}-4_{04}$ line of \deut\, is notable for being the only observed line that is stronger toward N2 than toward N1. 

\subsubsection{Methanol Isotopologues}

As with the \amiso\, and \deut\, lines, a single \meth\, or \methiso\, spectrum centered on either N1 or N2 generally also contains weak emission at the velocity of the second hot core, due both to the large apertures from which spectra are extracted, and to the spatially extended \meth\, emission that is present in Sgr B2(N). We therefore follow the same procedure as for the \am\, lines and fit these with two Gaussian profiles. No significant absorption is seen in the \meth\, spectra, simplifying the line fitting process. The brightness of the \methmain\, lines in N1 with $J\leq13$ is nearly constant, at around 30-35 K ( $\sim$65 K in the unconvolved images, comparable to the brightness temperature of the hyperfine satellite lines of \ammain, which are suggested to be optically thick), while \methmain\, lines in N2 are roughly a factor of three fainter (12 K). The weak J=25 and J=26 lines of \methmain\, are not detected with any significance toward N2. Interestingly, lines of \methiso\, in N1 and N2 have much more similar brightness temperatures. As discussed further in Section \ref{isotop}, this indicates that in these transitions, the \methmain\, emission in N2, though appearing weaker than that in N1, is likely more beam diluted and actually significantly more optically thick.

\section{Analysis}	
\label{ana}

\subsection{Column Densities}

For \deut, \methiso, \amiso, \ammain, and \methmain\, (the latter two of which are likely to have some optically thick transitions) we compute the upper-level column densities $N_u$ of each transition as follows from the fitted line properties: 
	
\begin{equation}
N_{u} = \frac{8 \pi k \nu^{2}}{h c^3 A_{ul}} \int \frac{T_B}{f} dv. 
\label{eq1}
\end{equation}

\noindent{Here}, $k$ is Boltzmann's constant, $\nu$ is the line frequency, $h$ is Planck's constant, $c$ is the speed of light, A$_{ul}$ is the Einstein A coefficient of spontaneous emission for the transition, and $\int T_B dv$ is the integrated main-beam brightness temperature. $f$ is the filling factor of the observed emission, which \citep[adopting the SMA source size from ][ for N2, which is unresolved, and assuming N1 to be $2''.8\times2''.5$]{Qin11} is taken to be 0.114 (0.076) for N1 (N2) for the smoothed VLA observations, and $\sim$0.002 (0.0013) for N1 (N2) for the GBT \amiso\, observations. Exact values of $f$ as well as values of $\int T_B dv$ for each measured transition are reported in Tables \ref{linesN1} and \ref{linesN2}. For \ammain, the determination of column density is slightly more complex, for as previously noted, the central satellite lines are absorbed (and in addition likely substantially optically thick). For these lines, we multiply measurements of the blended hyperfine satellite line strength by the theoretical hyperfine intensity ratio (given in Table \ref{hfines}), in order to compute the expected main line intensity. The column densities are then computed as usual from Equation \ref{eq1}. 
For the observed inversion transitions of \ammain, \amiso, and \deut\, the upper- and lower-level column densities are added together to yield the total column density of the (J,K) rotational state. The resulting column densities for each state are also reported in Table \ref{linesN1} and \ref{linesN2}. All column densities are computed under the assumption that the lines are optically thin. Significant deviations from these assumptions (for example, for the \ammain\, satellites or \methmain), which are visible in Boltzmann plots of multiple transitions from each molecule are noted in Tables \ref{linesN1}, \ref{linesN2} and \ref{abund}

For several of the nonmetastable \ammain\, lines in N1 we are able to determine the column density of the lines directly from their optical depth. The hyperfine satellites of these lines are intrinsically extremely faint for the $J$ transitions we observe, however toward N1 we detect hyperfine satellites for both the (10,9) and (12,11) lines. These satellites do not appear to be detected toward N2. Extrapolating from the hyperfine line properties for J$\leq$8 given by the TopModel catalog on Splatalogue,\footnotemark[1]\footnotetext{\href{http://www.cv.nrao.edu/php/splat/}{http://www.cv.nrao.edu/php/splat/}} we predict that the blended hyperfine satellites of the (10,9) line should be 0.0061 times the intensity of the main line, and those of the (12,11) line should be 0.0043 times the intensity of the main line. Given that the measured intensity ratios of hyperfine to main lines are 0.128 for the (10,9) line and 0.055 for the (12,11) line (unlike the metastable transitions, there is no apparent absorption in the main line), we would then infer optical depths of 22.2 and 12.9, respectively, for these transitions. For these two transitions, we then measure column densities using Equation \ref{eq1}, as if these lines were optically thin, and then apply a correction factor for the opacity from \cite{Goldsmith99}:

\begin{equation}
N_{u} = N^{thin}_{u}\; \frac{\tau}{1-exp(-\tau)}.
\label{eq4}
\end{equation}

The remaining nonmetastable lines in N1 and N2 are assumed to be optically thin, and their column densities are measured using Equation \ref{eq1}.

\subsection{Rotational Temperatures}

\begin{figure*}

\includegraphics[scale=0.25]{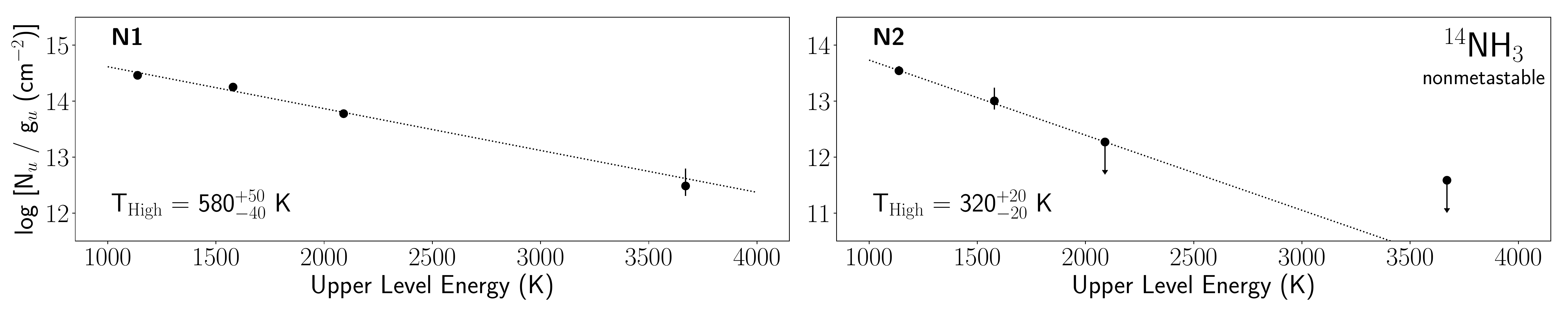}\\

\includegraphics[scale=0.25]{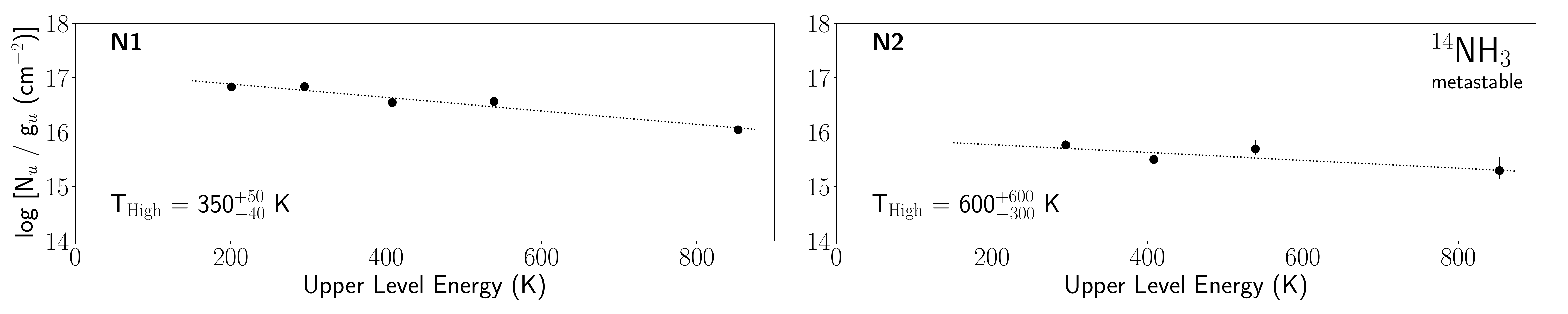}\\

\includegraphics[scale=0.25]{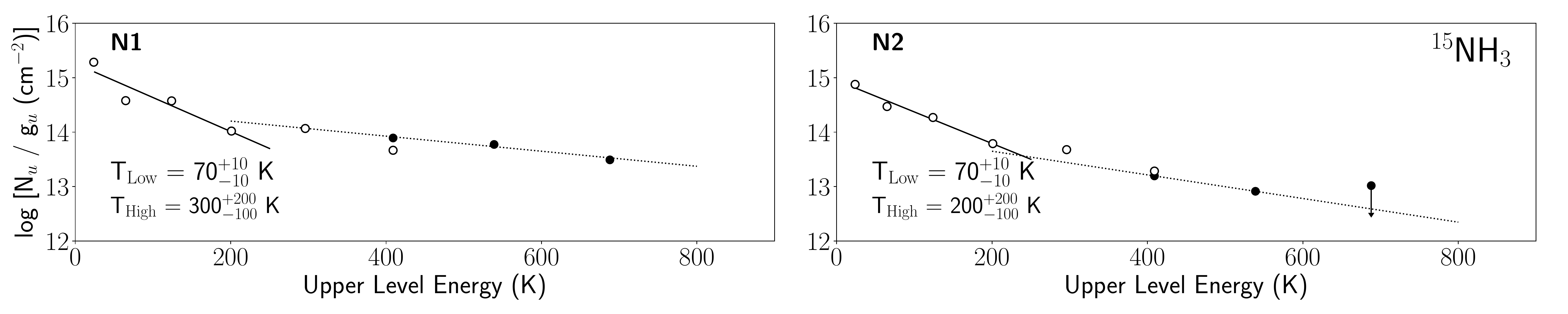}\\

\includegraphics[scale=0.25]{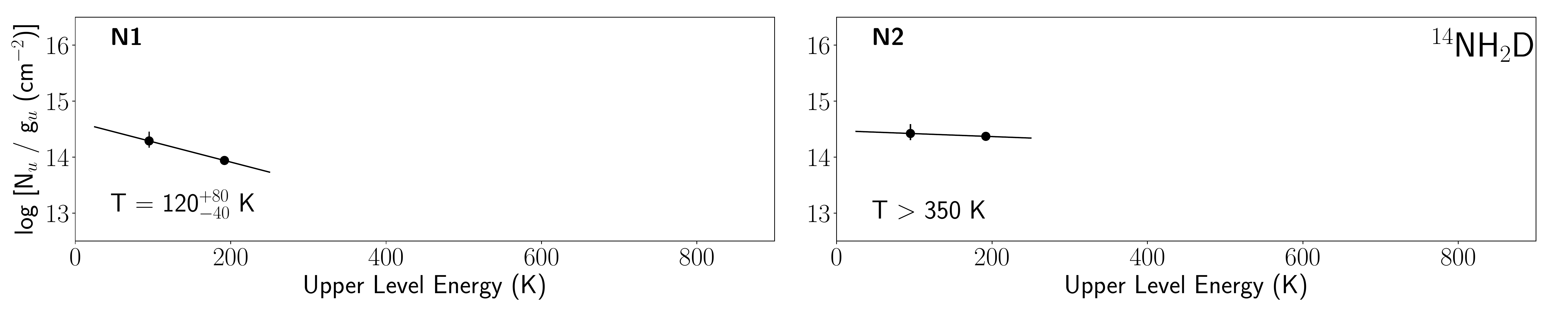}\\

\includegraphics[scale=0.25]{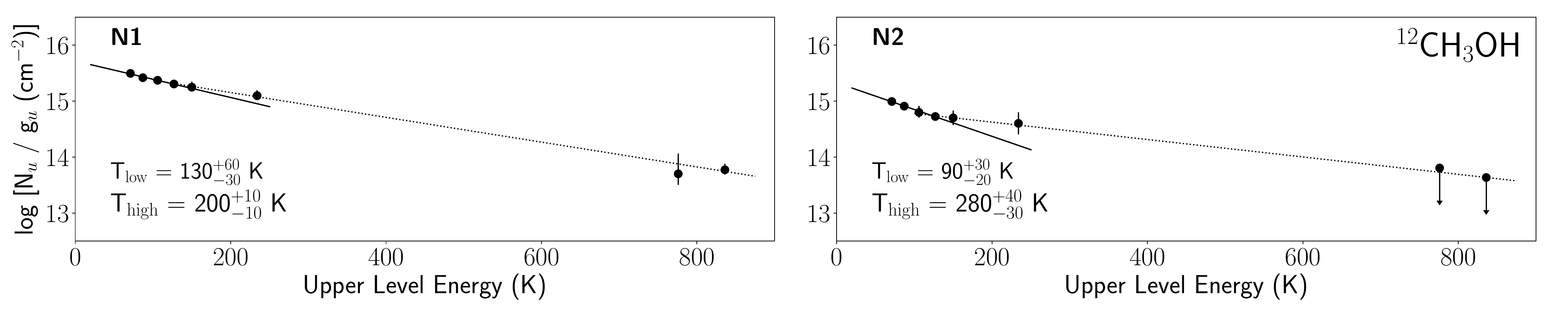}\\

\includegraphics[scale=0.25]{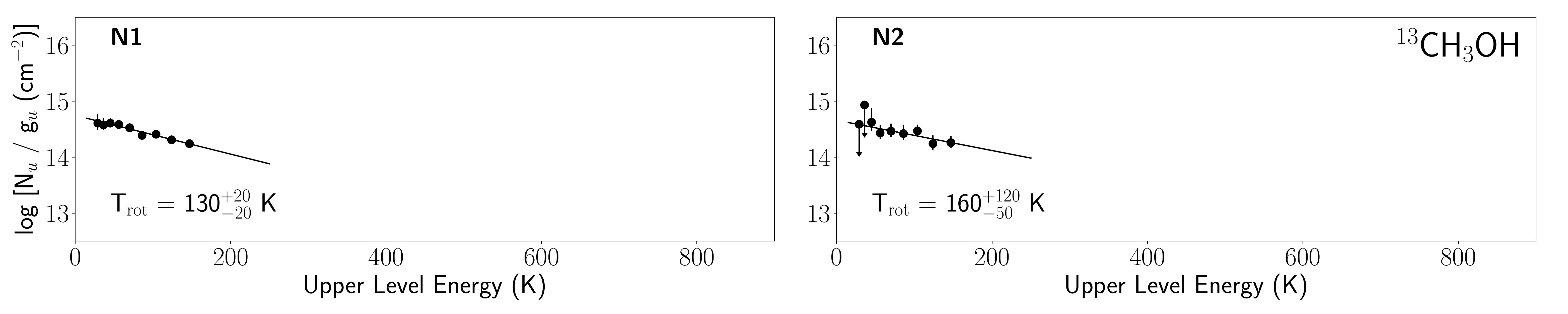}\\

\caption{Top to Bottom: Rotation temperature fits for non-metastable lines of \ammain, metastable lines of \ammain, \amiso, \deut, \methmain, and \methiso. All plots except those in the top row (non-metastable \ammain) are scaled to have an identical X-axis range and to have a Y-axis covering 4 orders of magnitude in column density, so that the temperature slopes derived for each molecule can be directly compared. For the \amiso\, population diagram, GBT data are shown as open symbols and VLA data as filled symbols. For all \am\, fits, the ortho to para ratio is fixed to be equal to its equilibrium value (2.0).} 
\label{plot}
\end{figure*}

After calculating column densities, we construct Boltzmann plots (or population diagrams) of \ammain, \amiso, \deut, \methmain, and \methiso. In these diagrams we plot log($N_u$/$g_u$), where $g_u$ is the total degeneracy of the observed state, including the $J$-dependent statistical weight, against the upper level energy E$_u$ of each transition (for the isotopologues of \am, we plot N$_J$/g$_u$). The adopted g$_u$ values are given in Tables \ref{linesN1} and \ref{linesN2}, and are taken from the JPL Molecular Spectroscopy Catalog \citep{Pickett98}. Technically, the $K\ne3$ (para) and $K=3$ (ortho) lines of \ammain\, and \amiso\, behave as separate species, however we fit these together assuming a standard ortho-to-para ratio of 3:1. Fitting a straight line in the Boltzmann plot then yields the rotational temperature of the species. Multiple slopes can be present in a Boltzmann diagram due, for example, to the presence of multiple kinetic temperature components in the gas, or to the subthermal excitation of the molecule. The low dipole moment of molecules like \ammain\, (1.42 D) and \methmain\, (1.69 D) indicate that they should be easily thermalized, especially in the high-density environment of the Sgr B2(N) cores. Further, for \ammain\, and \amiso\, which are symmetric tops, the rotational temperature should be a good approximation of the kinetic temperature of the gas.The presence of multiple slopes in Boltzmann plots of \am\, in Galactic center clouds like Sgr B2 are thus generally interpreted as being due to multiple gas temperatures, and we fit for two temperature components in both \meth\, and \am\, where it appears warranted. The Boltzmann plots for all species are shown in Figure \ref{plot}, and the rotation temperatures derived from fits to these plots are given in Table \ref{rot}.

\subsubsection{\ammain, \amiso, and \deut\, Temperatures}
\label{amtemp}

For N1, the (1,1) through (3,3) satellite lines of the metastable lines of \ammain\, do not have Gaussian shapes, due to absorption, optical thickness, or both. We do not perform fits to these lines. For N2 the hyperfine satellites do not even appear above the absorption until J,K=(5,5), so these lower-excitation lines are also not included in the fits. For both N1 and N2, the remaining metastable lines all have E$_{up}>$ 200 K, and we fit these with a single temperature component, finding T$_{rot}=350^{+50}_{-40}$ K for N1 and T$_{rot}=600^{+600}_{-300}$ K for N2. This is consistent with T$_{rot}\gtrsim$ 300 K for both N1 and N2. 

For the nonmetastable ammonia lines in N1, we measure temperatures of T$_{rot}=580^{+50}_{-40}$ K for N1 and T$_{rot}=320^{+20}_{-20}$ K for N2. This is the opposite of what is inferred from the metastable lines, where N2 appears warmer than N1. However, as we discuss further in Section \ref{density}, these transitions can be radiatively excited, and so this temperature may not reflect the gas kinetic temperature. This temperature may also be subject to additional uncertainty as the lines may not arise from an identical volume, due to the strongly increasing excitation requirements for progressively higher-$J$ transitions. 

Our observed transitions of \amiso\, span a large range of upper level energies ($\sim$24-700 K), and so we fit two temperature components to the column density data in the Boltzmann plots for \amiso: a cold and a hot component. Unlike the \ammain\, observations, the \amiso\, observations incorporate data from two telescopes: the VLA and GBT. When determining column densities, we adopt a filling factor for the GBT data that assumes that the bulk of the detected emission originates in the compact cores seen by the VLA. However as the GBT is a single-dish facility, these observations may also be sensitive to large-scale flux in the beam that is resolved out by VLA observations. We can assess the magnitude of this effect by comparing \amiso\, (6,6) emission from both the VLA and GBT, finding that the adopted filling factors yield good agreement between column densities for N2. However, the GBT column densities for N1 appear lower than expected, which cannot be explained by missing flux in the VLA observations. While one possibility is that there are additional uncertainties in subtracting complex baseline profiles in the GBT data, it seems likely that the (5,5) and (6,6) \amiso\, GBT spectra are affected by additional absorption in the larger beam (as noted in Section \ref{spec_amiso}), effectively hiding some of the emission that should be present. In the end however, any uncertainty in the relative flux scaling largely does not affect the temperatures we derive, as the cold-component temperatures is largely determined from the lowest-excitation GBT lines, and the hot-component temperature is largely determined from the higher-excitation VLA lines. 

Toward N1, we measure a cold component temperature of 70$^{+10}_{-10}$ K, and a hot-component temperature of 300$^{+200}_{-100}$ K. Toward N2, we measure a temperature of 70$^{+10}_{-10}$ K for the cold component, and a temperature $\sim$ 200$^{+200}_{-100}$ K for the hot component. Thus, the temperature of the cool component appears identical in both sources (though as we discuss in Section \ref{temps}, this `cool' component may not be a physically meaningful temperature), but the warmer component is slightly warmer in N1 than in N2, at odds with the larger temperature measured for N2 with the metastable lines of \ammain. 

Our \deut\, Boltzmann plots also incorporate data from multiple telescopes: \deut\, observations were made with both the VLA and ATCA. As we do not have any lines observed in common between the two telescopes, we cannot estimate any systematic difference in the flux calibration or the recovery of large scale structure between these data sets, and so the absolute temperatures derived from these data will be uncertain. Additionally, as the 3$_{1 3}$-3$_{0 3}$ and 4$_{1 4}$-4$_{0 4}$ lines are of the ortho-\deut\, species, while the 2$_{1 2}$-2$_{0 2}$ line is of the para-\deut\, species, we also cannot include the single para-\deut\, line in a temperature fit without a priori knowledge of the ortho-to-para ratio. As a result, we conduct our initial temperature fitting for only the two ortho-\deut\, lines. Because we are only concerned with the 3$_{1 3}$-3$_{0 3}$ and 4$_{1 4}$-4$_{0 4}$ lines for this initial fitting, we only need to smooth the 4$_{1 4}$-4$_{0 4}$ line to the resolution of the 3$_{1 3}$-3$_{0 3}$ line ($\sim4.4''$). From these fits, we find that the \deut\, in N2 has a higher rotational temperature ($>$350 K) than in N1 (120$^{+80}_{-40}$ K). While these uncertainties include estimates for the absolute calibration uncertainty of the ATCA (3$_{1 3}$-3$_{0 3}$) and VLA (4$_{1 4}$-4$_{0 4}$) data, it is possible that a larger offset in the relative flux calibration could exist. However, the significant temperature differential seen between N1 and N2 should be independent of this. 

Even with this large uncertainty, the lower limit on the temperature that we infer for \deut\, toward N2 is surprisingly high. We note that the 4$_{1 4}$-4$_{0 4}$ line is unlikely to be a maser, as its line width is not measurably narrower than other lines detected toward N2 with the VLA. Another possibility is that \deut\, is not in thermal equilibrium in this source. This could occur if (1) the \deut\, formed recently, not allowing time to reach thermal equilibrium and (2) it formed in a highly-excited state. However, as most processes suggested to cause (2) -- e.g. formation pumping via an exothermic reaction \citep{GonzalezAlfonso13,Lis14}, invoke gas-phase formation, this seems fairly unlikely. We discuss the relation between this temperature and mechanisms behind the observed \deut\, abundance in N2 in Section \ref{evo}. 

\subsubsection{\methmain\, and \methiso\, Temperatures}

Our \meth\, and \methiso\, observations were made entirely with the VLA, and largely in a single correlator setting, so there is much less concern about variations in the absolute flux scaling between the different observed transitions. Similar to \deut, the energies of the observed transitions of \methiso\, span a relatively small range of energies ($\sim$40-150K), and so we again fit only a single temperature component in the Boltzmann plot. We assume that these transitions are optically thin, which seems largely borne out by the ability to fit nearly all of the observed lines with a consistent temperature. The rotational temperatures of N1 (130$^{+20}_{-20}$ K) and N2 (160$^{+120}_{-50}$ K) in \methiso\, are not statistically distinguishable given the measurement uncertainties, particularly the larger scatter apparent in the \methiso\, columns derived for N2. %

For \methmain\, the observed transitions span a similar range of energy to the \amiso\, lines ($\sim$70-840 K), and so we fit these as well with two temperature components. However, given the significant optical depths in the majority of the observed \methmain\, lines (discussed further in Section \ref{isotop}), the low-temperature component determined from the Boltzmann plots may be significantly modified by the high optical depth of these transitions. The exact manifestation of this optical thickness is not immediately clear: \cite{Goldsmith99} show that for typically observed methanol transitions, large column densities will lead to a wide range of optical depths, and increased scatter in a rotation diagram. For the higher-excitation lines (J=13,25,26) we have no measurement of the optical depth. If the higher-J lines become optically thin, then these temperatures may better represent the core conditions. For these lines, we measure temperatures of T$_{rot}=200^{+10}_{-10}$ K for N1 and T$_{rot}=280^{+40}_{-30}$ K for N2. 

In general, we do not measure a consistent rotational temperature for either N1 or N2 for all of the observed transitions. This is not entirely surprising. All of these temperatures are rotational temperatures, and so may underestimate the kinetic temperature by various degrees (though as discussed earlier, given the high densities of the cores, most molecules should be thermalized). Maps of the emission from these molecules also show different spatial extents, suggesting that they could also probe different resolved (or unresolved) chemically-distinct layers in cores having strong temperature gradients. Finally, without direct measurements of the opacities of most of the observed transitions, it is possible that there may be residual optical depth effects. We will discuss these possibilities further in Section \ref{compare}.

\subsection{Column Densities and Abundances}
\label{column}

Using the measured rotation temperatures, we can determine the total column density for each species:

\begin{equation}
N_{T} = \frac{N_u}{g_u}Q(T_{rot})e^{\frac{E_u}{kT_{rot}}}
\label{eq2}
\end{equation}

\noindent{Here}, E$_u$ is the upper level energy in K, $g_u$ is the total degeneracy of the observed state, T$_{rot}$ is the rotational temperature for each species, as measured from fits to the Boltzman plots, and $Q_{rot}$ is the partition function of the molecule. For consistency, all $g_u$ values are obtained from the JPL Molecular Spectroscopy Catalog \citep{Pickett98}. The $Q_{rot}$ values are compiled from a variety of sources. For \ammain\, the partition function used was from the JPL Molecular Spectroscopy Catalog, where it was provided by Yu, Drouin, and Pearson (2010). For \amiso\, and \deut\, the partition function was taken from the Cologne Database for Molecular Spectroscopy \citep[CDMS;][]{Muller01,Muller05}. For \methmain\, and \methiso, the partition function was calculated using the rigid rotor approximation for a nonlinear molecule \citep{Herzberg45}:

\begin{equation}
Q = 2 \left(\frac{\pi\, (k\,T_{rot})^3}{ h^3\, c^3\, ABC}\right)^{0.5}
\label{eq3}
\end{equation}

where the rotational constants A, B, and C for \methmain\, and \methiso, were taken from \cite{Xu97}. The rotational temperatures adopted for each column density calculation are given in Table \ref{abund}, along with the computed column densities for each species for both N1 and N2.

For \ammain, we only detect a high-temperature component in both N1 and N2. As the temperature measured in N2 has a relatively large uncertainty, and the partition function becomes inaccurate above 300 K, we adopt an excitation temperature of 300 K for both. We then calculate $N_{T}$ for this component using $N_u$ from the (6,6) line. The column density of this component in N1 ($6\pm1\times10^{19}$ cm$^{-2}$) is roughly an order of magnitude higher than that measured for N2. 

For \amiso\, we detect both a high and a low temperature component in both sources. For both N1 and N2 we calculate $N_{T}$ for the low-excitation component using an excitation temperature of 75 K and $N_u$ from the (1,1) line. As with \ammain\, we calculate $N_{T}$ for the high-excitation component using $N_u$ from the VLA observations of the (6,6) line, and adopting a excitation temperatures of 300 K for both N1 and N2 (the measured rotation temperature in N2 is slightly lower, but is consistent with 300 K within the uncertainties, so we adopt this value for consistency with our other measurements). $N_{T}$ for \amiso\, is then the sum of these two components. We find a total \amiso\, column density in N1 of $3\pm1\times10^{17}$ cm$^{-2}$, roughly three times the total column density for this species in N2. 

We calculate $N_{T}$ for \deut\, using $N_u$ from the $J=3$ line observed with ATCA. For N1 we use an excitation temperature of 120 K, and for N2 we adopt an excitation temperature of 300 K (the maximum temperature for which the adopted partition function is valid). However the total column density is relatively insensitive to this choice of temperature, and only increases by 20\% if we extrapolate the partition function to 350 K. The \deut\, column density in N1 is $4.3\pm1.5\times10^{17}$ cm$^{-2}$, comparable to the \amiso\, column density in this source. However, unlike \amiso, the \deut\, column density in N2 is not smaller than that in N1, but is instead more than three times larger: $\sim1.4\pm0.5\times10^{18}$ cm$^{-2}$. We note that while we assume that the measured column densities here are the total \deut\, column densities, it is possible that (as the observed transitions are relatively low-$J$), there may be also be a high-temperature component like that seen with \amiso\, in higher $J$ transitions. If this were the case, the inferred \deut\, column densities for these sources would be underestimates. 

As the \methmain\, transitions appear to be significantly optically thick in both N1 and N2 (see Section \ref{isotop} below), we do not report a column density for this species derived directly from these lines. 
For the \meth\, isotopologues, we assume all transtions are well characterized with a single excitation temperature (as with \deut\, it may however be that we are missing a higher-temperature component like that seen in $J>$10 transitions in \methmain). For N1 we adopt an excitation temperature of 130 K and calculate $N_{T}$ for \methiso\, using $N_u$ from the $J=6$ line. For N2 we adopt $T_{rot}$=180 K, and again use $N_u$ from the $J=6$ line to determine N$_{T}$. The derived total column densities scale roughly proportionately with the adopted temperature. For example, setting the excitation temperature in N1 to 120 K (the temperature of \deut\, in this source) decreases the inferred column density by only 10\%. Setting the excitation temperature in N2 to 200 K, the temperature measured with \amiso\, in this source increases the column by 35\%.

\subsubsection{[$^{12}$C/$^{13}$C] and [$^{14}$N/$^{15}$N] Isotope Ratios}
\label{isotop}

If the observed isotopologues of \meth\, and \am\, are optically thin, then we can use these lines to measure the intrinsic isotope ratios from these species. We first measure the [$^{14}$N/$^{15}$N] isotope ratio using our \am\, observations. For this comparison we use only the high-excitation component seen in both \ammain\, and \amiso, as low-excitation component is not well detected in \ammain. We measure [$^{14}$N/$^{15}$N] to be $\sim$450 for N1 and to be $\sim$200 for N2. Note that the isotope ratio derived in this way could be different in the low-excitation component if there is some kind of selective fractionation, however we will assume for the purposes of this analysis that this ratio applies to both the low- and high-temperature \am. For N1, this is consistent with prior limits previously given in the literature for the Galactic center \citep[$400-600$;][]{WR94,Wannier81,Peng93}, and larger than extrapolation from a radial Galactic trend would suggest \citep[$124\pm37$;][]{Adande11}. The value for N2 could be considered to be more consistent with these latter two estimates. However, this could also be the result of a systematic error, either because (1) even these hyperfines are optically thick, as we suggest is true for the lower-$J$ hyperfine satellites in N1 or because (2) the absorption in the beam is hiding emission from the hyperfine satellites. Both of these scenarios would reduce the measured \ammain\, column, and could also lead to the apparently high \ammain\, rotational temperature observed for this source. We discuss further in Section \ref{deutdeut} what differences in [$^{14}$N/$^{15}$N] between N1 and N2 could mean, if real. 

\begin{figure*}
\includegraphics[scale=0.5]{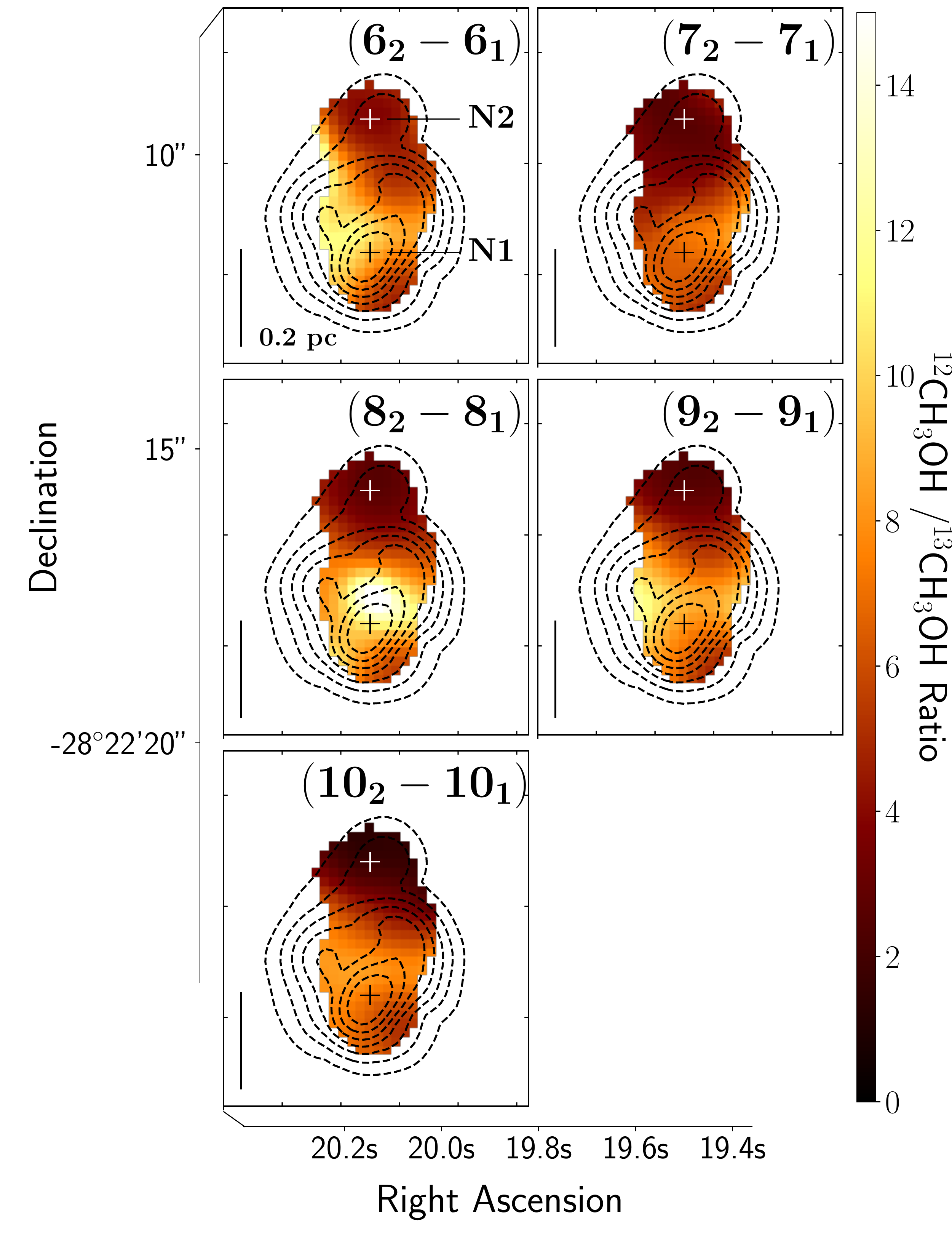}
\caption{Ratio maps for identical transitions of $^{12}$\meth\, and \methiso\, with contours of \meth\, (10$_2$-10$_1$) overlaid. Maps are convolved to identical resolutions and pixelizations prior to taking the ratio. For optically-thin emission, the ratio should reflect the $^{12}$C/$^{13}$C ratio of 25 in the Galactic center. Smaller values of this ratio correspond to large optical depths. A 0.2 pc scale bar is plotted in the bottom left of each panel.}
\label{ratiomap}
\end{figure*}

Applying the same procedure to our observations of \meth\, isotopologues yields apparent [$^{12}$C/$^{13}$C] ranging from 8 to 14 in N1 and from 2 to 4 in N2, as shown in Figure \ref{ratiomap}. As the [$^{12}$C/$^{13}$C] is well known to be 25 both in the Galactic center as a whole \citep{WR94,Riq10} and in Sgr B2 in particular \citep{Muller16,Halfen17}, this indicates that these lines are significantly optically thick. Adopting this ratio of 25, we can then use these measurements to calculate the optical depth of the \methmain\, transitions. We find that for N1, this corresponds to optical depths of 1.3 to 3. In contrast, the smaller values of this ratio in N2 correspond to much higher optical depths of 7-17. Although the observed \methmain\, optical depth in N1 is significantly lower than the average optical depth in N2, the column densities we measure for lines of \methiso\, in N2 are generally smaller than or comparable to those in N1. This suggests that the dominant source of \meth\, emission in N2 is more compact than the ($\sim2''.5\times3.0''$) VLA beam. Such a compact source size is consistent with the modeled source size of $\sim1''-1''.4$ for N2 in a variety of molecular lines \citep{Belloche16,Muller16}, and with the deconvolved source size of $1.44\times1.02''$ inferred for the submillimeter emission in N2 that is smaller than the inferred source size for N1 \citep[$1''.72\times1''.28$;][]{Qin11}. 

Isotope ratios of [$^{12}$C/$^{13}$C]= 25 and [$^{14}$N/$^{15}$N] = 450 (for both N1 and N2) and 200 (as an alternative value for N2) are then used to report total column densities for \ammain\, and \methmain\, in Table \ref{abund}. 

\subsubsection{Abundances of \methmain\, and \ammain\,}

The molecular column densities of each species in N1 and N2 are then compared to the total H$_2$ column density in each source in order to determine the molecular abundances. Values for the total H$_2$ column density in N1 and N2 are taken from the submillimeter dust continuum observations of \cite{Qin11}, who find that the source-averaged molecular gas column density (N$_{\mathrm{H}2}$) of N1 and N2 are 4.5$\times10^{25}$ cm$^{-2}$ (deconvolved source size = $1''.72\times1''.28$ / $0.07\times0.05$ pc) and $1.5\times10^{25}$ cm$^{-2}$ (deconvolved source size = $1''.44\times1''.02$ / $0.06\times0.04$ pc), respectively, based on the dust emission at 850 $\mu$m. The total column densities from \cite{Qin11}, and thus also the derived abundances, are then dependent upon the assumptions that the dust emission is optically-thin, the dust temperature is 150 K, the gas to dust ratio is 100, and typical grain sizes are 0.1 $\mu$m. The resulting abundances are reported in Table \ref{abund}. While all of the values we report are source-averaged abundances, over N1 and N2, the centroids of these sources do not perfectly coincide (being roughly consistent with the positions of K2 and K7 shown in Figure \ref{schematic}), and so there may be significant additional local variation in the abundance that is not captured in these averages. Indeed, much higher resolution ($0''.4$) ALMA continuum observations of \cite{SM17} show that N2 breaks up into three separate continuum peaks, while N1 exhibits a complex spiral structure.

As for the main isotopologues \ammain\, and \methmain\, we are unable to directly measure a total column density, the abundances reported for these species are based on their rare isotopologues and an assumed isotope ratio. For \methmain\, we adopt [$^{12}$C/$^{13}$C] = 25, and find abundances in N2 ($\sim3\times10^{-6}$) that are roughly 7 times higher than in N1 ($\sim4\times10^{-7}$), consistent with the higher optical depths measured in N2. For \ammain\, if we use the [$^{14}$N/$^{15}$N] ratio measured in N1 (450) for both sources, we find similar abundances of $\sim3\times10^{-6}$. However, if we use the measured ratio in N2 (200), we would find a lower abundance of $\sim1.3\times10^{-6}$ for this source. 

\subsubsection{The Deuteration Fraction}

For \deut, using the measured rotational temperature for this species toward each source we measure column densities of 4$\times10^{17}$ cm$^{-2}$ for N1 and 1$\times10^{18}$ cm$^{-2}$ for N2. 
We first compare these to the total column inferred from \amiso, assuming a [$^{14}$N/$^{15}$N] ratio of 450 for both sources. In this scenario, we measure [\am/\deut] of 0.3\% in N1, and 3\% in N2: an order of magnitude larger. However, if we adopt a [$^{14}$N/$^{15}$N] ratio of 200 for N2, the deuteration fraction in this source would increase to 7\%, or more than 20 times the deuteration fraction in N1. 
A deuteration fraction of $<$10\% is consistent with values found for young stellar objects by \citep{Busquet10}, much less than the deuteration fractions of 10-80\% seen in this study for younger, massive prestellar cores. In general, D and perhaps $^{15}N$ are more likely to be enhanced in cold environments of younger sources \citep[e.g.,][]{Aleon10,Fontani15b}, so finding signs of their enhancement in a hot core in the generally warm environment of the Galactic center \citep[e.g.,][]{Mills13a, Ao13,Ginsburg16} is somewhat surprising. We discuss this more in Section \ref{deutdeut}. 

\section{Discussion}
\label{dis}

We have presented new centimeter radio observations of molecular lines in Sgr B2(N), characterizing the spatial distribution and abundance of \meth, \ammain, and various of their rare isotopologues (\amiso, \deut, \methiso) toward N1 and N2, each of which is a hot core with an embedded \hii\, region. We find that both cores have high temperatures of at least 100-300 K, and relatively large abundances of \meth\, and \am\, of $10^{-7}$ to $10^{-6}$. Below, we discuss the physical conditions and abundances we derive for N1 and N2 in the context of other recent work, and assess whether the observed variations in core properties are indicative of different evolutionary states for these cores.

\subsection{Comparison of Physical Conditions with Prior Measurements}
\label{compare}

\subsubsection{Density}
\label{density}

Sgr B2 is the dominant molecular structure in the Galactic center, with a mass of $3-7\times10^6$ M$_\odot$ and an average density $>$ 2500 \cm\, in a 45 pc diameter region \citep{Scoville75,Morris76,Lis89}. A significant fraction of the mass, $8\times10^5$ M$_\odot$, is further concentrated into a $4\times11$ pc envelope, with 5\% of this mass (a few $10^4$ M$_\odot$) located in three massive star forming cores, each less than 0.4 pc in size \citep{Gordon93}. Sgr B2(N) is the most massive and also one of the densest of these concentrations. Modeling CO transitions from Herschel line observations with resolutions of $18''-35''$, \cite{Etxaluze13} find average gas densities of $10^6$ \cm\, for both the Sgr B2(N) and Sgr B2(M) cores. For Sgr B2(N), \cite{Lis91} measure an average density of $10^5$ \cm\, in a $5\times10$ pc region by modeling highly-excited rotational lines of HC$_3$N and suggest that the density could be as high as $>10^7$ \cm\, from the detection of the $J=39-38$ transitions of HC$_3$N. \cite{Huttem93a} also infer $2\times10^7$ \cm\, from their ammonia observations, assuming a source size of 0.16 pc. SMA observations show that Sgr B2(N) further breaks down into two cores; taking the reported mass and size of N1 and N2 from \cite{Qin11} and assuming a spherical shape and a mean molecular weight of 2.8 for the metallicity of Galactic center gas \citep{Kauffmann08}, we estimate average densities of $\sim4\times10^7$ \cm\, over a $\sim0.06$ pc size scale for N1 and $\sim1.6\times10^7$ \cm\, over a $\sim0.05$ pc size scale for N2 \citep[This can be compared to the peak density of N1, which is modeled to be $2\times10^{8}$ \cm;][]{Qin11}.

We detect multiple nonmetastable ($J \ne K$) \ammain\, lines toward both N1 and N2, including the (10,9) and (12,11) lines. Unlike metastable levels, for which radiative decay to lower $J$ levels can only proceed through slow $\Delta K$ = 3 octopole transitions \citep{Oka71}, the nonmetastable levels are able to decay quickly down the $K$-ladder through $\Delta J$=1 rotational transitions. Observing a significant population in nonmetastable levels then implies the existence of an excitation mechanism to continually populate these lines. Collisional excitation typically requires extremely large densities $\gtrsim10^9$ \cm. However, if the lines are optically thick, as we are able to measure for the (10,9) and (12,11) lines toward N1, the critical densities are lower, and are equal to n$_{\mathrm{crit}}$ / $\tau$ \citep[as discussed in the appendix of ][]{Pauls83}. Alternatively, the \am\, may be excited out of the metastable levels and into the observed nonmetastable levels by infrared emission. The wavelength of the rotational transition linking the (9,9) and (10,9) levels is 51 $\mu$m, and the wavelength of the (12,11)s-(11,11)a line is 42 $\mu$m. Interpreting the observed nonmetastable line emission then requires distinguishing between these two mechanisms. 

For optically thin emission, the critical density of the transition through which the (10,9) level decays to the (9,9) level is $1.2\times10^{10}$ \cm, and the corresponding critical density for the (12,11)s-(11,11)a transition is $1.8\times10^{10}$ \cm. Using the opacities we measure for N1 in these transitions (Section \ref{amtemp}) we would then infer effective critical densities of $5\times10^8$ and $1\times10^9$ \cm for these lines-- still nearly as much as an order of magnitude larger than the peak core density modeled by \cite{Qin11}. However, as can be seen in Figure \ref{schematic}, the \ammain\, (10,9) line does not just originate from an unresolved point source: (10,9) emission is extended across 9$''$ (0.34 pc), which is also significantly larger than the dense core traced by the SMA \citep{Qin11}. If all of the (10,9) emission is tracing gas with a density $>5\times10^8$ \cm\, then the inferred enclosed mass of a spherical core would be 7$\times10^5$ M$_\odot$. This is two orders of magnitude larger than the mass that should be enclosed within a radius of 0.17 pc \citep[$3-4\times10^3$ M$_\odot$;][]{Qin11,Walker16}. The extended nature of the (10,9) emission then favors infrared emission as the excitation mechanism for these lines. Infrared emission could also be invoked to explain the nonmetastable emission in N2, as this source also has an embedded hypercompact \hii\, region \citep{dePree15}. 

This picture is also consistent with a comparison of the temperature measured across $J$ levels in different $K$ ladders and the temperature between $J$ levels within a single $K$ ladder. For the former, we measure a temperature between different nonmetastable lines of $\sim$600 K in N1. This is higher than the temperature we can measure within a single $K$ ladder, which for the (10,9) and (9,9) lines is $\sim$230 K. This indicates that the nonmetastable line populations deviate somewhat from local thermodynamic equilibrium (LTE), and are not well thermalized, as might be expected for extremely large densities. Note however that these temperatures may have a substantial uncertainty due to varying volumes from which these transitions arise. Whether the nonmetastable lines are collisionally or radiatively excited, the excitation conditions necessary to produce emission in these lines are more stringent than for the metastable lines, and so they very likely do not arise from the same volume. This could raise the temperature that we measure within a single $K$ ladder. Further, the nonmetastable transitions also have different excitation requirements. The densities required for collisional excitation increase with increasing $J$, and the wavelength of IR radiation required for excitation decreases with increasing $J$. Both of these quantities should be radially dependent as the density should increase inward, and shorter-wavelength radiation should penetrate to a smaller volume than longer-wavelength emission. This effect could then raise the nonmetastable temperature measured between $K$ ladders. 

\subsubsection{Temperature}
\label{temps}

Overall, the temperatures we measure for \amiso, \methmain, and \methiso\, from 70-300 K, are basically consistent between N1 and N2. The temperature of N2 is potentially higher in \deut\, and metastable transitions of \ammain\, ($>$300-600 K) but is less well constrained. However, given typical densities in excess of $10^7$ \cm\, for the cores, all of the gas detected on compact size scales with the VLA is likely to be dense enough to be thermalized \citep[both with the other gas, as well as with the dust;][]{Clark13}. Over the entire Sgr B2 complex, dust temperatures measured by Herschel are 20-28 K \citep{Guzman15}, however dust temperatures for the individual hot cores are much higher. Peak brightness temperatures for the cores from SMA observations (a lower limit on the dust temperature) are 270 K for N1 and 200 K for N2 \citep{Qin11}. Comparable gas temperatures are seen toward Sgr B2(N) with metastable lines of \ammain\, \citep[200 K;][]{Vogel87} and vibrationally-excited \cyano\, \citep[250 K;][]{Lis91} and rotationally-excited \cyano\, \citep[230 K][]{Lis93}. So, what factors are causing the measured rotational gas temperatures to differ from the most likely kinetic gas temperatures for these sources? 

One possibility is that there are systematic issues with the rotational temperatures measured from population diagram analysis. For example, the rotational temperatures measured with \am\, are always a lower limit on the kinetic temperature, and are often sufficiently close to T$_{kin}$ to provide a useful approximation of this value. However, as shown by \cite{Danby88} and more recently \cite{Maret09} and \cite{Bouhafs17}, the rotational temperature measured by the (1,1) and (2,2) lines of \ammain\, essentially saturates at a maximum value of 60 K for kinetic temperatures up to 300 K. Similarly, the rotational temperature from the (4,4) and (5,5) lines is $<$ 150 K for the same kinetic temperature. This suggests that rotational temperatures we measure with \am\, could be substantially below the kinetic temperature. Using the optically-thick hyperfine \ammain\, satellite lines, the minimum kinetic temperature set by their brightness temperature is $>$65 K. However, if a sizable fraction of the ammonia emission originates from a core having size of the SMA source size for N1, consistent with modeling for other species by \cite{Belloche13}, the brightness temperature would be $\sim$200 K. Thus, while the spatial resolution of our observations does not allow us to rule out 70 K gas traced with \amiso\, , we suggest it is likely that the true kinetic temperature is warmer. 

For N2, the measured temperature of 160 K for \methiso\, is consistent with the temperatures derived from LTE modeling of a much larger number of lines of \methmain, \methiso, and CH$_3^{18}$OH observed at millimeter and submillimeter wavelengths \citep[140-160 K;][]{Muller16}. \cite{Pei00} in a BIMA interferometric study with $23''\times6''$ resolution also measure a (combined) temperature for N1 and N2 of 170 K by observing the same $K=2-1$ series of \methmain\, lines studied here, including transitions from $J=$13 to $J=$20. The measured temperature of 90 K from \methmain\, in this source is less consistent with these measurements, and is likely affected by the opacity of this line. Brightness temperatures of the \methmain\, lines in N2, which has a peak brightness temp of 44 K in unconvolved images of the $J=13$ line, are $>$ 150 K if the SMA source size is assumed (consistent with the size determined from modeling of lines in this source from ALMA data by \citealt{Belloche13} and \citealt{Muller16}). For \meth\, we thus favor temperatures $>$ 130-150 K in both cores. 

Finally, it is also possible that the variation in measured temperatures is real, as a spread in temperatures is not unexpected in a hot core. Variations of measured temperatures between 70 \citep[comparable to temperatures seen in non-star forming Galactic center gas;][]{Ao13,Ginsburg16,Krieger17} and 300 K \citep[and hotter: T$_{rot}$ 300-500 K are seen in Herschel observations of CO and in complex molecules in line surveys][]{Etxaluze13,Nummelin00} could be due to species probing material at different core radii, if there is a strong temperature gradient (as suggested by both sources having an embedded \hii\, region). 


\subsection{Comparison of Abundances with Prior Measurements}

\subsubsection{The \ammain\, Abundance}

We find that N1 and N2 have similarly high \ammain\, abundances, of a few $10^{-6}$, consistent with the abundance of $3\times10^{-6}$ derived by \cite{Peng93} for N1. However, this agreement may be coincidence, as \cite{Peng93} find an \ammain\, column density for this source of $9\times10^{18}$ cm$^{-2}$, which is an order of magnitude lower than the column density we infer for N1. Our column density for N1 is more comparable to the value found by \cite{Vogel87}, who used observations of hyperfine structure in the (7,6) nonmetastable \ammain\, line to infer a source-averaged column density equal to $1\times10^{20}$ cm$^{-2}$. We replicate their method for the observed (10,9) and (12,11) lines in N1, adopting the partition function used in Section \ref{column}, and assuming a temperature of 300 K. The source-averaged total column densities determined from the (10,9) and (12,11) lines are 7.5$\times10^{19}$ and 2$\times10^{20}$, respectively. These values are comparable to those we have derived from the metastable \am\, lines. 

As a cautionary note, we also find when examining the \ammain\, (10,9) emission closely that there is a very slight ($\sim0''.5$) spatial offset between the centroids of the red-shifted and blue-shifted hyperfine satellite emission, which should not be the case if this emission is solely from the hyperfine satellite lines. This suggests that the observed line profile may also be tracing the inner parts of the outflow in N1, like that seen in wings of the \cyano\, 25-24 line over nearly the same velocity range \citep{Lis93}. High-velocity wings are also observed in both the GBT and VLA spectra of \amiso. However, they are narrower in extent than the wings in the nonmetastable lines, and we suspect that the bulk of the detected emission in the wings of the (10,9) and (12,11) lines is due to hyperfine satellites, as the peak brightness temperature in the unconvolved (10,9) line is 105 K, consistent with other lines suspected to be optically thick. This is similar to the situation in Orion, where hyperfine satellite lines are detected over a velocity range that is comparable to a broad `plateau' component from the outflow seen in HC$_3$N and other tracers \citep{Genzel82,Blake87}. We conclude then that this method of measuring the \ammain\, column density in N1 is likely valid, but should be treated with care. 

Finally, \cite{Huttem93a} measure a much higher \ammain\, column density toward N1 ($10^{21}$ cm$^{-2}$) from observations of nonmetastable \ammain\, emission, assuming the observed population distribution to be thermalized. If we correct their column density for our adopted source size (they assume a source size of 4$''$), we find that it is nearly 90 times larger than the value we infer. The simplest explanation for this discrepancy is that the lower-$J$ nonmetastable line emission observed by \cite{Huttem93a} is significantly extended, and that a large fraction of the emission detected with a single dish in a 42$''$ beam is resolved out with the VLA. This would be consistent with an infrared excitation scenario, as the wavelengths required to excite the lower $J$ transitions are significantly longer (e.g., 257 $\mu$m for the 2,1 line) and would be expected to penetrate to larger volumes around the core, and to even be present due to surrounding cloud emission for the lowest-$J$ transitions.

\subsubsection{The \methmain\, Abundance}

We compare our derived \methmain\, column density in N2 ($1.7\times10^{19}$ cm$^{-2}$) to recent measurements of the \methmain\, column in N2 by \cite{Muller16}. By conducting LTE modeling of a greater number of \methmain\, transitions, they report a higher column of $4\times10^{19}$ cm$^{-2}$. We suspect that the factor of two difference may be either from uncertainties in the optical thickness (requiring LTE modeling) or possibly from our use of an older partition function, as improvements to the existing partition function data are mentioned in \cite{Muller16}. Alternatively, as the resolution of these ALMA data is slightly higher than our VLA observations ($1''.8$, compared to $2''.7$), it could be possible that there is unresolved absorption against the K2 \hii\, region, lowering our average column densities. We can also compare our derived abundances ($4\times10^{-7}$ for N1, $1\times1-^{-6}$ for N2) with the abundance of $10^{-8}$ determined by \cite{Pei00}. We suggest that the slightly higher abundance we infer for N1 could be partly because the lines observed by \cite{Pei00} are optically thick. 

\subsubsection{[D/H] and $^{15}$N Fractionation in N2}
\label{deutdeut}

We find that N2 has a higher \am\, deuteration fraction ($\sim$3-7\%) than the more well-studied N1 core ($\sim$ 0.3\%). Further, the value we infer for N1 is several times larger than the deuteration fraction previously inferred for Sgr B2(N) using the ($1_{11}-1_{01}$) transition of \deut\, \citep[0.1\%;][]{Peng93}. The \am\, deuteration fraction in N2 is also significantly larger than the deuteration fractions recently observed in complex organic molecules for this source \citep[0.05-0.38\%, with upper limits on additional observed species all $<$ 2\%;][]{Belloche16}. In particular, the deuteration fraction for methanol in the CH$_2$DOH isotopologue in N2 is observed only to be 0.12\%, and $<$0.07\% for CH$_3$OD \citep{Belloche16}. For all of these species, the level of deuteration is less than predicted by current chemical models, which is suggested to be due either to high temperatures or an intrinsically low value of the [D/H] ratio in the Galactic center. The latter scenario is supported by a measured [DCN/HCN] ratio of $\sim10^{-4}$ toward Sgr B2, which is interpreted as an indication that deuterium is an order of magnitude less abundant in the Galactic center than in the solar neighborhood \citep{Jacq1990,Jacq1999}. However, observations of even lower \meth\, deuteration fractions in NGC 6334I outside of the Galactic center support the former scenario \citep{Bogelund18}.

Our observed \am\, deuteration fraction in N2 is then much larger than that measured for \meth\, by \cite{Belloche16}, even though \meth\, is a molecule that has been seen to be highly deuterated \citep[up to 40 \%][]{Parise06} in prestellar cores. This could indicate that the processes leading to large \deut\, values in N2 are different than those for organic molecules \citep[e.g.,][possibly due to different formation routes, partially or wholly on grain surfaces]{Cazaux11,Fontani15a}, or that the grain mantles have chemically distinct layers \citep[e.g.;][]{RC08,Taquet12}, and organic species either desorbed earlier (giving them more time to return to a chemical equilibrium), or have not yet fully desorbed. \meth\, does desorb from water ice-coated grains at temperatures of 130 K, while \am\, will desorb earlier at 80 K \citep{Collings04}. Given the temperatures ($\sim$ 70-300 K) we measure for the cores, this scenario favors temperatures $<$150 K, in which case there might still be a little water on the surface trapping some material, and relatively young ages for grain mantles not to have been fully thermally desorbed \citep[e.g., $<10^4$ years; ][]{Viti99}. 

Alternatively, the difference in observed deuteration fractions could mean that \deut\, survives longer than deuterated \meth\, at high temperatures after the initial grain mantle evaporation. Models of \cite{RM96} argue against this latter scenario, predicting that the processes that disrupt deuterated `parent' species like \deut, CH$_3$OD and HDCO, should similarly disrupt other isotopologues, essentially preserving the initial deuteration fraction of the grain mantle in the gas phase for up to $10^5$ years. However, \cite{Fontani15a} find that, instead of an observed decrease in the deuteration fraction of N$_2$H$^+$ and HCN as a function of evolution, there is no evidence for changing deuteration fraction of \deut\, (insufficient lines of deuterated \meth\, were observed to assess how its deuteration fraction evolved), implying that \deut\,may indeed persist longer in the gas phase at high temperatures. This suggests that changing deuteration fractions as species chemically equilibrate at varying rates could explain some of the differences in deuteration between \meth\, and \am\, that is seen in N2.

Large \am\, deuteration fractions are typically understood as resulting from gas-phase chemical reactions, in a cold (T$<$ 20 K) environment where CO is depleted from the gas phase, and are most often seen in young, starless sources such as prestellar cores \citep{Shah01,Hatchell03,Busquet10,Fontani15a}. However, measured gas temperatures of 70-300 K in N1 and N2 are too hot for \deut\, to have formed (recently) via gas-phase formation routes. This leaves two possibilities (which are not mutually exclusive): either the \deut\, observed here formed on grain surfaces, or it formed in the gas phase at T$<$20 K, likely at an earlier time (and possibly at a larger Galactocentric radius), and then accreted onto grain mantles, from which it has recently evaporated. A third possibility, a high temperature pathway to \am\, deuteration-- seems unlikely given that [\deut/\am] appears to decrease for more evolved sources \citep{Busquet10,Fontani15a}. Formation from gas with T$<$ 20 K does not seem likely, as there is no compelling evidence for T$<$ 20 K gas in Sgr B2 or globally in the Galactic center. Typical gas temperatures in the central 200 parsecs of the Galaxy are significantly elevated compared to temperatures in the Galactic disk, having T $>$ 60-70 K (\citealt{Ao13}; \citealt{Ginsburg16}; though \am\, measurements favor somewhat lower temperatures of 30-50 K; \citealt{Krieger17}). Dust grain temperatures in the Galactic center however are much less elevated compared to the Galactic disk: typical dust temperatures in Galactic center clouds are 20-25 K \citep{Molinari11,Etx11,Longmore12}. With future observations of multiply-deuterated \am\, it should be possible to distinguish between gas phase vs. grain surface formation of \deut, as these two formation methods predict different abundance ratios of the multiply deuterated species NHD$_2$ and ND$_3$ \citep{BM89}. 

Not only do warm temperatures inhibit deuteration via gas-phase reactions, they also favor the destruction of \deut\, via a return to equilibrium chemistry through gas phase reactions, e.g., 

\begin{equation}
\mathrm{NH}_2\mathrm{D} + \mathrm{HCO}^+\, \rightarrow \mathrm{CO} + \mathrm{NH}_3\mathrm{D}^+. 
\end{equation}

For the Orion hot core, \cite{Walmsley87} estimate the timescale (in seconds) on which \deut\, will be destroyed through this reaction as $10^9$ / (X[i] $n$), where $n$ is the number density of H$_2$ and X[i] is the sum of the abundance of HCO$^+$ and other relevant ions like H$_3^+$, and estimate that disequilibrium can persist in this source only for $\sim10^3$ years. For the cores in Sgr B2(N), with a similar density of n=$10^7$ \cm, and adopting an HC$^{18}$O$^+$ abundance of $2\times10^{-12}$ from \cite{Nummelin00} and assuming [$^{16}$O/$^{18}$O] = 180, as measured from \cite{Muller16}, we would estimate a timescale of $\sim10^4$ years. 

While the [D/H] values determined for N2 from \am\, are consistent with typical low values seen in young stellar objects \citep[$<$0.1;][]{Busquet10}, the [$^{14}$N/$^{15}$N] value inferred for N2 from \am\, ($\sim200$) is among the lowest measured in the ISM: typical values for low-mass prestellar cores are [$^{14}$N/$^{15}$N] $\sim$ 400 \citep{Lis10}, with values as low as $\sim$180 (but also as large as 1300) measured from N$_2$H$^+$ in a sample of massive star-forming cores \citep{Fontani15b}. While \cite{Adande11} show that [$^{14}$N/$^{15}$N] exhibits a radial trend with Galactocentric distance, consistent with a value of $\sim$ 125 in the Galactic center (and measured in Sgr B2, though a second velocity component in this source gives [$^{14}$N/$^{15}$N] = 230), other work has found much larger values in the Galactic center \citep[$400-600$;][]{WR94,Wannier81,Peng93}.  Thus, the [$^{14}$N/$^{15}$N] value observed in N2, if it is not the result of a systematic error (see Section \ref{isotop}) is likely a result of fractionation \citep[for example isotope-selective photodissociation of N$_2$, which can operate at relatively high temperatures in UV-irradiated gas;][]{Visser18}, and is not representative of the typical Galactic center value. 

An example of a low [$^{14}$N/$^{15}$N] value is potentially interesting for understanding the origins of material in our own solar system. The observed value in N2 of $\sim$ 200 is lower than values measured for both atmospheres of Earth and Venus \citep[$\sim$270;][]{Junk58,Hoffman79} and inferred for the protosolar nebula \citep[$\sim$440;][]{Owen01,Fouchet04,Meibom07,Marty10}. It is more similar to the extremely low values of [$^{14}$N/$^{15}$N] found in primitive solar system material, like cometary CN and HCN \citep[$\sim$140;][]{Arpigny03,BM08,Manfroid09} and in organic material in chondritic meteorites \citep[$\sim$50-150;][]{Bonal10,Briani09}, and that recently observed in HCN in a sample of T Tauri and Herbig disks \citep{Guzman17}. Low [$^{14}$N/$^{15}$N] values are seen to be correlated with high D abundances in carbonaceous chondrites, `cluster' interplanetary dust particles, and comets Wild2 and Hale-Bopp \citep[][]{Messenger00,Floss06,Busemann06,Aleon10}, an enrichment which occurred either in the protosolar disk or in the precursor cloud core. Recent ISM modeling suggests this correlation may not occur in the core stage, as a correlation between D and $^{15}$N is not expected \citep[due to e.g., sensitive dependence of chemistry on ortho-to-para variations in collisional partners like H$_2$;][]{Wirstrom12,deSimone18}. This is supported by observations showing a hint of anticorrelation between the abundances of D and $^{15}$N using N$_2$H$^+$ in a sample of massive prestellar cores \citep{Fontani15b}. However, the potential presence of simultaneously high $^{15}$N and D in N2 suggests that it could still be possible for this to occur in an individual core, allowing for an interstellar origin for this solar system feature. As the solar system is suggested to have formed in a rich stellar cluster experiencing a nearby supernova \citep{Adams10,Dukes12,Pfalzner13}, the environment of a massive protocluster like Sgr B2 may actually be quite relevant for understanding the neighborhood of the protosolar nebula. 
 
\subsection{The Relative Evolutionary States of N1 and N2}
\label{evo}

\begin{figure*}
\includegraphics[scale=0.6]{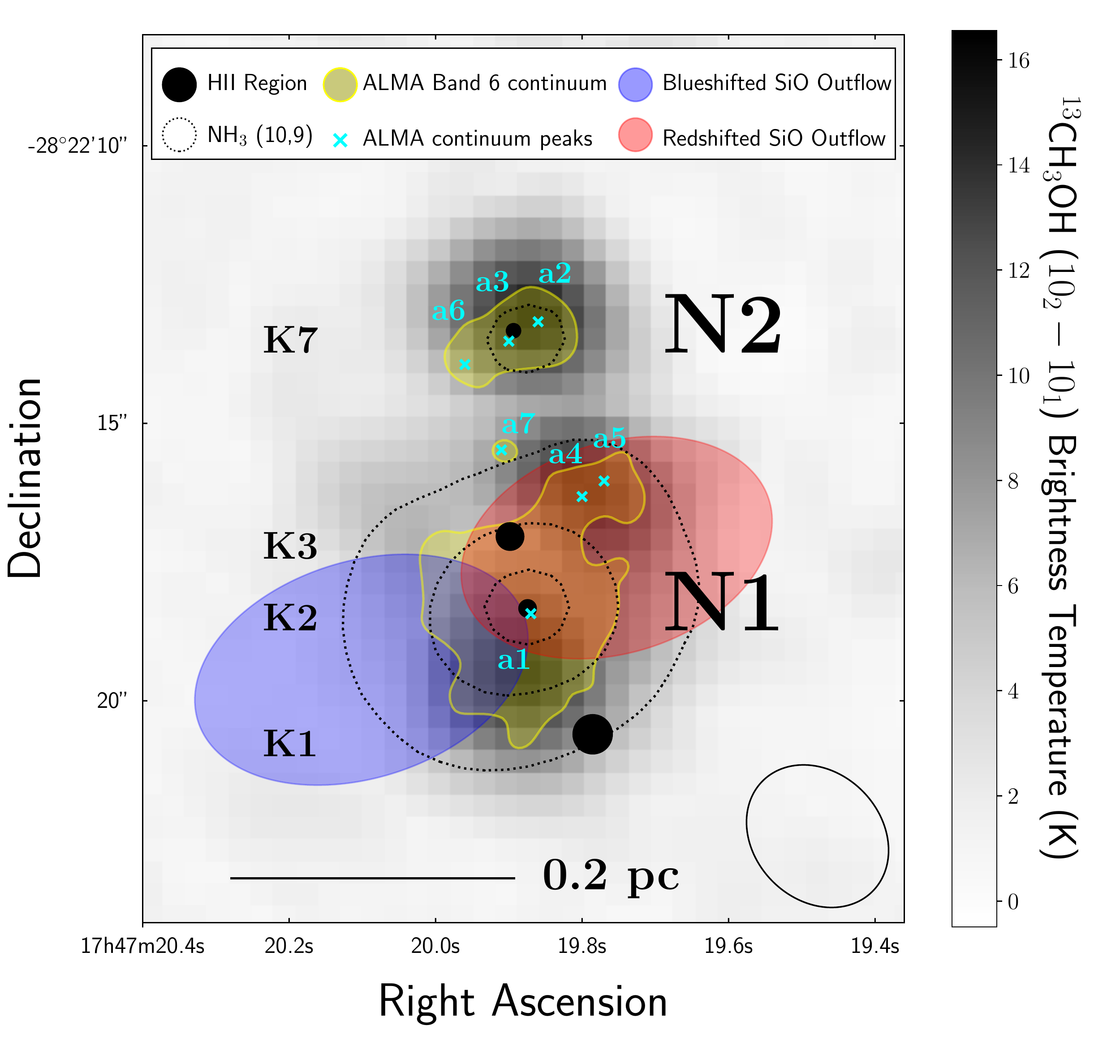}
\caption{A map of the currently identified components of Sgr B2(N). The background greyscale is the peak emission from \methiso\, (10$_2$-10$_1$) from the VLA data. A 0.2 pc scale bar is plotted at the bottom. Dotted contours show the emission from the \ammain\, (10,9) line in N2 and N1. The beam for both of these maps is displayed in the bottom right corner. The black circles represent the sizes and positions of compact \hii\, regions seen with the VLA at 8.4 GHz and 22 GHz by \cite{Gaume95} and \cite{dePree15}. The extent of the dust continuum emission from this region as seen in the Band 6 ALMA observations of \cite{SM17} is shown in yellow, as well as the seven strongest individual continuum peaks identified in that study, which are overplotted as cyan `x's and numbered a1-a7. Also shown are the approximate positions of the red- and blue-shifted lobes of the outflow detected in SiO with ALMA by \cite{Higuchi15}}
\label{schematic}
\end{figure*}

Overall, N1 and N2 appear quite similar: both are chemically-rich hot cores \citep{Corby15} with an associated hypercompact \hii\, region \citep{dePree15}. {Whether the \hii region K7 is actually embedded in N2 is a subject of some uncertainty:  \cite{Bonfand17} suggest that the \hii\, region in N2 may be separate from the hot core source, and \cite{SM17} find that N2 breaks up into three continuum sources in their high-resolution ALMA observations, though as seen in Figure \ref{schematic}, it is not clear which is associated with K7}. We find that N1 and N2 have similar temperatures in most species, except for \ammain\, and \deut\, in which the temperature of N2 is more uncertain, but may be higher. However, N1 appears to have a somewhat more rich chemistry than N2  \citep{Corby15,Bonfand17}. N1 is also known to be more massive \citep[2700 M$_\odot$ compared to 600 M$_\odot$;][]{Qin11} and hosts a fast molecular outflow, seen originally in the kinematics of water masers in this region \citep{Reid88} and detected in \cyano\, \citep{Lis93} and more recently, ALMA observations of SiO \citep{Higuchi15}. In our observations, the \amiso\, lines have broad wings which may trace the inner part of this outflow. \cite{Bonfand17} suggest that this outflow may not be driven by N1, due to the high densities ($\sim10^9$ cm$^{-3}$) and temperatures ($\sim$ 400 K) required to excite the water masers. However, while \cite{Bonfand17} only see temperatures of $\sim$ 150 K in N1, we see much higher temperatures of at least 300-350 K in \am, with volume densities measured by \cite{SM17} to be $\sim10^9$ cm$^{-3}$, consistent with associating this outflow with N1. In contrast, for N2,  we do not see any line wings or other indications of an outflow being driven by this source, consistent with the nondetection of an outflow in this source in ALMA data \citep{Bonfand17}. Overall then, the richer chemistry and the presence of a fast outflow suggest N1 is older than N2. \citep[Note that while the presence of a hypercompact \hii\, region might suggest that N1 and N2 are a similar age, the size of an \hii\, region in these early stages is not an indicator of age, e.g.,][]{dePree15}. 

We propose an evolutionary scenario that matches these constraints, while also explaining the observed abundances of \am\, and \meth\, isotopologues in both sources. In our scenario, N1 is a normal hot core that formed via gradual warm-up due to infrared emission from the embedded stellar source. Its chemistry is representative of three stages of molecular processing:
\begin{enumerate}
\item Cold ice mantle surface chemistry at T = 10-25 K, for T $\sim$1 Myr. 
\item Ice-grain chemistry as the grains warm from 25 to 70 K on a timescale on the order of a few 10$^4$ years. 
\item Gas-phase chemistry after liberation from grains in the dense, warm material.
\end{enumerate}
Although the warm-up grain chemistry stage is relatively short, it is an important contribution to increased chemical diversity in N1 as the chemical reactions are quite efficient during this phase. As an older hot core, N1 has also likely spent time in stage 3, allowing efficient gas phase chemistry to proceed in dense, warm material, leading to its observed rich chemical diversity; though perhaps not enough time for the [\deut/\ammain] ratio to fully equilibrate. As further support for an extended heating history, we see an extended halo of highly-abundant \meth\, in N1 that is similar to those seen in W51 \citep{Ginsburg17}, which have been attributed to internal heating rather than shocks. 

In contrast, we propose that N2 formed after N1 and has not gone through the same stages of molecular processing. During its formation, this source may have experienced an early shock, for example from the expanding K5 shell \hii\ region, an outflow from the N1 source, or even a collision, as the velocity of N2 is intermediate between N1 and the 80-90 \kms\, cloud \citep[just as N1 is intermediate in velocity between the 40-50 \kms\, cloud and the 80-90 \kms\, cloud, which are suggested to have collided;][]{Hasegawa94}. This shock liberated cold grain species en masse from the ice mantles of the grains (skipping the slow warmup phase of grain-surface chemistry on warm dust grains at 25-60 K). 
The shock would then also be responsible for a faster warmup of this source, bringing it to a similar temperature with N1. Being younger however, there has been less time for gas-phase processing, which would act to decrease the deuteration fraction. Without the grain warmup phase, and without as much time for gas-phase processing, the chemistry of N2 would be somewhat less rich than that of N1, consistent with observations of \cite{Corby15}. This scenario would explain the higher abundances of \meth, and \deut\, in N2, while also reproducing the similar temperatures of N1 and N2, and their observed differences in overall chemical richness. 

As determined in Section \ref{deutdeut}, the relatively high abundances of \deut\, in N2 despite temperatures of $\sim$130 K require the shock to have occurred in last $10^4$ years such that the chemistry has not had time to reach gas-phase equilibrium. This could be consistent with the shock originating in the N1 outflow, as \cite{Higuchi15} estimate its age to be 5$\times10^3$ years. Depending on its strength, the passage of a shock through N2 might also predict that this source would have higher SiO abundances. However, currently, SiO is only observed in absorption toward both cores, though there may be a $v=2$ maser observed in emission toward the K2 \hii\, region in N1 \citep[e.g.,][]{Higuchi15}. 

\section{Conclusions}

We have analyzed VLA, ATCA, and GBT observations of centimeter-wavelength molecular line emission in isotopologues of \am\, and \meth\, from Sgr B2(N). The main conclusions are as follows:

\begin{itemize}

\item We observe the highest-$J$ lines of metastable \ammain\, yet detected in emission from Sgr B2(N), and present the first interferometric maps of emission in \ammain, \amiso, \deut, \methmain, and \methiso\, with sub-parsec resolution. For the first time, we resolve the \am\, emission in this source into two spatially and kinematically-separate cores: N1 (63.5 \kms) and N2 (75 \kms). 

\item We use isotopologues of \am\, and \meth\, in order to measure the temperature in N1 and N2, finding that N1 and N2 have largely consistent rotational temperatures ranging from 70-300 K. A possibly hotter component $>350-600$ K is seen toward N2 in \ammain\, and \deut\, though this temperature component is poorly constrained. 

\item We find similar [\ammain/H$_2$] abundances in both cores ($\sim1-3\times10^{-6}$), and higher [\methmain/H$_2$] abundances in N2 than N1 ($3\times10^{-6}$ compared to $4\times10^{-7}$ ). [\deut/\am] is also larger in N2 than in N1: 3-7\% vs 0.3\%. 

\item Finally, [\ammain/\amiso] is smaller in N2 than in N1 (we derive [$^{14}$N/$^{15}$N] $\sim$200 in N2 and $\sim$450 in N1). If this is not a result of optical depth or absorption in the \ammain\, line, the value derived for N2 would be among the lowest values seen in the ISM, and is lower than values inferred for both the terrestrial atmosphere ($\sim$270) and the protosolar nebula ($\sim$440). Such low values approach the lowest values ([$^{14}$N/$^{15}$N] $\sim$50-150) measured for primitive solar system material. This makes further study of the chemistry in N2 relevant for probing links between interstellar and cometary material, particularly given recent suggestions that the sun may have formed in a rich cluster environment \citep{Adams10}. 

\item We propose an evolutionary model in which the abundance variations between N1 and N2 are due to a shock and fast warmup in the younger N2 source, that liberated \deut\, and \meth\, from the ice mantles of cold dust grains. We suggest that, combined with the correspondingly smaller amount of time available for grain warmup and to reach an equilibrium in gas phase chemistry, this has led to the observed large abundances of \deut\, and \meth\, in N2, and overall, less chemical richness than seen in N1.

\end{itemize}

\section{Acknowledgements}
We thank the anonymous referee for their detailed comments, which improved the presentation and interpretation of the results in this paper. We also thank \'{A}lvaro Sanchez-Monge for sharing his band 6 ALMA continuum map of Sgr B2(N).

\bibliographystyle{hapj}
\bibliography{NH2D_Sgr_B2}

\begin{thebibliography}{}
\expandafter\ifx\csname natexlab\endcsname\relax\def\natexlab#1{#1}\fi

\bibitem[{{Adams}(2010)}]{Adams10}
{Adams}, F.~C. 2010, \araa, 48, 47

\bibitem[{Adande \& Ziurys(2011)}]{Adande11}
Adande, G.~R., \& Ziurys, L.~M. 2011, \apj, 744, 194

\bibitem[{{Al{\'e}on}(2010)}]{Aleon10}
{Al{\'e}on}, J. 2010, \apj, 722, 1342

\bibitem[{{Ao} {et~al.}(2013){Ao}, {Henkel}, {Menten}, {Requena-Torres},
  {Stanke}, {Mauersberger}, {Aalto}, {M{\"u}hle}, \& {Mangum}}]{Ao13}
{Ao}, Y., {Henkel}, C., {Menten}, K.~M., {et~al.} 2013, \aap, 550, A135

\bibitem[{{Arpigny} {et~al.}(2003){Arpigny}, {Jehin}, {Manfroid},
  {Hutsem{\'e}kers}, {Schulz}, {St{\"u}we}, {Zucconi}, \& {Ilyin}}]{Arpigny03}
{Arpigny}, C., {Jehin}, E., {Manfroid}, J., {et~al.} 2003, Science, 301, 1522

\bibitem[{Belloche {et~al.}(2013)Belloche, M~ller, Menten, Schilke, \&
  Comito}]{Belloche13}
Belloche, A., M~ller, H. S.~P., Menten, K.~M., Schilke, P., \& Comito, C. 2013,
  \aap, 559, A47

\bibitem[{Belloche {et~al.}(2008)Belloche, Menten, Comito, M{\"u}ller, Schilke,
  Ott, Thorwirth, \& Hieret}]{Belloche08}
Belloche, A., Menten, K.~M., Comito, C., {et~al.} 2008, \aap, 482, 179

\bibitem[{Belloche {et~al.}(2016)Belloche, M{\"u}ller, Garrod, \&
  Menten}]{Belloche16}
Belloche, A., M{\"u}ller, H. S.~P., Garrod, R.~T., \& Menten, K.~M. 2016, \aap,
  587, A91

\bibitem[{Blake {et~al.}(1987)Blake, Sutton, Masson, \& Phillips}]{Blake87}
Blake, G.~A., Sutton, E.~C., Masson, C.~R., \& Phillips, T.~G. 1987, \apj, 315,
  621

\bibitem[{{Bockel{\'e}e-Morvan} {et~al.}(2008){Bockel{\'e}e-Morvan}, {Biver},
  {Jehin}, {Cochran}, {Wiesemeyer}, {Manfroid}, {Hutsem{\'e}kers}, {Arpigny},
  {Boissier}, {Cochran}, {Colom}, {Crovisier}, {Milutinovic}, {Moreno},
  {Prochaska}, {Ramirez}, {Schulz}, \& {Zucconi}}]{BM08}
{Bockel{\'e}e-Morvan}, D., {Biver}, N., {Jehin}, E., {et~al.} 2008, \apjl, 679,
  L49

\bibitem[{Boehle {et~al.}(2016)Boehle, Ghez, Sch{\"o}del, Meyer, Yelda, Albers,
  Martinez, Becklin, Do, Lu, Matthews, Morris, Sitarski, \& Witzel}]{Boehle16}
Boehle, A., Ghez, A.~M., Sch{\"o}del, R., {et~al.} 2016, \apj, 830, 1

\bibitem[{{B{\o}gelund} {et~al.}(2018){B{\o}gelund}, {McGuire}, {Ligterink},
  {Taquet}, {Brogan}, {Hunter}, {Pearson}, {Hogerheijde}, \& {van
  Dishoeck}}]{Bogelund18}
{B{\o}gelund}, E.~G., {McGuire}, B.~A., {Ligterink}, N.~F.~W., {et~al.} 2018,
  \aap, 615, A88

\bibitem[{{Bonal} {et~al.}(2010){Bonal}, {Huss}, {Krot}, {Nagashima}, {Ishii},
  \& {Bradley}}]{Bonal10}
{Bonal}, L., {Huss}, G.~R., {Krot}, A.~N., {et~al.} 2010, \gca, 74, 6590

\bibitem[{{Bonfand} {et~al.}(2017){Bonfand}, {Belloche}, {Menten}, {Garrod}, \&
  {M{\"u}ller}}]{Bonfand17}
{Bonfand}, M., {Belloche}, A., {Menten}, K.~M., {Garrod}, R.~T., \&
  {M{\"u}ller}, H.~S.~P. 2017, \aap, 604, A60

\bibitem[{{Bouhafs} {et~al.}(2017){Bouhafs}, {Rist}, {Daniel}, {Dumouchel},
  {Lique}, {Wiesenfeld}, \& {Faure}}]{Bouhafs17}
{Bouhafs}, N., {Rist}, C., {Daniel}, F., {et~al.} 2017, \mnras, 470, 2204

\bibitem[{{Briani} {et~al.}(2009){Briani}, {Gounelle}, {Marrocchi},
  {Mostefaoui}, {Leroux}, {Quirico}, \& {Meibom}}]{Briani09}
{Briani}, G., {Gounelle}, M., {Marrocchi}, Y., {et~al.} 2009, Proceedings of
  the National Academy of Science, 106, 10522

\bibitem[{{Brown} \& {Millar}(1989)}]{BM89}
{Brown}, P.~D., \& {Millar}, T.~J. 1989, \mnras, 240, 25P

\bibitem[{{Busemann} {et~al.}(2006){Busemann}, {Young}, {O'D.~Alexander},
  {Hoppe}, {Mukhopadhyay}, \& {Nittler}}]{Busemann06}
{Busemann}, H., {Young}, A.~F., {O'D.~Alexander}, C.~M., {et~al.} 2006,
  Science, 312, 727

\bibitem[{{Busquet} {et~al.}(2010){Busquet}, {Palau}, {Estalella}, {Girart},
  {S{\'a}nchez-Monge}, {Viti}, {Ho}, \& {Zhang}}]{Busquet10}
{Busquet}, G., {Palau}, A., {Estalella}, R., {et~al.} 2010, \aap, 517, L6

\bibitem[{{Butterfield} {et~al.}(2018){Butterfield}, {Lang}, {Morris}, {Mills},
  \& {Ott}}]{Butterfield18}
{Butterfield}, N., {Lang}, C.~C., {Morris}, M., {Mills}, E.~A.~C., \& {Ott}, J.
  2018, \apj, 852, 11

\bibitem[{{Cazaux} {et~al.}(2011){Cazaux}, {Caselli}, \& {Spaans}}]{Cazaux11}
{Cazaux}, S., {Caselli}, P., \& {Spaans}, M. 2011, \apjl, 741, L34

\bibitem[{Ceccarelli {et~al.}(2002)Ceccarelli, Baluteau, Walmsley, Swinyard,
  Caux, Sidher, Cox, Gry, Kessler, \& Prusti}]{Ceccarelli02}
Ceccarelli, C., Baluteau, J.~P., Walmsley, M., {et~al.} 2002, \aap, 383, 603

\bibitem[{{Clark} {et~al.}(2013){Clark}, {Glover}, {Ragan}, {Shetty}, \&
  {Klessen}}]{Clark13}
{Clark}, P.~C., {Glover}, S.~C.~O., {Ragan}, S.~E., {Shetty}, R., \& {Klessen},
  R.~S. 2013, \apjl, 768, L34

\bibitem[{{Collings} {et~al.}(2004){Collings}, {Anderson}, {Chen}, {Dever},
  {Viti}, {Williams}, \& {McCoustra}}]{Collings04}
{Collings}, M.~P., {Anderson}, M.~A., {Chen}, R., {et~al.} 2004, \mnras, 354,
  1133

\bibitem[{Corby {et~al.}(2015)Corby, Jones, Cunningham, Menten, Belloche,
  Schwab, Walsh, Balnozan, Bronfman, Lo, \& Remijan}]{Corby15}
Corby, J.~F., Jones, P.~A., Cunningham, M.~R., {et~al.} 2015, \mnras, 452, 3969

\bibitem[{{Danby} {et~al.}(1988){Danby}, {Flower}, {Valiron}, {Schilke}, \&
  {Walmsley}}]{Danby88}
{Danby}, G., {Flower}, D.~R., {Valiron}, P., {Schilke}, P., \& {Walmsley},
  C.~M. 1988, \mnras, 235, 229

\bibitem[{{De Pree} {et~al.}(1998){De Pree}, {Goss}, \& {Gaume}}]{dePree98}
{De Pree}, C.~G., {Goss}, W.~M., \& {Gaume}, R.~A. 1998, \apj, 500, 847

\bibitem[{{De Pree} {et~al.}(2015){De Pree}, {Peters}, {Mac Low}, {Wilner},
  {Goss}, {Galv{\'a}n-Madrid}, {Keto}, {Klessen}, \& {Monsrud}}]{dePree15}
{De Pree}, C.~G., {Peters}, T., {Mac Low}, M.~M., {et~al.} 2015, \apj, 815, 123

\bibitem[{{De Simone} {et~al.}(2018){De Simone}, {Fontani}, {Codella},
  {Ceccarelli}, {Lefloch}, {Bachiller}, {L{\'o}pez-Sepulcre}, {Caux}, {Vastel},
  \& {Soldateschi}}]{deSimone18}
{De Simone}, M., {Fontani}, F., {Codella}, C., {et~al.} 2018, \mnras, 476, 1982

\bibitem[{de~Vicente {et~al.}(2000)de~Vicente, Mart{\'\i}n-Pintado, Neri, \&
  Colom}]{deVicente00}
de~Vicente, P., Mart{\'\i}n-Pintado, J., Neri, R., \& Colom, P. 2000, \aap,
  361, 1058

\bibitem[{{Dukes} \& {Krumholz}(2012)}]{Dukes12}
{Dukes}, D., \& {Krumholz}, M.~R. 2012, \apj, 754, 56

\bibitem[{{Etxaluze} {et~al.}(2011){Etxaluze}, {Smith}, {Tolls}, {Stark}, \&
  {Gonz{\'a}lez-Alfonso}}]{Etx11}
{Etxaluze}, M., {Smith}, H.~A., {Tolls}, V., {Stark}, A.~A., \&
  {Gonz{\'a}lez-Alfonso}, E. 2011, \aj, 142, 134

\bibitem[{{Etxaluze} {et~al.}(2013){Etxaluze}, {Goicoechea}, {Cernicharo},
  {Polehampton}, {Noriega-Crespo}, {Molinari}, {Swinyard}, {Wu}, \&
  {Bally}}]{Etxaluze13}
{Etxaluze}, M., {Goicoechea}, J.~R., {Cernicharo}, J., {et~al.} 2013, \aap,
  556, A137

\bibitem[{{Floss} {et~al.}(2006){Floss}, {Stadermann}, {Bradley}, {Dai},
  {Bajt}, {Graham}, \& {Lea}}]{Floss06}
{Floss}, C., {Stadermann}, F.~J., {Bradley}, J.~P., {et~al.} 2006, \gca, 70,
  2371

\bibitem[{Flower {et~al.}(1995)Flower, Pineau~des Forets, \&
  Walmsley}]{Flower95}
Flower, D.~R., Pineau~des Forets, G., \& Walmsley, C.~M. 1995, \aap, 294, 815

\bibitem[{{Fontani} {et~al.}(2015{\natexlab{a}}){Fontani}, {Busquet}, {Palau},
  {Caselli}, {S{\'a}nchez-Monge}, {Tan}, \& {Audard}}]{Fontani15a}
{Fontani}, F., {Busquet}, G., {Palau}, A., {et~al.} 2015{\natexlab{a}}, \aap,
  575, A87

\bibitem[{{Fontani} {et~al.}(2015{\natexlab{b}}){Fontani}, {Caselli}, {Palau},
  {Bizzocchi}, \& {Ceccarelli}}]{Fontani15b}
{Fontani}, F., {Caselli}, P., {Palau}, A., {Bizzocchi}, L., \& {Ceccarelli}, C.
  2015{\natexlab{b}}, \apjl, 808, L46

\bibitem[{{Fouchet} {et~al.}(2004){Fouchet}, {Irwin}, {Parrish}, {Calcutt},
  {Taylor}, {Nixon}, \& {Owen}}]{Fouchet04}
{Fouchet}, T., {Irwin}, P.~G.~J., {Parrish}, P., {et~al.} 2004, \icarus, 172,
  50

\bibitem[{Gaume {et~al.}(1995)Gaume, Claussen, de~Pree, Goss, \&
  Mehringer}]{Gaume95}
Gaume, R.~A., Claussen, M.~J., de~Pree, C.~G., Goss, W.~M., \& Mehringer, D.~M.
  1995, \apj, 449, 663

\bibitem[{{Genzel} {et~al.}(1982){Genzel}, {Ho}, {Bieging}, \&
  {Downes}}]{Genzel82}
{Genzel}, R., {Ho}, P.~T.~P., {Bieging}, J., \& {Downes}, D. 1982, \apjl, 259,
  L103

\bibitem[{{Ginsburg} \& {Mirocha}(2011)}]{Ginsburg11}
{Ginsburg}, A., \& {Mirocha}, J. 2011, {PySpecKit: Python Spectroscopic
  Toolkit}, Astrophysics Source Code Library, ascl:1109.001

\bibitem[{{Ginsburg} {et~al.}(2016){Ginsburg}, {Henkel}, {Ao}, {Riquelme},
  {Kauffmann}, {Pillai}, {Mills}, {Requena-Torres}, {Immer}, {Testi}, {Ott},
  {Bally}, {Battersby}, {Darling}, {Aalto}, {Stanke}, {Kendrew}, {Kruijssen},
  {Longmore}, {Dale}, {Guesten}, \& {Menten}}]{Ginsburg16}
{Ginsburg}, A., {Henkel}, C., {Ao}, Y., {et~al.} 2016, \aap, 586, A50

\bibitem[{{Ginsburg} {et~al.}(2017){Ginsburg}, {Goddi}, {Kruijssen}, {Bally},
  {Smith}, {Galv{\'a}n-Madrid}, {Mills}, {Wang}, {Dale}, {Darling},
  {Rosolowsky}, {Loughnane}, {Testi}, \& {Bastian}}]{Ginsburg17}
{Ginsburg}, A., {Goddi}, C., {Kruijssen}, J.~M.~D., {et~al.} 2017, \apj, 842,
  92

\bibitem[{{Goldsmith} \& {Langer}(1999)}]{Goldsmith99}
{Goldsmith}, P.~F., \& {Langer}, W.~D. 1999, \apj, 517, 209

\bibitem[{{Gonz{\'a}lez-Alfonso} {et~al.}(2013){Gonz{\'a}lez-Alfonso},
  {Fischer}, {Bruderer}, {M{\"u}ller}, {Graci{\'a}-Carpio}, {Sturm}, {Lutz},
  {Poglitsch}, {Feuchtgruber}, {Veilleux}, {Contursi}, {Sternberg},
  {Hailey-Dunsheath}, {Verma}, {Christopher}, {Davies}, {Genzel}, \&
  {Tacconi}}]{GonzalezAlfonso13}
{Gonz{\'a}lez-Alfonso}, E., {Fischer}, J., {Bruderer}, S., {et~al.} 2013, \aap,
  550, A25

\bibitem[{{Gordon} {et~al.}(1993){Gordon}, {Berkermann}, {Mezger}, {Zylka},
  {Haslam}, {Kreysa}, {Sievers}, \& {Lemke}}]{Gordon93}
{Gordon}, M.~A., {Berkermann}, U., {Mezger}, P.~G., {et~al.} 1993, \aap, 280,
  208

\bibitem[{{Gravity Collaboration} {et~al.}(2018){Gravity Collaboration},
  {Abuter}, {Amorim}, {Anugu}, {Baub{\"o}ck}, {Benisty}, {Berger}, {Blind},
  {Bonnet}, {Brandner}, {Buron}, {Collin}, {Chapron}, {Cl{\'e}net}, {Coud{\'e}
  Du Foresto}, {de Zeeuw}, {Deen}, {Delplancke-Str{\"o}bele}, {Dembet},
  {Dexter}, {Duvert}, {Eckart}, {Eisenhauer}, {Finger}, {F{\"o}rster
  Schreiber}, {F{\'e}dou}, {Garcia}, {Garcia Lopez}, {Gao}, {Gendron},
  {Genzel}, {Gillessen}, {Gordo}, {Habibi}, {Haubois}, {Haug}, {Hau{\ss}mann},
  {Henning}, {Hippler}, {Horrobin}, {Hubert}, {Hubin}, {Jimenez Rosales},
  {Jochum}, {Jocou}, {Kaufer}, {Kellner}, {Kendrew}, {Kervella}, {Kok},
  {Kulas}, {Lacour}, {Lapeyr{\`e}re}, {Lazareff}, {Le Bouquin}, {L{\'e}na},
  {Lippa}, {Lenzen}, {M{\'e}rand}, {M{\"u}ler}, {Neumann}, {Ott}, {Palanca},
  {Paumard}, {Pasquini}, {Perraut}, {Perrin}, {Pfuhl}, {Plewa}, {Rabien},
  {Ram{\'{\i}}rez}, {Ramos}, {Rau}, {Rodr{\'{\i}}guez-Coira}, {Rohloff},
  {Rousset}, {Sanchez-Bermudez}, {Scheithauer}, {Sch{\"o}ller}, {Schuler},
  {Spyromilio}, {Straub}, {Straubmeier}, {Sturm}, {Tacconi}, {Tristram},
  {Vincent}, {von Fellenberg}, {Wank}, {Waisberg}, {Widmann}, {Wieprecht},
  {Wiest}, {Wiezorrek}, {Woillez}, {Yazici}, {Ziegler}, \& {Zins}}]{Gravity18}
{Gravity Collaboration}, {Abuter}, R., {Amorim}, A., {et~al.} 2018, \aap, 615,
  L15

\bibitem[{{Guzm{\'a}n} {et~al.}(2015){Guzm{\'a}n}, {Sanhueza}, {Contreras},
  {Smith}, {Jackson}, {Hoq}, \& {Rathborne}}]{Guzman15}
{Guzm{\'a}n}, A.~E., {Sanhueza}, P., {Contreras}, Y., {et~al.} 2015, \apj, 815,
  130

\bibitem[{{Guzm{\'a}n} {et~al.}(2017){Guzm{\'a}n}, {{\"O}berg}, {Huang},
  {Loomis}, \& {Qi}}]{Guzman17}
{Guzm{\'a}n}, V.~V., {{\"O}berg}, K.~I., {Huang}, J., {Loomis}, R., \& {Qi}, C.
  2017, \apj, 836, 30

\bibitem[{{Hacar} {et~al.}(2016){Hacar}, {Alves}, {Burkert}, \&
  {Goldsmith}}]{Hacar16}
{Hacar}, A., {Alves}, J., {Burkert}, A., \& {Goldsmith}, P. 2016, \aap, 591,
  A104

\bibitem[{{Halfen} {et~al.}(2017){Halfen}, {Woolf}, \& {Ziurys}}]{Halfen17}
{Halfen}, D.~T., {Woolf}, N.~J., \& {Ziurys}, L.~M. 2017, \apj, 845, 158

\bibitem[{{Hasegawa} {et~al.}(1994){Hasegawa}, {Sato}, {Whiteoak}, \&
  {Miyawaki}}]{Hasegawa94}
{Hasegawa}, T., {Sato}, F., {Whiteoak}, J.~B., \& {Miyawaki}, R. 1994, \apjl,
  429, L77

\bibitem[{{Hatchell}(2003)}]{Hatchell03}
{Hatchell}, J. 2003, \aap, 403, L25

\bibitem[{Herzberg \& Crawford(1946)}]{Herzberg45}
Herzberg, G., \& Crawford, B.~L. 1946, The Journal of Physical Chemistry, 50,
  288

\bibitem[{Higuchi {et~al.}(2015)Higuchi, Hasegawa, Saigo, Sanhueza, \&
  Chibueze}]{Higuchi15}
Higuchi, A.~E., Hasegawa, T., Saigo, K., Sanhueza, P., \& Chibueze, J.~O. 2015,
  \apj, 815, 1

\bibitem[{{Hoffman} {et~al.}(1979){Hoffman}, {Hodges}, {McElroy}, {Donahue}, \&
  {Kolpin}}]{Hoffman79}
{Hoffman}, J.~H., {Hodges}, R.~R., {McElroy}, M.~B., {Donahue}, T.~M., \&
  {Kolpin}, M. 1979, Science, 205, 49

\bibitem[{Hollis {et~al.}(2003)Hollis, Pedelty, Boboltz, Liu, Snyder, Palmer,
  Lovas, \& Jewell}]{Hollis03}
Hollis, J.~M., Pedelty, J.~A., Boboltz, D.~A., {et~al.} 2003, \apj, 596, L235

\bibitem[{Huettemeister {et~al.}(1995)Huettemeister, Wilson, Mauersberger,
  Lemme, Dahmen, \& Henkel}]{Huttem95}
Huettemeister, S., Wilson, T.~L., Mauersberger, R., {et~al.} 1995, \aap, 294,
  667

\bibitem[{H{\"u}ttemeister {et~al.}(1993)H{\"u}ttemeister, Wilson, Henkel, \&
  Mauersberger}]{Huttem93a}
H{\"u}ttemeister, S., Wilson, T.~L., Henkel, C., \& Mauersberger, R. 1993,
  \aap, 276, 445

\bibitem[{{Jacq} {et~al.}(1999){Jacq}, {Baudry}, {Walmsley}, \&
  {Caselli}}]{Jacq1999}
{Jacq}, T., {Baudry}, A., {Walmsley}, C.~M., \& {Caselli}, P. 1999, \aap, 347,
  957

\bibitem[{{Jacq} {et~al.}(1990){Jacq}, {Walmsley}, \& {Henkel}}]{Jacq1990}
{Jacq}, T., {Walmsley}, C.~M., \& {Henkel}, C. 1990, \aap, 228

\bibitem[{{Junk} \& {Svec}(1958)}]{Junk58}
{Junk}, G., \& {Svec}, H.~J. 1958, \gca, 14, 234

\bibitem[{{Kaifu} {et~al.}(1975){Kaifu}, {Morris}, {Palmer}, \&
  {Zuckerman}}]{Kaifu75}
{Kaifu}, N., {Morris}, M., {Palmer}, P., \& {Zuckerman}, B. 1975, \apj, 201, 98

\bibitem[{{Kauffmann} {et~al.}(2008){Kauffmann}, {Bertoldi}, {Bourke}, {Evans},
  \& {Lee}}]{Kauffmann08}
{Kauffmann}, J., {Bertoldi}, F., {Bourke}, T.~L., {Evans}, II, N.~J., \& {Lee},
  C.~W. 2008, \aap, 487, 993

\bibitem[{{Krieger} {et~al.}(2017){Krieger}, {Ott}, {Beuther}, {Walter},
  {Kruijssen}, {Meier}, {Mills}, {Contreras}, {Edwards}, {Ginsburg}, {Henkel},
  {Henshaw}, {Jackson}, {Kauffmann}, {Longmore}, {Mart{\'{\i}}n}, {Morris},
  {Pillai}, {Rickert}, {Rosolowsky}, {Shinnaga}, {Walsh}, {Yusef-Zadeh}, \&
  {Zhang}}]{Krieger17}
{Krieger}, N., {Ott}, J., {Beuther}, H., {et~al.} 2017, \apj, 850, 77

\bibitem[{{Lis} \& {Goldsmith}(1989)}]{Lis89}
{Lis}, D.~C., \& {Goldsmith}, P.~F. 1989, \apj, 337, 704

\bibitem[{Lis \& Goldsmith(1991)}]{Lis91}
Lis, D.~C., \& Goldsmith, P.~F. 1991, \apj, 369, 157

\bibitem[{Lis {et~al.}(1993)Lis, Goldsmith, Carlstrom, \& Scoville}]{Lis93}
Lis, D.~C., Goldsmith, P.~F., Carlstrom, J.~E., \& Scoville, N.~Z. 1993, \apj,
  402, 238

\bibitem[{{Lis} {et~al.}(1990){Lis}, {Goldsmith}, {Hills}, \&
  {Lasenby}}]{Lis90}
{Lis}, D.~C., {Goldsmith}, P.~F., {Hills}, R., \& {Lasenby}, J. 1990, in
  Astrophysics and Space Science Library, Vol. 158, Submillimetre Astronomy,
  ed. G.~D. {Watt} \& A.~S. {Webster}, 183--184

\bibitem[{{Lis} {et~al.}(2010){Lis}, {Wootten}, {Gerin}, \& {Roueff}}]{Lis10}
{Lis}, D.~C., {Wootten}, A., {Gerin}, M., \& {Roueff}, E. 2010, \apjl, 710, L49

\bibitem[{Lis {et~al.}(2014)Lis, Schilke, Bergin, Gerin, Black, Comito,
  De~Luca, Godard, Higgins, Le~Petit, Pearson, Pellegrini, Phillips, \&
  Yu}]{Lis14}
Lis, D.~C., Schilke, P., Bergin, E.~A., {et~al.} 2014, \apj, 785, 135

\bibitem[{Longmore {et~al.}(2012)Longmore, Rathborne, Bastian, Alves, Ascenso,
  Bally, Testi, Longmore, Battersby, Bressert, Purcell, Walsh, Jackson, Foster,
  Molinari, Meingast, Amorim, Lima, Marques, Moitinho, Pinhao, Rebordao, \&
  Santos}]{Longmore12}
Longmore, S.~N., Rathborne, J., Bastian, N., {et~al.} 2012, \apj, 746, 117

\bibitem[{{Loomis} {et~al.}(2013){Loomis}, {Zaleski}, {Steber}, {Neill},
  {Muckle}, {Harris}, {Hollis}, {Jewell}, {Lattanzi}, {Lovas}, {Martinez},
  {McCarthy}, {Remijan}, {Pate}, \& {Corby}}]{Loomis13}
{Loomis}, R.~A., {Zaleski}, D.~P., {Steber}, A.~L., {et~al.} 2013, \apjl, 765,
  L9

\bibitem[{{Ludovici} {et~al.}(2016){Ludovici}, {Lang}, {Morris}, {Mutel},
  {Mills}, {Toomey}, \& {Ott}}]{Ludovici16}
{Ludovici}, D.~A., {Lang}, C.~C., {Morris}, M.~R., {et~al.} 2016, \apj, 826,
  218

\bibitem[{{Manfroid} {et~al.}(2009){Manfroid}, {Jehin}, {Hutsem{\'e}kers},
  {Cochran}, {Zucconi}, {Arpigny}, {Schulz}, {St{\"u}we}, \&
  {Ilyin}}]{Manfroid09}
{Manfroid}, J., {Jehin}, E., {Hutsem{\'e}kers}, D., {et~al.} 2009, \aap, 503,
  613

\bibitem[{{Maret} {et~al.}(2009){Maret}, {Faure}, {Scifoni}, \&
  {Wiesenfeld}}]{Maret09}
{Maret}, S., {Faure}, A., {Scifoni}, E., \& {Wiesenfeld}, L. 2009, \mnras, 399,
  425

\bibitem[{{Marty} {et~al.}(2010){Marty}, {Zimmermann}, {Burnard}, {Wieler},
  {Heber}, {Burnett}, {Wiens}, \& {Bochsler}}]{Marty10}
{Marty}, B., {Zimmermann}, L., {Burnard}, P.~G., {et~al.} 2010, \gca, 74, 340

\bibitem[{Mehringer {et~al.}(1994)Mehringer, Goss, \& Palmer}]{Mehringer94}
Mehringer, D.~M., Goss, W.~M., \& Palmer, P. 1994, \apj, 434, 237

\bibitem[{Mehringer \& Menten(1997)}]{MM97}
Mehringer, D.~M., \& Menten, K.~M. 1997, \apj, 474, 346

\bibitem[{{Meibom} {et~al.}(2007){Meibom}, {Krot}, {Robert}, {Mostefaoui},
  {Russell}, {Petaev}, \& {Gounelle}}]{Meibom07}
{Meibom}, A., {Krot}, A.~N., {Robert}, F., {et~al.} 2007, \apjl, 656, L33

\bibitem[{Menten {et~al.}(1986)Menten, Walmsley, Henkel, \& Wilson}]{Menten86}
Menten, K.~M., Walmsley, C.~M., Henkel, C., \& Wilson, T.~L. 1986, \aap, 157,
  318

\bibitem[{{Messenger}(2000)}]{Messenger00}
{Messenger}, S. 2000, \nat, 404, 968

\bibitem[{Miao {et~al.}(1995)Miao, Mehringer, Kuan, \& Snyder}]{Miao95}
Miao, Y., Mehringer, D.~M., Kuan, Y.-J., \& Snyder, L.~E. 1995, \apj, 445, L59

\bibitem[{{Mills} {et~al.}(2015){Mills}, {Butterfield}, {Ludovici}, {Lang},
  {Ott}, {Morris}, \& {Schmitz}}]{Mills15}
{Mills}, E.~A.~C., {Butterfield}, N., {Ludovici}, D.~A., {et~al.} 2015, \apj,
  805, 72

\bibitem[{{Mills} {et~al.}(2014){Mills}, {Lang}, {Morris}, {Ott},
  {Butterfield}, {Ludovici}, {Schmitz}, \& {Schmiedeke}}]{Mills14}
{Mills}, E.~A.~C., {Lang}, C.~C., {Morris}, M.~R., {et~al.} 2014, in IAU
  Symposium, Vol. 303, IAU Symposium, ed. L.~O. {Sjouwerman}, C.~C. {Lang}, \&
  J.~{Ott}, 139--143

\bibitem[{{Mills} \& {Morris}(2013)}]{Mills13a}
{Mills}, E.~A.~C., \& {Morris}, M.~R. 2013, \apj, 772, 105

\bibitem[{{Molinari} {et~al.}(2011){Molinari}, {Bally}, {Noriega-Crespo},
  {Compi{\`e}gne}, {Bernard}, {Paradis}, {Martin}, {Testi}, {Barlow}, {Moore},
  {Plume}, {Swinyard}, {Zavagno}, {Calzoletti}, {Di Giorgio}, {Elia},
  {Faustini}, {Natoli}, {Pestalozzi}, {Pezzuto}, {Piacentini}, {Polenta},
  {Polychroni}, {Schisano}, {Traficante}, {Veneziani}, {Battersby}, {Burton},
  {Carey}, {Fukui}, {Li}, {Lord}, {Morgan}, {Motte}, {Schuller},
  {Stringfellow}, {Tan}, {Thompson}, {Ward-Thompson}, {White}, \&
  {Umana}}]{Molinari11}
{Molinari}, S., {Bally}, J., {Noriega-Crespo}, A., {et~al.} 2011, \apjl, 735,
  L33

\bibitem[{{Morris} {et~al.}(1983){Morris}, {Polish}, {Zuckerman}, \&
  {Kaifu}}]{Morris83}
{Morris}, M., {Polish}, N., {Zuckerman}, B., \& {Kaifu}, N. 1983, \aj, 88, 1228

\bibitem[{{Morris} {et~al.}(1976){Morris}, {Turner}, {Palmer}, \&
  {Zuckerman}}]{Morris76}
{Morris}, M., {Turner}, B.~E., {Palmer}, P., \& {Zuckerman}, B. 1976, \apj,
  205, 82

\bibitem[{{M{\"u}ller} {et~al.}(2005){M{\"u}ller}, {Schl{\"o}der}, {Stutzki},
  \& {Winnewisser}}]{Muller05}
{M{\"u}ller}, H.~S.~P., {Schl{\"o}der}, F., {Stutzki}, J., \& {Winnewisser}, G.
  2005, Journal of Molecular Structure, 742, 215

\bibitem[{{M{\"u}ller} {et~al.}(2001){M{\"u}ller}, {Thorwirth}, {Roth}, \&
  {Winnewisser}}]{Muller01}
{M{\"u}ller}, H.~S.~P., {Thorwirth}, S., {Roth}, D.~A., \& {Winnewisser}, G.
  2001, \aap, 370, L49

\bibitem[{{M{\"u}ller} {et~al.}(2016){M{\"u}ller}, {Belloche}, {Xu}, {Lees},
  {Garrod}, {Walters}, {van Wijngaarden}, {Lewen}, {Schlemmer}, \&
  {Menten}}]{Muller16}
{M{\"u}ller}, H.~S.~P., {Belloche}, A., {Xu}, L.-H., {et~al.} 2016, \aap, 587,
  A92

\bibitem[{{Neill} {et~al.}(2012){Neill}, {Muckle}, {Zaleski}, {Steber}, {Pate},
  {Lattanzi}, {Spezzano}, {McCarthy}, \& {Remijan}}]{Neill12}
{Neill}, J.~L., {Muckle}, M.~T., {Zaleski}, D.~P., {et~al.} 2012, \apj, 755,
  153

\bibitem[{{Nummelin} {et~al.}(2000){Nummelin}, {Bergman}, {Hjalmarson},
  {Friberg}, {Irvine}, {Millar}, {Ohishi}, \& {Saito}}]{Nummelin00}
{Nummelin}, A., {Bergman}, P., {Hjalmarson}, {\AA}., {et~al.} 2000, \apjs, 128,
  213

\bibitem[{Oka {et~al.}(1971)Oka, Shimizu, Shimizu, \& Watson}]{Oka71}
Oka, T., Shimizu, F.~O., Shimizu, T., \& Watson, J. K.~G. 1971, \aj, 165, L15

\bibitem[{{Owen} {et~al.}(2001){Owen}, {Mahaffy}, {Niemann}, {Atreya}, \&
  {Wong}}]{Owen01}
{Owen}, T., {Mahaffy}, P.~R., {Niemann}, H.~B., {Atreya}, S., \& {Wong}, M.
  2001, \apjl, 553, L77

\bibitem[{{Parise} {et~al.}(2006){Parise}, {Ceccarelli}, {Tielens}, {Castets},
  {Caux}, {Lefloch}, \& {Maret}}]{Parise06}
{Parise}, B., {Ceccarelli}, C., {Tielens}, A.~G.~G.~M., {et~al.} 2006, \aap,
  453, 949

\bibitem[{{Pauls} {et~al.}(1983){Pauls}, {Wilson}, {Bieging}, \&
  {Martin}}]{Pauls83}
{Pauls}, A., {Wilson}, T.~L., {Bieging}, J.~H., \& {Martin}, R.~N. 1983, \aap,
  124, 23

\bibitem[{Pei {et~al.}(2000)Pei, Liu, \& Snyder}]{Pei00}
Pei, C.~C., Liu, S.-Y., \& Snyder, L.~E. 2000, \apj, 530, 800

\bibitem[{{Peng} {et~al.}(1993){Peng}, {Vogel}, \& {Carlstrom}}]{Peng93}
{Peng}, Y., {Vogel}, S.~N., \& {Carlstrom}, J.~E. 1993, \apj, 418, 255

\bibitem[{{Pfalzner}(2013)}]{Pfalzner13}
{Pfalzner}, S. 2013, \aap, 549, A82

\bibitem[{{Phillips} {et~al.}(1979){Phillips}, {Huggins}, {Wannier}, \&
  {Scoville}}]{Phillips79}
{Phillips}, T.~G., {Huggins}, P.~J., {Wannier}, P.~G., \& {Scoville}, N.~Z.
  1979, \apj, 231, 720

\bibitem[{{Pickett} {et~al.}(1998){Pickett}, {Poynter}, {Cohen}, {Delitsky},
  {Pearson}, \& {M{\"u}ller}}]{Pickett98}
{Pickett}, H.~M., {Poynter}, R.~L., {Cohen}, E.~A., {et~al.} 1998, \jqsrt, 60,
  883

\bibitem[{{Qin} {et~al.}(2011){Qin}, {Schilke}, {Rolffs}, {Comito}, {Lis}, \&
  {Zhang}}]{Qin11}
{Qin}, S.-L., {Schilke}, P., {Rolffs}, R., {et~al.} 2011, \aap, 530, L9

\bibitem[{Reid {et~al.}(2009)Reid, Menten, Zheng, Brunthaler, \& Xu}]{Reid09}
Reid, M.~J., Menten, K.~M., Zheng, X.~W., Brunthaler, A., \& Xu, Y. 2009, \apj,
  705, 1548

\bibitem[{{Reid} {et~al.}(1988){Reid}, {Schneps}, {Moran}, {Gwinn}, {Genzel},
  {Downes}, \& {Roennaeng}}]{Reid88}
{Reid}, M.~J., {Schneps}, M.~H., {Moran}, J.~M., {et~al.} 1988, \apj, 330, 809

\bibitem[{{Riquelme} {et~al.}(2010){Riquelme}, {Amo-Baladr{\'o}n},
  {Mart{\'{\i}}n-Pintado}, {Mauersberger}, {Mart{\'{\i}}n}, \&
  {Bronfman}}]{Riq10}
{Riquelme}, D., {Amo-Baladr{\'o}n}, M.~A., {Mart{\'{\i}}n-Pintado}, J.,
  {et~al.} 2010, \aap, 523, A51

\bibitem[{{Rodgers} \& {Charnley}(2008)}]{RC08}
{Rodgers}, S.~D., \& {Charnley}, S.~B. 2008, \mnras, 385, L48

\bibitem[{{Rodgers} \& {Millar}(1996)}]{RM96}
{Rodgers}, S.~D., \& {Millar}, T.~J. 1996, \mnras, 280, 1046

\bibitem[{{S{\'a}nchez-Monge} {et~al.}(2017){S{\'a}nchez-Monge}, {Schilke},
  {Schmiedeke}, {Ginsburg}, {Cesaroni}, {Lis}, {Qin}, {M{\"u}ller}, {Bergin},
  {Comito}, \& {M{\"o}ller}}]{SM17}
{S{\'a}nchez-Monge}, {\'A}., {Schilke}, P., {Schmiedeke}, A., {et~al.} 2017,
  \aap, 604, A6

\bibitem[{Schmiedeke {et~al.}(2016)Schmiedeke, Schilke, M{\"o}ller,
  S{\'a}nchez-Monge, Bergin, Comito, Csengeri, Lis, Molinari, Qin, \&
  Rolffs}]{Schmiedeke16}
Schmiedeke, A., Schilke, P., M{\"o}ller, T., {et~al.} 2016, \aap, 588, A143

\bibitem[{{Scoville} {et~al.}(1975{\natexlab{a}}){Scoville}, {Solomon}, \&
  {Penzias}}]{Scoville1975}
{Scoville}, N.~Z., {Solomon}, P.~M., \& {Penzias}, A.~A. 1975{\natexlab{a}},
  \apj, 201, 352

\bibitem[{{Scoville} {et~al.}(1975{\natexlab{b}}){Scoville}, {Solomon}, \&
  {Penzias}}]{Scoville75}
---. 1975{\natexlab{b}}, \apj, 201, 352

\bibitem[{{Shah} \& {Wootten}(2001)}]{Shah01}
{Shah}, R.~Y., \& {Wootten}, A. 2001, \apj, 554, 933

\bibitem[{Snyder {et~al.}(1994)Snyder, Kuan, \& Miao}]{Snyder94}
Snyder, L.~E., Kuan, Y.~J., \& Miao, Y. 1994, The Structure and Content of
  Molecular Clouds, 439, 187

\bibitem[{Sweitzer {et~al.}(1979)Sweitzer, Palmer, Morris, Turner, \&
  Zuckerman}]{Sweitzer79}
Sweitzer, J.~S., Palmer, P., Morris, M., Turner, B.~E., \& Zuckerman, B. 1979,
  \apj, 227, 415

\bibitem[{{Taquet} {et~al.}(2012){Taquet}, {Ceccarelli}, \&
  {Kahane}}]{Taquet12}
{Taquet}, V., {Ceccarelli}, C., \& {Kahane}, C. 2012, \apjl, 748, L3

\bibitem[{{Visser} {et~al.}(2018){Visser}, {Bruderer}, {Cazzoletti},
  {Facchini}, {Heays}, \& {van Dishoeck}}]{Visser18}
{Visser}, R., {Bruderer}, S., {Cazzoletti}, P., {et~al.} 2018, \aap, 615, A75

\bibitem[{{Viti} \& {Williams}(1999)}]{Viti99}
{Viti}, S., \& {Williams}, D.~A. 1999, \mnras, 305, 755

\bibitem[{Vogel {et~al.}(1987)Vogel, Genzel, \& Palmer}]{Vogel87}
Vogel, S.~N., Genzel, R., \& Palmer, P. 1987, \apj, 316, 243

\bibitem[{{Walker} {et~al.}(2016){Walker}, {Longmore}, {Bastian}, {Kruijssen},
  {Rathborne}, {Galv{\'a}n-Madrid}, \& {Liu}}]{Walker16}
{Walker}, D.~L., {Longmore}, S.~N., {Bastian}, N., {et~al.} 2016, \mnras, 457,
  4536

\bibitem[{{Walmsley} {et~al.}(1987){Walmsley}, {Hermsen}, {Henkel},
  {Mauersberger}, \& {Wilson}}]{Walmsley87}
{Walmsley}, C.~M., {Hermsen}, W., {Henkel}, C., {Mauersberger}, R., \&
  {Wilson}, T.~L. 1987, \aap, 172, 311

\bibitem[{{Wannier} {et~al.}(1981){Wannier}, {Linke}, \& {Penzias}}]{Wannier81}
{Wannier}, P.~G., {Linke}, R.~A., \& {Penzias}, A.~A. 1981, \apj, 247, 522

\bibitem[{Wilson {et~al.}(2006)Wilson, Henkel, \& H{\"u}ttemeister}]{Wilson06}
Wilson, T.~L., Henkel, C., \& H{\"u}ttemeister, S. 2006, \aap, 460, 533

\bibitem[{{Wilson} \& {Rood}(1994)}]{WR94}
{Wilson}, T.~L., \& {Rood}, R. 1994, \araa, 32, 191

\bibitem[{{Wilson} {et~al.}(2011){Wilson}, {Ferris}, {Axtens}, {Brown},
  {Davis}, {Hampson}, {Leach}, {Roberts}, {Saunders}, {Koribalski}, {Caswell},
  {Lenc}, {Stevens}, {Voronkov}, {Wieringa}, {Brooks}, {Edwards}, {Ekers},
  {Emonts}, {Hindson}, {Johnston}, {Maddison}, {Mahony}, {Malu}, {Massardi},
  {Mao}, {McConnell}, {Norris}, {Schnitzeler}, {Subrahmanyan}, {Urquhart},
  {Thompson}, \& {Wark}}]{Wilson11}
{Wilson}, W.~E., {Ferris}, R.~H., {Axtens}, P., {et~al.} 2011, \mnras, 416, 832

\bibitem[{{Wirstr{\"o}m} {et~al.}(2012){Wirstr{\"o}m}, {Charnley}, {Cordiner},
  \& {Milam}}]{Wirstrom12}
{Wirstr{\"o}m}, E.~S., {Charnley}, S.~B., {Cordiner}, M.~A., \& {Milam}, S.~N.
  2012, \apjl, 757, L11

\bibitem[{{Xu} \& {Lovas}(1997)}]{Xu97}
{Xu}, L.-H., \& {Lovas}, F.~J. 1997, Journal of Physical and Chemical Reference
  Data, 26, 17

\bibitem[{{Zaleski} {et~al.}(2013){Zaleski}, {Seifert}, {Steber}, {Muckle},
  {Loomis}, {Corby}, {Martinez}, {Crabtree}, {Jewell}, {Hollis}, {Lovas},
  {Vasquez}, {Nyiramahirwe}, {Sciortino}, {Johnson}, {McCarthy}, {Remijan}, \&
  {Pate}}]{Zaleski13}
{Zaleski}, D.~P., {Seifert}, N.~A., {Steber}, A.~L., {et~al.} 2013, \apjl, 765,
  L10

\bibitem[{Zhao \& Wright(2011)}]{Zhao11}
Zhao, J.-H., \& Wright, M. C.~H. 2011, \apj, 742, 50

\end{thebibliography}

\clearpage

\begin{table}
\caption{Properties of Imaged Lines}
\centering
\begin{tabular}{l l l l l l l l l}
\hline\hline
Species &Transition & Rest Frequency\footnotemark[1] & Beam size & Spectral Resolution & Telescope & E$_{u}$ & log A$_{ul}$ & Notes \\
              &                & (GHz)                                   &                   & (km s$^{-1}$ )           &                  &  (K)        & log (s$^{-1}$) & \\ 
\hline
$^{14}$NH$_3$ & (1,1) & 23.6944955 & 2\arcsec.83 $\times$ 2\arcsec.57 & 1.58 & VLA & 23.3 & -6.77534 & \\
$^{14}$NH$_3$ & (2,2) & 23.7226333 & 2\arcsec.83 $\times$ 2\arcsec.56 & 1.58 & VLA & 64.4 & -6.64906 & \\
$^{14}$NH$_3$ & (3,3) & 23.8701292 & 2\arcsec.86 $\times$ 2\arcsec.49 & 3.14 & VLA & 124 & -6.59019 & \\
$^{14}$NH$_3$ & (4,4) & 24.1394163 & 2\arcsec.82 $\times$ 2\arcsec.46 & 3.10 & VLA & 201 & -6.54820 & \\
$^{14}$NH$_3$ & (5,5) & 24.5329887& 2\arcsec.73 $\times$ 2\arcsec.48 & 3.05 & VLA & 295 & -6.51013 &  \\
$^{14}$NH$_3$ & (6,6) & 25.056025 & 2\arcsec.74 $\times$ 2\arcsec.37 & 2.99 & VLA & 408 & -6.47143 & \\
$^{14}$NH$_3$ & (7,7) & 25.715182 & 2\arcsec.67 $\times$ 2\arcsec.32 & 2.91 & VLA & 539 & -6.42979 & \\
$^{14}$NH$_3$ & (9,9) & 27.477943 & 2\arcsec.84 $\times$ 2\arcsec.32 & 2.73 & VLA & 853 & -6.33416 & \\
$^{14}$NH$_3$ & (10,9) & 24.205287 & 2\arcsec.81 $\times$ 2\arcsec.45 & 3.10 & VLA & 1138 & -6.58442 & \\
$^{14}$NH$_3$ & (12,11)\footnotemark[2] & 25.695230 & 2\arcsec.39 $\times$ 1\arcsec.95 & 2.92 & VLA & 1579 & -6.48740 & \\
$^{14}$NH$_3$ & (14,13)& 27.772294 & 2\arcsec.65 $\times$ 2\arcsec.08 & 2.70 & VLA & 2090 & -6.37444 & \footnotemark[3] \\
$^{14}$NH$_3$ & (19,18)& 36.287873 & 1\arcsec.97 $\times$ 1\arcsec.77 & 2.07 & VLA & 3671 & -6.01578 & \footnotemark[3] \\
\hline
$^{15}$NH$_3$ & (1,1) & 22.6249295 &  33\arcsec.4 & 0.32 & GBT & 23.8 & -6.83648 &    \\
$^{15}$NH$_3$ & (2,2) & 22.6498434 &  33\arcsec.4 & 0.32 & GBT & 64.9 & -6.71019 & \\
$^{15}$NH$_3$ & (3,3) & 22.7894217 &  33\arcsec.2 & 0.32 & GBT & 124 & -6.65097 & \\
$^{15}$NH$_3$ & (4,4) & 23.0460158 &  32\arcsec.8 & 0.31 & GBT & 201 & -6.60839 & \\
$^{15}$NH$_3$ & (5,5) & 23.4219824 &  32\arcsec.3 & 0.31 & GBT & 296 & -6.56956 & \\
$^{15}$NH$_3$ & (6,6) & 23.9223132 &  31\arcsec.6 & 0.30 & GBT & 409 & -6.53296 &  \\
                          &          &                 & 2\arcsec.81 $\times$ 2\arcsec.54 & 3.13 & VLA  &        &    &\\
$^{15}$NH$_3$ & (7,7) & 24.5534251 & 2\arcsec.73 $\times$ 2\arcsec.48  & 3.05 & VLA & 539 & -6.49129 & \\
$^{15}$NH$_3$ & (8,8) & 25.3235100 & 2\arcsec.71 $\times$ 2\arcsec.35 & 2.96 & VLA & 688 & -6.44556 & \footnotemark[3]  \\
\hline
$^{14}$NH$_2$D & 2$_{1 2}$-2$_{0 2}$ & 49.9627602 & 11\arcsec.23 $\times$ 8\arcsec.46 & 6.00 & ATCA & 50 & -6.21977 &  \\
$^{14}$NH$_2$D & 3$_{1 3}$-3$_{0 3}$ & 43.0422277 & 4\arcsec.40 $\times$ 3\arcsec.38 & 6.96 & ATCA & 95 & -6.42485 &  \\
$^{14}$NH$_2$D & 4$_{1 4}$-4$_{0 4}$ & 25.0237707 & 2\arcsec.71 $\times$ 2\arcsec.40 & 2.99 & VLA & 152 & -7.01363 &  \\
\hline
$^{12}$CH$_3$OH & 6$_2$-6$_1$     & 25.018176 & 2\arcsec.74 $\times$ 2\arcsec.38 & 3.00 & VLA  & 71 & -7.06583 & \\
$^{12}$CH$_3$OH & 7$_2$-7$_1$     & 25.124932 & 2\arcsec.73 $\times$ 2\arcsec.37 & 2.98 & VLA  & 87 & -7.04964 & \\
$^{12}$CH$_3$OH & 8$_2$-8$_1$     & 25.294483 & 2\arcsec.68 $\times$ 2\arcsec.37 & 2.96 & VLA  & 106 & -7.03163 &   \\
$^{12}$CH$_3$OH & 9$_2$-9$_1$     & 25.541467 & 2\arcsec.69 $\times$ 2\arcsec.33 & 2.93 & VLA  & 127 & -7.01078 & \\
$^{12}$CH$_3$OH & 10$_2$-10$_1$ & 25.878337 & 2\arcsec.62 $\times$ 2\arcsec.33 & 2.90 & VLA  & 150 & -6.98647 & \\
$^{12}$CH$_3$OH & 13$_2$-13$_1$ & 27.472583 & 2\arcsec.52 $\times$ 2\arcsec.00 & 2.73 & VLA  &  234 & -6.89622 &  \\
$^{12}$CH$_3$OH & 25$_2$-25$_1$ & 27.470897 & 2\arcsec.57 $\times$ 2\arcsec.02 & 2.73 & VLA  &  776 & -7.26138 &  \footnotemark[3] \\
$^{12}$CH$_3$OH & 26$_2$-26$_1$ & 25.787059 & 2\arcsec.52 $\times$ 2\arcsec.00 & 2.91 & VLA  &  836 & -7.40872 &  \footnotemark[3] \\
\hline
$^{13}$CH$_3$OH & 2$_2$-2$_1$     & 27.05303 & 2\arcsec.90 $\times$ 2\arcsec.36 & 2.77 & VLA  & 29 & -7.13418 &  \footnotemark[3] \\
$^{13}$CH$_3$OH & 3$_2$-3$_1$     & 27.04728 & 2\arcsec.90 $\times$ 2\arcsec.35 & 2.77 & VLA  & 36 & -7.03518 &  \footnotemark[3] \\
$^{13}$CH$_3$OH & 4$_2$-4$_1$     & 27.05054 & 2\arcsec.90 $\times$ 2\arcsec.36 & 2.77 & VLA  & 45 & -6.99847 &  \\
$^{13}$CH$_3$OH & 5$_2$-5$_1$     & 27.07193 & 2\arcsec.90 $\times$ 2\arcsec.35 & 2.77 & VLA  & 56 & -6.97779 &  \\
$^{13}$CH$_3$OH & 6$_2$-6$_1$     & 27.12272 & 2\arcsec.90 $\times$ 2\arcsec.36 & 2.76 & VLA  & 70 & -6.96210 &  \\
$^{13}$CH$_3$OH & 7$_2$-7$_1$     & 27.21557 & 2\arcsec.89 $\times$ 2\arcsec.33 & 2.75 & VLA  & 86 & -6.94729 &  \\
$^{13}$CH$_3$OH & 8$_2$-8$_1$     & 27.36409 & 3\arcsec.07 $\times$ 2\arcsec.41 & 2.74 & VLA  & 104 & -6.93126 &  \\
$^{13}$CH$_3$OH & 9$_2$-9$_1$     & 27.58163 & 2\arcsec.82 $\times$ 2\arcsec.31 & 2.72 & VLA  & 124 & -6.91305 &  \\
$^{13}$CH$_3$OH & 10$_2$-10$_1$ & 27.88003 & 2\arcsec.81 $\times$ 2\arcsec.30 & 2.69 & VLA  & 147 & -6.89190 &  \\
\hline\hline
\end{tabular}
\footnotetext[1]{Frequencies are from the JPL Submillimeter, Millimeter, and Microwave Spectral Line Catalog \citep{Pickett98}}
\footnotetext[2]{This transition is blended with the H63-$\alpha$ line.}
\footnotetext[3]{This transition was not detected toward N2}
\label{data}
\end{table}

\clearpage

\begin{table}
\caption{Rotational Temperatures}
\centering
\begin{tabular}{l l l }
\hline\hline
& N1 &  N2  \\
& T(K)  & T (K)  \rule[-1.2ex]{0pt}{0pt} \\
\hline
$^{14}$NH$_3$ (metastable) & 350$^{+50}_{-40}$ & 600$^{+600}_{-300}$ \rule{0pt}{2.6ex} \\
$^{14}$NH$_3$ (nonmetastable) & 580$^{+50}_{-40}$ & 320$^{+20}_{-20}$ \rule{0pt}{2.6ex} \\
$^{15}$NH$_3$ & 70$^{+10}_{-10}$ & 70$^{+10}_{-10}$ \rule{0pt}{2.6ex} \\
& 300$^{+200}_{-100}$ & 200$^{+200}_{-100}$ \rule{0pt}{2.6ex} \\
$^{14}$NH$_2$D & 120$^{+80}_{-40}$ &  $>$350  \rule[-1.2ex]{0pt}{0pt} \\
\hline
$^{12}$CH$_3$OH & 130$^{+60}_{-30}$  & 90$^{+30}_{-20}$ \rule{0pt}{2.6ex} \\
& 200$^{+10}_{-10}$  & 280$^{+40}_{-30}$ \rule{0pt}{2.6ex} \\
$^{13}$CH$_3$OH & 130$^{+20}_{-20}$ & 160$^{+120}_{-50}$ \rule{0pt}{2.6ex} \rule[-1.2ex]{0pt}{0pt}\\
\hline\hline
\end{tabular}
\label{rot}
\end{table}

\clearpage 

\begin{table}
\caption{Total Column Densities and Abundances}
\centering
\begin{tabular}{l c c c c | c c c c}
\hline\hline
\multicolumn{9}{l}{ }\\
\multicolumn{9}{l}{\bf Column Densities} \\
 & \multicolumn{4}{c}{N1} &  \multicolumn{4}{c}{N2}  \\
\hline
 Molecule & $J_u$ & $N_u$ & T$_{rot}$ & N$_{T}$ & $J_u$ & $N_u$ & T$_{rot}$ & N$_{T}$ \\
 & & cm$^{-2}$ & (K) & cm$^{-2}$ & & cm$^{-2}$ (K) & cm$^{-2}$ \\
\hline
$^{14}$NH$_3$ & 6 & 1.8$\pm$0.2$\times10^{18}$ & 300 & 5.6$\pm$1.2$\times10^{19}$ & 6 & 1.5$\pm$0.5$\times10^{17}$ & 300 & 5.0$\pm$1.5$\times10^{18}$ \\
 & 1 & 5.2$\pm$0.1$\times10^{18}$ & 75 & \footnotemark[1]9.0$\pm$2.0$\times10^{19}$ & 1 & 1.9$\pm$0.2$\times10^{18}$ & 75 &  \footnotemark[1]3.3$\pm$1.0$\times10^{19}$ \\
 & 6 & 1.8$\pm$0.2$\times10^{18}$ & 300 & \footnotemark[1]5.5$\pm$1.2$\times10^{19}$ & 6 & 3.7$\pm$0.1$\times10^{17}$ & 300 &  \footnotemark[1]1.1$\pm$0.3$\times10^{19}$ \\
 & & & 75 \& 300 & \footnotemark[1]1.5$\pm$0.3$\times10^{20}$ & & &  75 \& 300 &  \footnotemark[1]4.4$\pm$1.3$\times10^{19}$ \\
& & & & & 1 & 8.4$\pm$0.1$\times10^{17}$ & 75 &  \footnotemark[2]1.5$\pm$0.4$\times10^{19}$ \\
& & & & & 6 & 1.6$\pm$0.5$\times10^{17}$ & 300 &  \footnotemark[2]4.9$\pm$1.5$\times10^{18}$ \\
& & & & & & & 75 \& 300 &  \footnotemark[2]2.0$\pm$0.6$\times10^{19}$ \\
\hline
$^{15}$NH$_3$ & 1 & 1.2$\pm$0.1$\times10^{16}$ & 75 &  2.0$\pm$0.4$\times10^{17}$ & 1 & 4.2$\pm$0.4$\times10^{15}$ & 75  & 7.3$\pm$2.2$\times10^{16}$ \\
& 6 & 4.1$\pm$0.4$\times10^{15}$ & 300 & 1.2$\pm$0.3$\times10^{17}$ & 6 & 8.2$\pm$0.3$\times10^{14}$ & 300 & 2.5$\pm$0.7$\times10^{16}$ \\
& & & 75 \& 300 &  3.2$\pm$0.7$\times10^{17}$ & & & 75 \& 300 &  9.8$\pm$2.9$\times10^{16}$ \\
\hline
$^{14}$NH$_2$D & 3 & 8.2$\pm$2.5$\times10^{15}$ & 120 &  4.3$\pm$1.5$\times10^{17}$ & 3 & 1.1$\pm$0.3$\times10^{16}$ & 300 &  1.4$\pm$0.5$\times10^{18}$ \\
\hline
$^{12}$CH$_3$OH & 6 & 1.1$\pm$0.1$\times10^{17}$ & 130 & \footnotemark[3]1.9$\pm$0.4$\times10^{19}$ & 6 & 9.5$\pm$2.5$\times10^{16}$ & 160 & \footnotemark[3]2.7$\pm$0.8$\times10^{19}$ \\
\hline
$^{13}$CH$_3$OH & 6 & 4.4$\pm$0.6$\times10^{15}$ & 130 & 7.6$\pm$1.7$\times10^{17}$ & 6 & 3.8$\pm$1.0$\times10^{15}$ & 160 & 1.1$\pm$0.3$\times10^{18}$ \\
\hline
\hline
\multicolumn{9}{l}{ }\\
\multicolumn{9}{l}{\bf Abundances} \\
 & \multicolumn{4}{c}{N1} &  \multicolumn{4}{c}{N2}  \\
\hline
[$^{14}$NH$_3$/H$_2$] & & & & \footnotemark[1]3.2$\pm$0.7$\times10^{-6}$ &  & & & \footnotemark[1]3.0$\pm$0.9$\times10^{-6}$ \\
 & & & & & & & &\footnotemark[2]1.3$\pm$0.4$\times10^{-6}$ \\
 {[$^{14}$NH$_3$/$^{15}$NH$_3$]} & & & & \footnotemark[4]460$\pm$140 & & & & \footnotemark[4]210$\pm$90 \\
{[$^{14}$NH$_2$D/$^{14}$NH$_3$]} & & & & \footnotemark[1]0.003$\pm$0.001 & & & & \footnotemark[1]0.03$\pm$0.01 \\
  & & & & & & & &  \footnotemark[2]0.07$\pm$0.03 \\
\hline
{[$^{12}$CH$_3$OH/H$_2$]} & & & & \footnotemark[3]4.2$\pm$0.9$\times10^{-7}$ & & & & \footnotemark[3]1.8$\pm$0.6$\times10^{-6}$ \\
\hline\hline
\end{tabular}
\label{abund}
\footnotetext[1]{Value computed from the isotopologue assuming [14N/15N]=450}
\footnotetext[2]{Value computed from the isotopologue assuming [14N/15N]=200}
\footnotetext[3]{Value computed from the isotopologue assuming [12C/13C]=25}
\footnotetext[4]{Value computed using only a T=300 K component}
\end{table}

\clearpage

\clearpage

\clearpage

\clearpage

\clearpage

\clearpage

\clearpage

\appendix

\begin{figure*}
\hspace{-1cm}\includegraphics[scale=0.5]{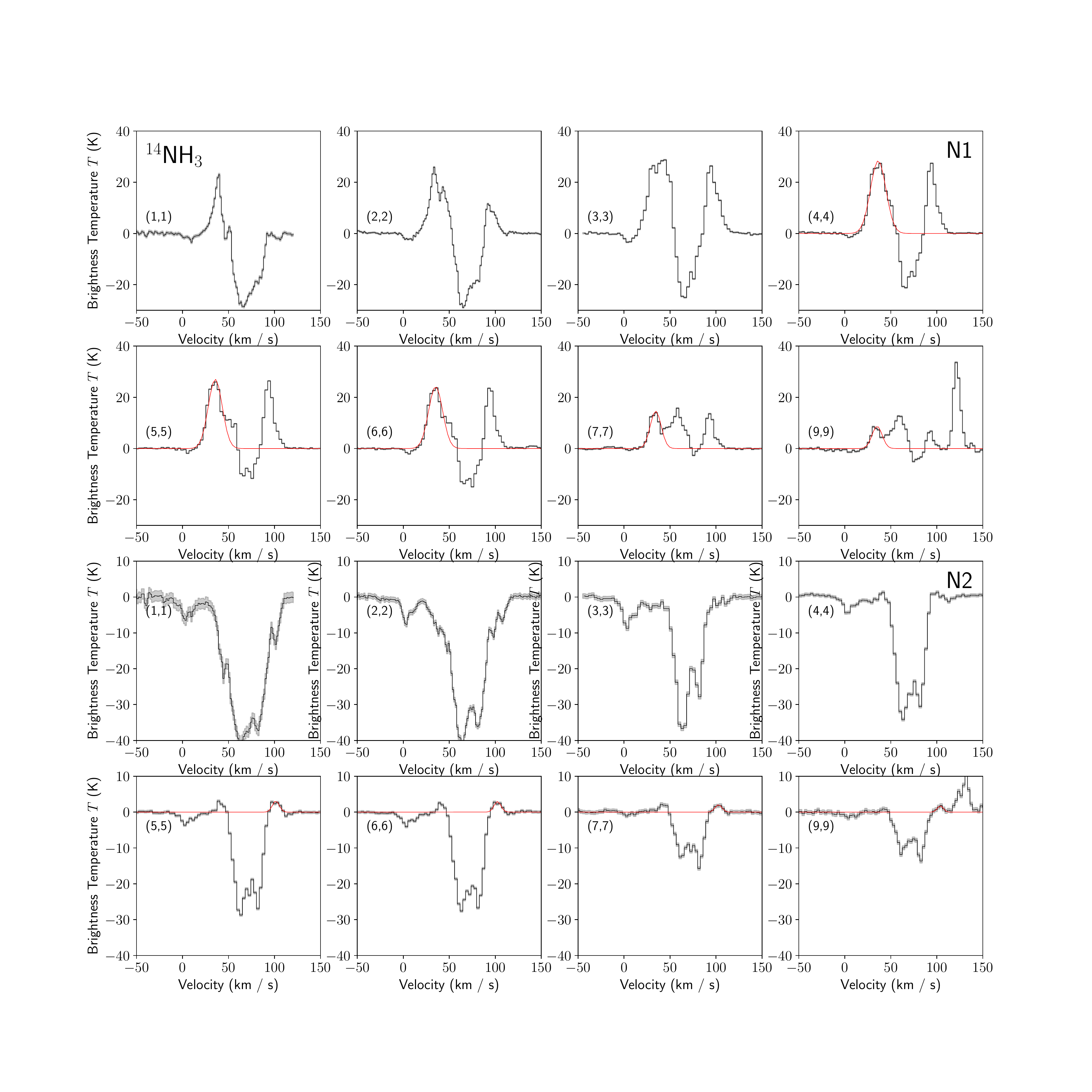}
\caption{Fitted spectra of \ammain\, metastable transitions. Where possible ($J>4$ for the N2 source) a single Gaussian is fit to the blended hyperfine satellite lines that appear in emission. Red curves show Gaussian fits to each line. For N1 (v=63.5 \kms) fits are performed for the blueshifted satellite lines, as these are subject to less absorption. For N2 (v=73 \kms), fits are performed for the redshifted satellite lines, for the same reason. }
\label{am_fit}
\end{figure*}

\begin{figure*}
\hspace{-1cm}\includegraphics[scale=0.5]{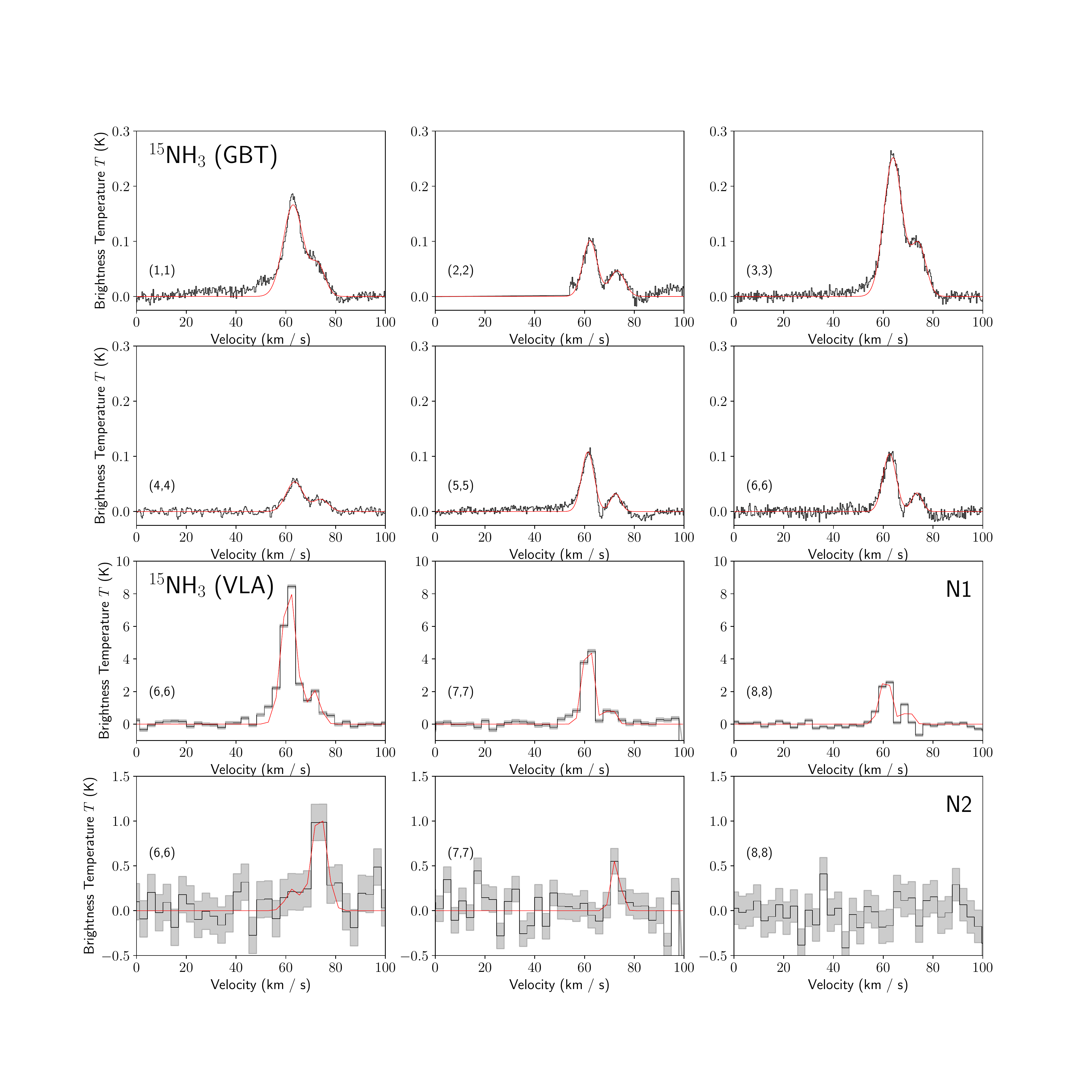}
\caption{Fitted spectra of \amiso\, metastable transitions. Spectra from GBT and VLA observations are labeled. For each line, two Gaussian components are fit, for the contribution from the N1 source at 63.5 \kms\, and the N2 source at 73 \kms. }
\label{amiso_fit}
\end{figure*}

\begin{figure*}
\hspace{-1cm}\includegraphics[scale=0.5]{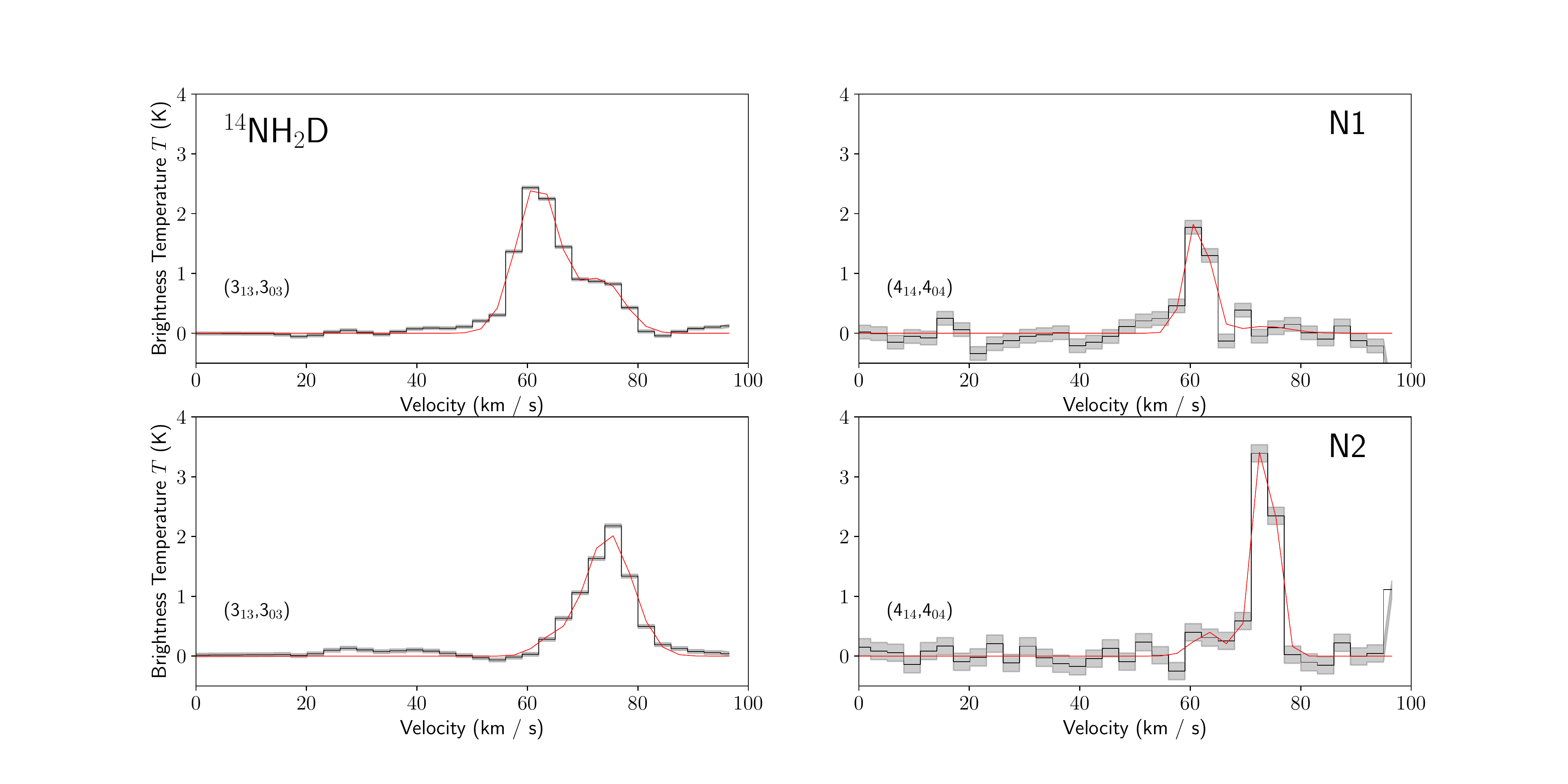}
\caption{Fitted spectra of \deut\, transitions. The $3_{13}-3_{03}$ transition is from ATCA observations, while the $4_{14}-4_{04}$ transition is from VLA observations. For each line, two Gaussian components are fit, for the contribution from the N1 source at 63.5 \kms\, and the N2 source at 73 \kms. }
\label{deut_fit}
\end{figure*}

\begin{figure*}
\hspace{-1cm}\includegraphics[scale=0.5]{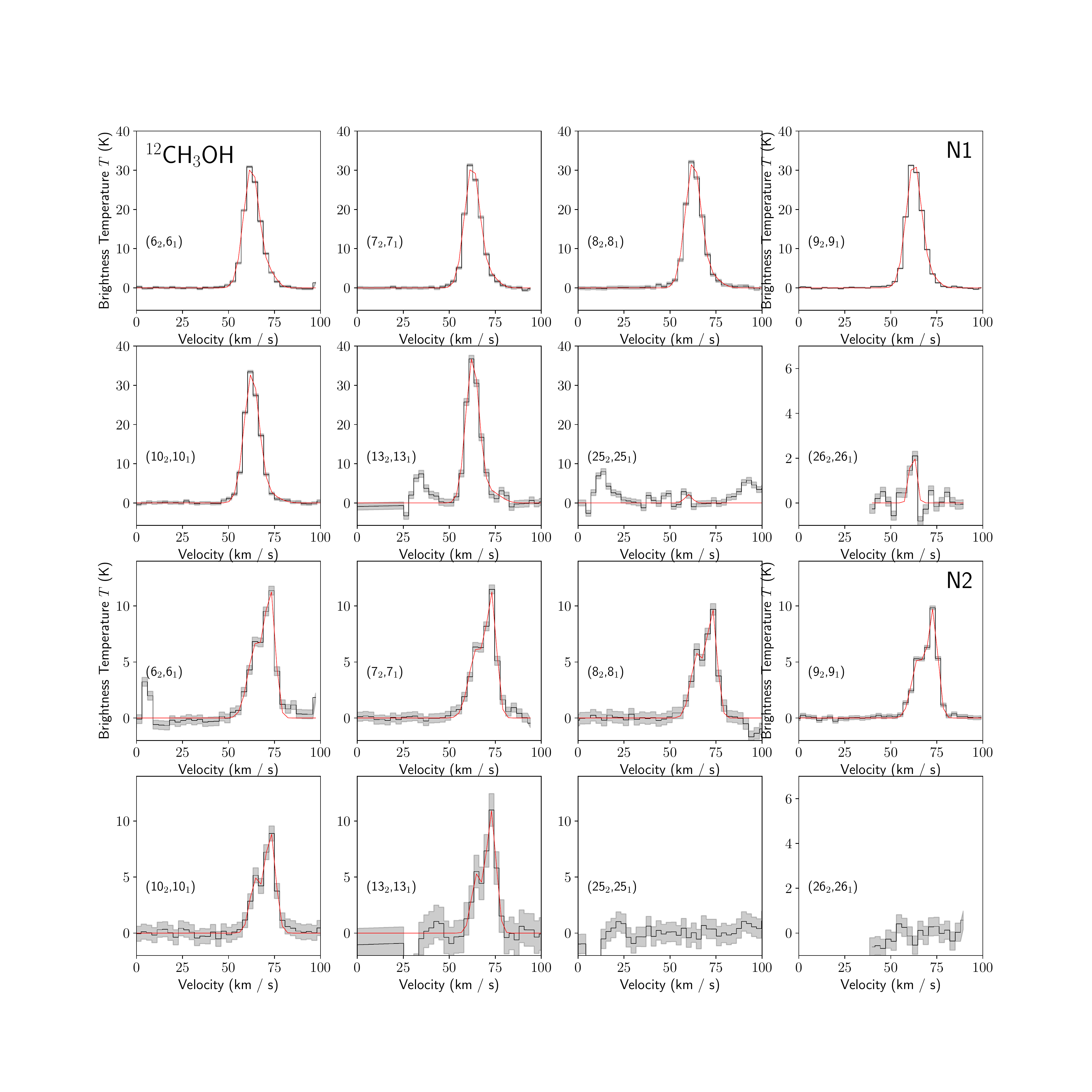}
\caption{Fitted spectra of \methmain\, transitions. The $25_2-25_1$ transition is fit after subtracting the nearby and much stronger $13_2-13_1$ line from the spectrum. For each line, two Gaussian components are fit, for the contribution from the N1 source at 63.5 \kms\, and the N2 source at 73 \kms. }
\label{meth_fit}
\end{figure*}

\begin{figure*}
\vspace{-4cm}
\hspace{-1cm}\includegraphics[scale=0.45]{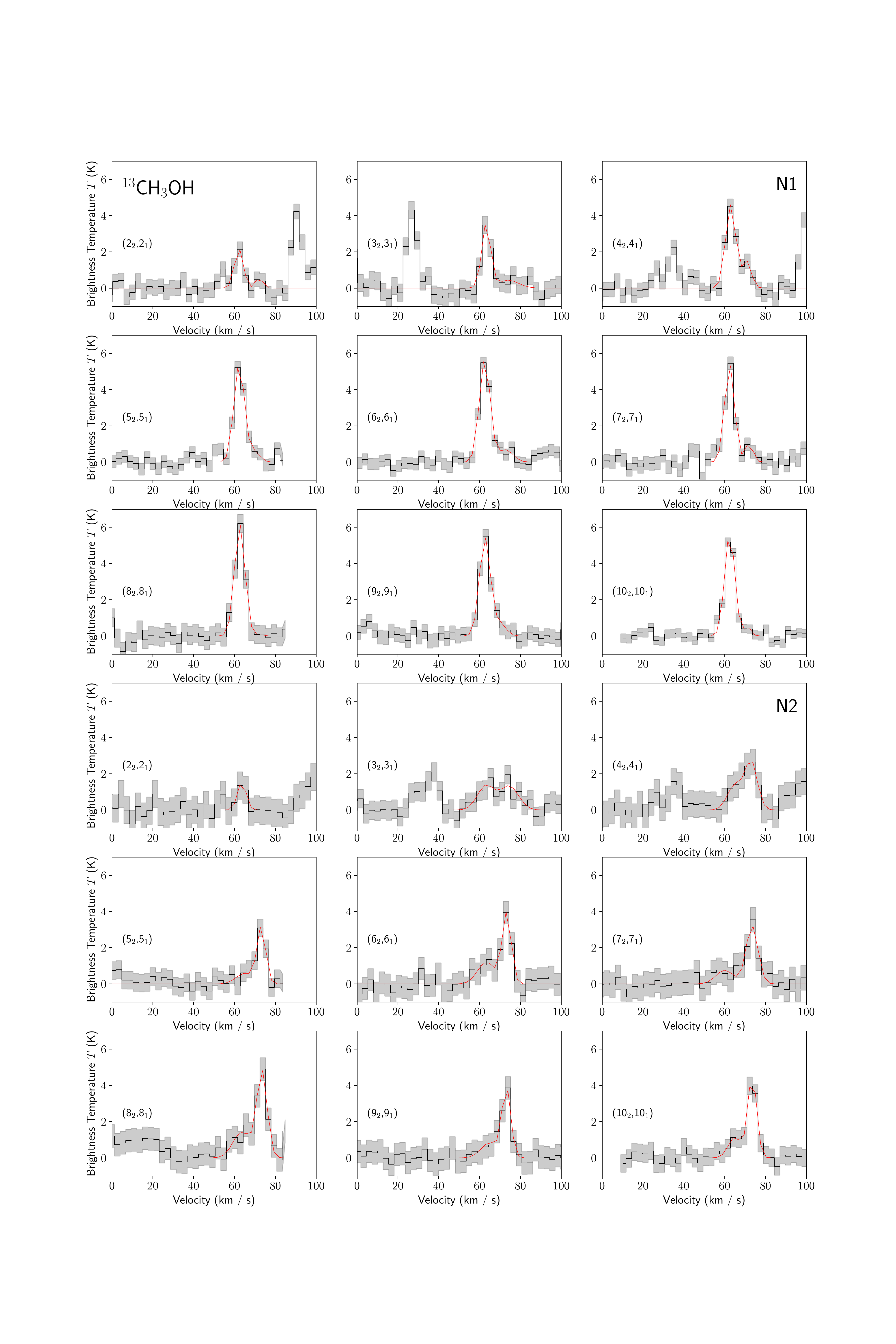}
\caption{Fitted spectra of \methiso\, transitions. Note that the $2_2-2_1$, $4_2-4_1$ and $3_2-3_1$ lines are closely spaced, and each can be seen in the spectra of the other two lines. For each line, two Gaussian components are fit, for the contribution from the N1 source at 63.5 \kms\, and the N2 source at 73 \kms. }
\label{methiso_fit}
\end{figure*}

Fitting of all spectral lines is accomplished using PySpecKit \citep{Ginsburg11}, a python-based spectroscopic analysis toolkit for astronomy. As the combination of smoothed data and relatively large apertures leads to each pointing having some contribution from both N2 and N1, we fit each spectrum with two Gaussian components at the central velocities of N1 and N2. We then report the fit parameters of the Gaussian corresponding to the source (either N1 or N2) centered in the aperture from which the spectrum was extracted. For the (12,11) line of \ammain\, toward N2, we fit for an additional component due to blending with the H63 $\alpha$ line. This component dominates the fitting, and leads to a larger uncertainty on the fitted line parameters. Fits are shown in Figures \ref{am_fit}, \ref{amiso_fit}, \ref{deut_fit}, \ref{meth_fit}, and \ref{methiso_fit} and the results of the fitting are reported in Tables \ref{linesN1} and \ref{linesN2}. The reported errors on the peak and integrated brightness temperatures given in these tables originate both from the fitting errors and the uncertainty on the absolute flux calibration for each telescope (given in Section \ref{obs}). These errors are added in quadrature, with the uncertainty due to the flux calibration typically dominating. Uncertainties on further derived quantities (e.g., column densities) are determined via propagation of these errors.

Unlike the other observed lines, the \ammain\, line exhibits substantial absorption. However, this absorption is primarily confined to the central or main hyperfine component of the line. All of the hyperfine satellites in N1, and the J$\geq$4 hyperfine satellites in N2, appear in emission. We thus fit the hyperfine satellite lines in order to constrain the rotational temperature and column density of this molecule. We assume that the two hyperfine satellite components (with separations of $\sim$2-10 \kms, decreasing with increasing $J$) are blended, and fit them with a single Gaussian profile. For N1, we fit only the blueshifted (lower-velocity) satellite lines, as the redshifted satellite lines are subject to increased absorption. In contrast, for N2, the redshifted satellite lines appear more prominent, and we fit to just these lines (fits to the blueshifted satellite lines yield a lower column density, but a consistent rotational temperature). In general, above $J$=3 ($J$=4 for N2), the hyperfine satellites are well fit by a single Gaussian profile. The profiles of the J$<$4 lines in N1 are non-Gaussian, and appear to be affected by some combination of absorption from the main line and self-absorption in the satellite lines. The hyperfine satellites of the J$<5$ lines in N2 are not apparent, and are likely obscured by absorption. In Table \ref{hfines} we report the line frequencies adopted for the blueshifted (N1) and redshifted (N2) blended hyperfine components. The velocities reported in Tables \ref{linesN1} and \ref{linesN2} are determined relative to these frequencies, and well match the expected source velocities of 64 \kms\, for N1 and 73 \kms\, for N2. Note that the velocity widths represent an average of the blended hyperfine components. While it is clear from the detection of the hyperfine satellites that the central line components are optically thick, we cannot measure the optical depth of individual hyperfine satellite lines (though as noted in Section \ref{spec_amiso}, it seems likely that at least the lower-$J$ satellites are optically thick).

\clearpage 
\begin{table}
\caption{Adopted \ammain\, Hyperfine Line Parameters}
\centering
\begin{tabular}{l l l l}
\hline\hline
Transition &  Blueshifted Hyperfines & Redshifted Hyperfines & Blended Main-to-Hyperfine \\
                &  Blended Frequency      & Blended Frequency &  Intensity Ratio \\
              &                (GHz)                        & (GHz)  & \\
\hline
(4,4) & 24.141622 & & 28.8 \\
(5,5) & 24.535328 & 24.530642 & 43.8 \\
(6,6) & 25.058470 & 25.053579 & 61.8 \\
(7,7) & 25.717706 & 25.712658 & 82.8 \\
(9,9) & 27.478247 & 27.477633 &  125\\
\hline\hline
\end{tabular}
\label{hfines}
\end{table}

\begin{table}
\caption{Measured Line Parameters and Column Densities}
\centering
\begin{tabular}{l l l l l  l l l l}
\hline\hline
Species & Transition   & v$_{\mathrm{cen}}$ & v$_{\mathrm{fwhm}}$ & Peak T$_{\mathrm{MB}}$ & $\int T_{\mathrm{MB}} dv$ & N$_u$ & $g_u$ & $f$\\
        &              &(km s$^{-1}$)      & (km s$^{-1}$)       & (K)                     &  (K km s$^{-1}$)            & ( 10$^{15}$ cm$^{-2}$) & &\\
\hline
{\bf Source N1} &&&&&&&& \\
\hline
\hline
$^{14}$NH$_3$ &  (4,4) &  \footnotemark[1]63.7 $\pm$ 0.1 &  \footnotemark[1]19.9 $\pm$ 0.1 &  \footnotemark[2]28.42 $\pm$ 2.85 &  \footnotemark[2]602.3 $\pm$ 60.4 & \footnotemark[2]1216.6 $\pm$ 122.0 & 9 & 0.114 \\
$^{14}$NH$_3$ &  (5,5) &  \footnotemark[1]64.3 $\pm$ 0.1 &  \footnotemark[1]18.0 $\pm$ 0.1 &  \footnotemark[2]27.07 $\pm$ 2.71 &  \footnotemark[2]518.6 $\pm$ 51.9 & \footnotemark[2]1507.7 $\pm$ 150.9 & 11 & 0.114 \\
$^{14}$NH$_3$ &  (6,6) &  \footnotemark[1]64.4 $\pm$ 0.1 &  \footnotemark[1]18.0 $\pm$ 0.1 &  \footnotemark[2]24.23 $\pm$ 2.42 &  \footnotemark[2]464.2 $\pm$ 46.5 & \footnotemark[2]1817.3 $\pm$ 181.9 & 26 & 0.114 \\
$^{14}$NH$_3$ &  (7,7) &  \footnotemark[1]63.8 $\pm$ 0.1 &  \footnotemark[1]14.0 $\pm$ 0.1 &  \footnotemark[2]14.62 $\pm$ 1.47 &  \footnotemark[2]217.8 $\pm$ 21.9 & \footnotemark[2]1093.6 $\pm$ 109.9 & 15 & 0.114 \\
$^{14}$NH$_3$ &  (9,9) &  \footnotemark[1]64.5 $\pm$ 0.2 &  \footnotemark[1]13.0 $\pm$ 0.1 &  \footnotemark[2]8.72 $\pm$ 0.89 &  \footnotemark[2]120.6 $\pm$ 12.3 & \footnotemark[2]837.8 $\pm$ 85.7 & 38 & 0.114 \\
$^{14}$NH$_3$ &  (10,9)  & 60.0 $\pm$ 0.1 &  10.0 $\pm$ 0.1 &  30.70 $\pm$ 3.21 &  326.8 $\pm$ 34.4 & 24.3 $\pm$ 2.6 & 38 & 0.114 \\
$^{14}$NH$_3$ &  (12,11)  & 61.0 $\pm$ 0.2 &  10.0 $\pm$ 0.1 &  12.59 $\pm$ 1.32 &  134.0 $\pm$ 14.1 & 8.9 $\pm$ 0.9 & 23 & 0.114 \\
$^{14}$NH$_3$ &  (14,13)  & 61.3 $\pm$ 0.5 &  8.7 $\pm$ 1.0 &  6.06 $\pm$ 0.68 &  56.1 $\pm$ 8.7 & 3.5 $\pm$ 0.5 & 27 & 0.114 \\
$^{14}$NH$_3$ &  (19,18)  & 61.3 $\pm$ 1.9 &  9.5 $\pm$ 4.2 &  1.01 $\pm$ 0.24 &  10.2 $\pm$ 5.2 & 0.5 $\pm$ 0.2 & 74 & 0.114 \\
\hline
$^{15}$NH$_3$ &  (1,1)  & 62.9 $\pm$ 0.1 &  9.4 $\pm$ 0.1 &  0.17 $\pm$ 0.02 &  1.7 $\pm$ 0.2 & \footnotemark[3]11.5 $\pm$ 1.2 & 3 & 0.0020\\
$^{15}$NH$_3$ &  (2,2)  & 62.2 $\pm$ 0.1 &  6.8 $\pm$ 0.1 &  0.10 $\pm$ 0.01 &  0.7 $\pm$ 0.1 & \footnotemark[3]3.8 $\pm$ 0.4 & 5 & 0.0020\\
$^{15}$NH$_3$ &  (3,3)  & 63.9 $\pm$ 0.1 &  8.7 $\pm$ 0.1 &  0.25 $\pm$ 0.03 &  2.3 $\pm$ 0.2 & \footnotemark[3]10.5 $\pm$ 1.0 & 14 & 0.0020\\
$^{15}$NH$_3$ &  (4,4)  & 63.5 $\pm$ 0.1 &  8.0 $\pm$ 0.1 &  0.05 $\pm$ 0.01 &  0.5 $\pm$ 0.0 & \footnotemark[3]1.9 $\pm$ 0.2 & 9 & 0.0020\\
$^{15}$NH$_3$ &  (5,5)  & 61.3 $\pm$ 0.1 &  5.9 $\pm$ 0.1 &  0.11 $\pm$ 0.01 &  0.7 $\pm$ 0.1 & \footnotemark[3]2.6 $\pm$ 0.3 & 11 & 0.0021\\
$^{15}$NH$_3$ &  (6,6)  & 62.5 $\pm$ 0.1 &  6.3 $\pm$ 0.1 &  0.10 $\pm$ 0.01 &  0.7 $\pm$ 0.1 & \footnotemark[3]2.4 $\pm$ 0.2 & 26 & 0.0022\\
$^{15}$NH$_3$ &  (6,6)& 61.3 $\pm$ 0.1 &  6.7 $\pm$ 0.1 &  8.55 $\pm$ 0.86 &  61.0 $\pm$ 6.2 & 4.1 $\pm$ 0.4 & 26 & 0.114 \\
$^{15}$NH$_3$ &  (7,7)& 61.5 $\pm$ 0.1 &  4.8 $\pm$ 0.2 &  5.49 $\pm$ 0.57 &  28.0 $\pm$ 3.1 & 1.8 $\pm$ 0.2 & 15 & 0.114 \\
$^{15}$NH$_3$ &  (8,8)& 61.0 $\pm$ 0.1 &  5.4 $\pm$ 0.2 &  2.98 $\pm$ 0.31 &  17.3 $\pm$ 1.9 & 1.1 $\pm$ 0.1 & 17 & 0.114 \\
\hline
$^{14}$NH$_2$D & 3$_{1 3}$-3$_{0 3}$  & 61.9 $\pm$ 0.1 &  9.1 $\pm$ 0.2 &  2.52 $\pm$ 0.76 &  24.3 $\pm$ 7.3 & 8.2 $\pm$ 2.5 & 21 & 0.114\\
$^{14}$NH$_2$D & 4$_{1 4}$-4$_{0 4}$  & 61.4 $\pm$ 0.2 &  5.1 $\pm$ 0.5 &  1.97 $\pm$ 0.24 &  10.7 $\pm$ 1.6 & 4.7 $\pm$ 0.7 & 27 & 0.114\\
\hline
\hline
$^{12}$CH$_3$OH & 6$_2$-6$_1$  & 62.5 $\pm$ 0.1 &  9.9 $\pm$ 0.1 &  31.10 $\pm$ 3.12 &  327.7 $\pm$ 33.0 & \footnotemark[4]40.7 $\pm$ 4.1 & 13 & 0.114\\
$^{12}$CH$_3$OH & 7$_2$-7$_1$  & 62.5 $\pm$ 0.1 &  9.7 $\pm$ 0.2 &  31.55 $\pm$ 3.17 &  325.9 $\pm$ 32.9 & \footnotemark[4]39.3 $\pm$ 4.0 & 15 & 0.114\\
$^{12}$CH$_3$OH & 8$_2$-8$_1$  & 62.6 $\pm$ 0.1 &  10.0 $\pm$ 0.2 &  32.21 $\pm$ 3.24 &  341.5 $\pm$ 34.7 & \footnotemark[4]40.1 $\pm$ 4.1 & 17 & 0.114\\
$^{12}$CH$_3$OH & 9$_2$-9$_1$  & 62.6 $\pm$ 0.1 &  9.8 $\pm$ 0.1 &  32.31 $\pm$ 3.23 &  337.5 $\pm$ 33.8 & \footnotemark[4]38.5 $\pm$ 3.9 & 19 & 0.114\\
$^{12}$CH$_3$OH & 10$_2$-10$_1$  & 62.6 $\pm$ 0.1 &  9.6 $\pm$ 0.2 &  33.00 $\pm$ 3.32 &  337.3 $\pm$ 34.2 & \footnotemark[4]37.3 $\pm$ 3.8 & 21 & 0.114\\
$^{12}$CH$_3$OH & 13$_2$-13$_1$  & 62.6 $\pm$ 0.1 &  8.5 $\pm$ 0.3 &  36.87 $\pm$ 3.76 &  333.6 $\pm$ 35.1 & \footnotemark[4]33.8 $\pm$ 3.6 & 27 & 0.114\\
$^{12}$CH$_3$OH & 25$_2$-25$_1$ & 60.0 $\pm$ 0.1 &  4.7 $\pm$ 1.8 &  2.19 $\pm$ 0.80 &  10.9 $\pm$ 5.9 & 2.6 $\pm$ 1.4 & 51 & 0.114\\
$^{12}$CH$_3$OH & 26$_2$-26$_1$ & 62.0 $\pm$ 0.2 &  4.0 $\pm$ 0.1 &  2.54 $\pm$ 0.33 &  10.8 $\pm$ 1.4 & 3.1 $\pm$ 0.4 & 53 & 0.114\\
\hline
$^{13}$CH$_3$OH & 2$_2$-2$_1$  & 62.1 $\pm$ 0.6 &  5.1 $\pm$ 1.2 &  2.17 $\pm$ 0.47 &  11.9 $\pm$ 3.8 & 2.0 $\pm$ 0.6 & 5 & 0.114\\
$^{13}$CH$_3$OH & 3$_2$-3$_1$  & 63.2 $\pm$ 0.4 &  5.1 $\pm$ 0.9 &  3.58 $\pm$ 0.61 &  19.4 $\pm$ 4.6 & 2.6 $\pm$ 0.6 & 7 & 0.114\\
$^{13}$CH$_3$OH & 4$_2$-4$_1$  & 62.9 $\pm$ 0.3 &  6.0 $\pm$ 0.7 &  4.61 $\pm$ 0.60 &  29.3 $\pm$ 5.2 & 3.6 $\pm$ 0.6 & 9 & 0.114\\
$^{13}$CH$_3$OH & 5$_2$-5$_1$  & 62.3 $\pm$ 0.2 &  6.2 $\pm$ 0.5 &  5.42 $\pm$ 0.63 &  35.5 $\pm$ 4.9 & 4.2 $\pm$ 0.6 & 11 & 0.114\\
$^{13}$CH$_3$OH & 6$_2$-6$_1$  & 62.5 $\pm$ 0.3 &  6.3 $\pm$ 0.5 &  5.67 $\pm$ 0.64 &  37.9 $\pm$ 5.2 & 4.4 $\pm$ 0.6 & 13 & 0.114\\
$^{13}$CH$_3$OH & 7$_2$-7$_1$  & 62.4 $\pm$ 0.2 &  5.7 $\pm$ 0.5 &  5.42 $\pm$ 0.65 &  32.8 $\pm$ 4.7 & 3.7 $\pm$ 0.5 & 15 & 0.114\\
$^{13}$CH$_3$OH & 8$_2$-8$_1$  & 62.6 $\pm$ 0.3 &  6.1 $\pm$ 0.6 &  6.14 $\pm$ 0.78 &  40.0 $\pm$ 6.3 & 4.4 $\pm$ 0.7 & 17 & 0.114\\
$^{13}$CH$_3$OH & 9$_2$-9$_1$  & 62.6 $\pm$ 0.3 &  6.3 $\pm$ 0.6 &  5.45 $\pm$ 0.67 &  36.5 $\pm$ 5.5 & 3.9 $\pm$ 0.6 & 19 & 0.114\\
$^{13}$CH$_3$OH & 10$_2$-10$_1$  & 62.5 $\pm$ 0.2 &  5.9 $\pm$ 0.3 &  5.67 $\pm$ 0.61 &  35.3 $\pm$ 4.2 & 3.7 $\pm$ 0.4 & 21 & 0.114\\
\hline\hline
\end{tabular}
\label{linesN1}
\footnotetext[1]{Values are for the (blended) blueshifted hyperfine satellite lines}
\footnotetext[2]{Values determined using (optically thin) hyperfine satellite lines}
\footnotetext[3]{Column density calculated from GBT data corrected for filling factor}
\footnotetext[4]{Column density is a lower limit due to optically thick transition}
\end{table}
\clearpage

\begin{table}
\caption{Measured Line Parameters and Column Densities}
\centering
\begin{tabular}{l l l l l  l l l l}
\hline\hline
Species & Transition & v$_{\mathrm{cen}}$ & v$_{\mathrm{fwhm}}$ & Peak T$_{\mathrm{MB}}$ & $\int T_{\mathrm{MB}} dv$ & N$_u$ & $g_u$ & $f$\\
        &            & (km s$^{-1}$)      & (km s$^{-1}$)       & (K)                     &  (K km s$^{-1}$)            & ( 10$^{14}$ cm$^{-2}$) & & \\
\hline
{\bf Source N2} &&&&&&& \\
\hline
\hline
$^{14}$NH$_3$ &  (5,5) & \footnotemark[1]72.8 $\pm$ 0.5 &  \footnotemark[1]8.7 $\pm$ 1.0 &  \footnotemark[2]3.14 $\pm$ 0.43 &  \footnotemark[2]29.2 $\pm$ 5.3 & \footnotemark[2]1271.6 $\pm$ 229.1 & 11 & 0.076\\
$^{14}$NH$_3$ &  (6,6) & \footnotemark[1]73.1 $\pm$ 0.5 &  \footnotemark[1]8.8 $\pm$ 1.0 &  \footnotemark[2]2.98 $\pm$ 0.41 &  \footnotemark[2]27.9 $\pm$ 5.1 & \footnotemark[2]1638.0 $\pm$ 299.2 & 26 & 0.076\\
$^{14}$NH$_3$ &  (7,7) & \footnotemark[1]73.3 $\pm$ 0.9 &  \footnotemark[1]8.9 $\pm$ 2.0 &  \footnotemark[2]2.07 $\pm$ 0.45 &  \footnotemark[2]19.6 $\pm$ 6.3 & \footnotemark[2]1473.0 $\pm$ 474.5 & 15 & 0.076\\
$^{14}$NH$_3$ &  (9,9) & \footnotemark[1]74.4 $\pm$ 1.1 &  \footnotemark[1]7.6 $\pm$ 2.4 &  \footnotemark[2]1.77 $\pm$ 0.51 &  \footnotemark[2]14.4 $\pm$ 6.3 & \footnotemark[2]1497.8 $\pm$ 660.6 & 38 & 0.076\\
$^{14}$NH$_3$ &  (10,9) & 74.2 $\pm$ 0.4 &  7.7 $\pm$ 0.9 &  3.20 $\pm$ 0.42 &  26.3 $\pm$ 4.7 & 2.9 $\pm$ 0.5 & 38 & 0.076\\
$^{14}$NH$_3$ &  (12,11) & 72.8 $\pm$ 0.7 &  5.7 $\pm$ 1.8 &  0.83 $\pm$ 0.23 &  5.1 $\pm$ 2.1 & 0.5 $\pm$ 0.2 & 23 & 0.076\\
$^{14}$NH$_3$ & (14,13)   & & & $<$ 0.25 & $<$ 2.2 & $<$ 0.1 & 27 & 0.076\\
$^{14}$NH$_3$ & (19,18)   & & & $<$ 0.60 & $<$ 2.5 & $<$ 0.1 & 74 & 0.076\\
\hline
$^{15}$NH$_3$ &  (1,1)  & 72.9 $\pm$ 0.1 &  6.9 $\pm$ 0.1 &  0.06 $\pm$ 0.01 &  0.4 $\pm$ 0.0 & \footnotemark[3]42.0 $\pm$ 4.2 & 3 & 0.0013\\
$^{15}$NH$_3$ &  (2,2)  & 73.1 $\pm$ 0.1 &  7.1 $\pm$ 0.1 &  0.05 $\pm$ 0.00 &  0.4 $\pm$ 0.0 & \footnotemark[3]27.5 $\pm$ 2.7 & 5 & 0.0013\\
$^{15}$NH$_3$ &  (3,3)  & 74.2 $\pm$ 0.1 &  7.1 $\pm$ 0.1 &  0.09 $\pm$ 0.01 &  0.7 $\pm$ 0.1 & \footnotemark[3]48.4 $\pm$ 4.8 & 14 & 0.0013\\
$^{15}$NH$_3$ &  (4,4)  & 74.1 $\pm$ 0.1 &  7.2 $\pm$ 0.1 &  0.02 $\pm$ 0.00 &  0.2 $\pm$ 0.0 & \footnotemark[3]10.3 $\pm$ 1.0 & 9 & 0.0014\\
$^{15}$NH$_3$ &  (5,5)  & 72.4 $\pm$ 0.1 &  5.0 $\pm$ 0.1 &  0.03 $\pm$ 0.00 &  0.2 $\pm$ 0.0 & \footnotemark[3]9.8 $\pm$ 1.0 & 11 & 0.0014\\
$^{15}$NH$_3$ &  (6,6)  & 73.5 $\pm$ 0.1 &  5.0 $\pm$ 0.1 &  0.03 $\pm$ 0.00 &  0.2 $\pm$ 0.0 & \footnotemark[3]9.3 $\pm$ 0.9 & 26 & 0.0015\\
$^{15}$NH$_3$ &  (6,6)   & 73.5 $\pm$ 0.7 &  6.8 $\pm$ 1.6 &  1.13 $\pm$ 0.23 &  8.2 $\pm$ 2.6 & 8.2 $\pm$ 2.6 & 26 & 0.076\\
$^{15}$NH$_3$ &  (7,7)   & 72.6 $\pm$ 0.9 &  4.1 $\pm$ 1.6 &  0.59 $\pm$ 0.19 &  2.6 $\pm$ 1.3 & 2.5 $\pm$ 1.3 & 15 & 0.076\\
$^{15}$NH$_3$ & (8,8)  & & & $<$ 0.52 & $<$ 2.2 & $<$ 3.5 & 17 & 0.076 \\
\hline
$^{14}$NH$_2$D & 3$_{1 3}$-3$_{0 3}$  & 74.7 $\pm$ 0.1 &  10.1 $\pm$ 0.2 &  2.05 $\pm$ 0.62 &  22.0 $\pm$ 6.6 & 111.1 $\pm$ 33.4 & 21 & 0.076 \\
$^{14}$NH$_2$D & 4$_{1 4}$-4$_{0 4}$  & 73.5 $\pm$ 0.2 &  4.7 $\pm$ 0.4 &  3.86 $\pm$ 0.43 &  19.2 $\pm$ 2.5 & 126.9 $\pm$ 16.8 & 27 & 0.076 \\
\hline
\hline
$^{12}$CH$_3$OH & 6$_2$-6$_1$   & 72.9 $\pm$ 0.2 &  6.1 $\pm$ 0.4 &  10.53 $\pm$ 1.14 &  68.7 $\pm$ 8.3 & \footnotemark[4]128.0 $\pm$ 15.5 & 13 & 0.076 \\
$^{12}$CH$_3$OH & 7$_2$-7$_1$   & 73.1 $\pm$ 0.2 &  6.2 $\pm$ 0.3 &  10.28 $\pm$ 1.10 &  67.4 $\pm$ 8.1 & \footnotemark[4]122.0 $\pm$ 14.7 & 15 & 0.076 \\
$^{12}$CH$_3$OH & 8$_2$-8$_1$   & 73.0 $\pm$ 0.4 &  6.0 $\pm$ 0.6 &  9.47 $\pm$ 1.18 &  60.8 $\pm$ 14.1 & \footnotemark[4]107.0 $\pm$ 24.8 & 17 & 0.076 \\
$^{12}$CH$_3$OH & 9$_2$-9$_1$   & 73.0 $\pm$ 0.1 &  6.0 $\pm$ 0.2 &  9.23 $\pm$ 0.94 &  59.0 $\pm$ 6.8 & \footnotemark[4]101.0 $\pm$ 11.6 & 19 & 0.076 \\
$^{12}$CH$_3$OH & 10$_2$-10$_1$   & 72.8 $\pm$ 0.4 &  6.6 $\pm$ 0.9 &  9.06 $\pm$ 1.12 &  63.6 $\pm$ 19.2 & \footnotemark[4]105.6 $\pm$ 31.9 & 21 & 0.076 \\
$^{12}$CH$_3$OH & 13$_2$-13$_1$   & 72.8 $\pm$ 0.7 &  6.2 $\pm$ 1.4 &  10.87 $\pm$ 1.77 &  71.2 $\pm$ 38.9 & \footnotemark[4]108.2 $\pm$ 59.2 & 27 & 0.076 \\
$^{12}$CH$_3$OH & 25$_2$-25$_1$ & & & $<$ 0.65 & $<$ 3.1 & $<$ 10.9 & 51 & 0.076 \\
$^{12}$CH$_3$OH & 26$_2$-26$_1$ & & & $<$ 0.41 & $<$ 1.8 & $<$ 7.6 & 53 & 0.076 \\
\hline
$^{13}$CH$_3$OH & 2$_2$-2$_1$   & & & $<$ 1.02 & $<$ 7.6 & $<$ 19.3 & 5 & 0.076 \\
$^{13}$CH$_3$OH & 3$_2$-3$_1$   & & & $<$ 3.96 & $<$ 44.7 & $<$ 59.8 & 7 & 0.076 \\
$^{13}$CH$_3$OH & 4$_2$-4$_1$   & 72.9 $\pm$ 1.0 &  7.1 $\pm$ 2.3 &  2.67 $\pm$ 0.79 &  20.2 $\pm$ 8.9 & 37.7 $\pm$ 16.5 & 9 & 0.076 \\
$^{13}$CH$_3$OH & 5$_2$-5$_1$   & 73.1 $\pm$ 0.5 &  5.0 $\pm$ 1.0 &  3.13 $\pm$ 0.59 &  16.7 $\pm$ 4.6 & 29.7 $\pm$ 8.2 & 11 & 0.076 \\
$^{13}$CH$_3$OH & 6$_2$-6$_1$   & 73.3 $\pm$ 0.5 &  5.2 $\pm$ 1.0 &  3.94 $\pm$ 0.72 &  22.0 $\pm$ 5.9 & 37.9 $\pm$ 10.1 & 13 & 0.076 \\
$^{13}$CH$_3$OH & 7$_2$-7$_1$   & 73.4 $\pm$ 0.7 &  6.8 $\pm$ 1.5 &  3.23 $\pm$ 0.69 &  23.5 $\pm$ 7.4 & 39.4 $\pm$ 12.4 & 15 & 0.076 \\
$^{13}$CH$_3$OH & 8$_2$-8$_1$   & 73.4 $\pm$ 0.5 &  6.0 $\pm$ 1.0 &  4.81 $\pm$ 0.77 &  30.7 $\pm$ 7.1 & 50.1 $\pm$ 11.5 & 17 & 0.076 \\
$^{13}$CH$_3$OH & 9$_2$-9$_1$   & 73.1 $\pm$ 0.5 &  5.1 $\pm$ 1.1 &  3.84 $\pm$ 0.76 &  20.7 $\pm$ 6.0 & 33.0 $\pm$ 9.6 & 19 & 0.076 \\
$^{13}$CH$_3$OH & 10$_2$-10$_1$   & 73.5 $\pm$ 0.4 &  5.1 $\pm$ 1.0 &  4.51 $\pm$ 0.69 &  24.6 $\pm$ 6.1 & 38.1 $\pm$ 9.5 & 21 & 0.076 \\
\hline\hline
\end{tabular}
\label{linesN2}
\footnotetext[1]{Values are for the (blended) redshifted hyperfine satellite lines}
\footnotetext[2]{Values determined using (optically thin) hyperfine satellite lines}
\footnotetext[3]{Column density calculated from GBT data corrected for filling factor}
\footnotetext[4]{Column density is a lower limit due to optically thick transition}
\end{table}

\begin{figure*}
\hspace{-1cm}\includegraphics[scale=0.5]{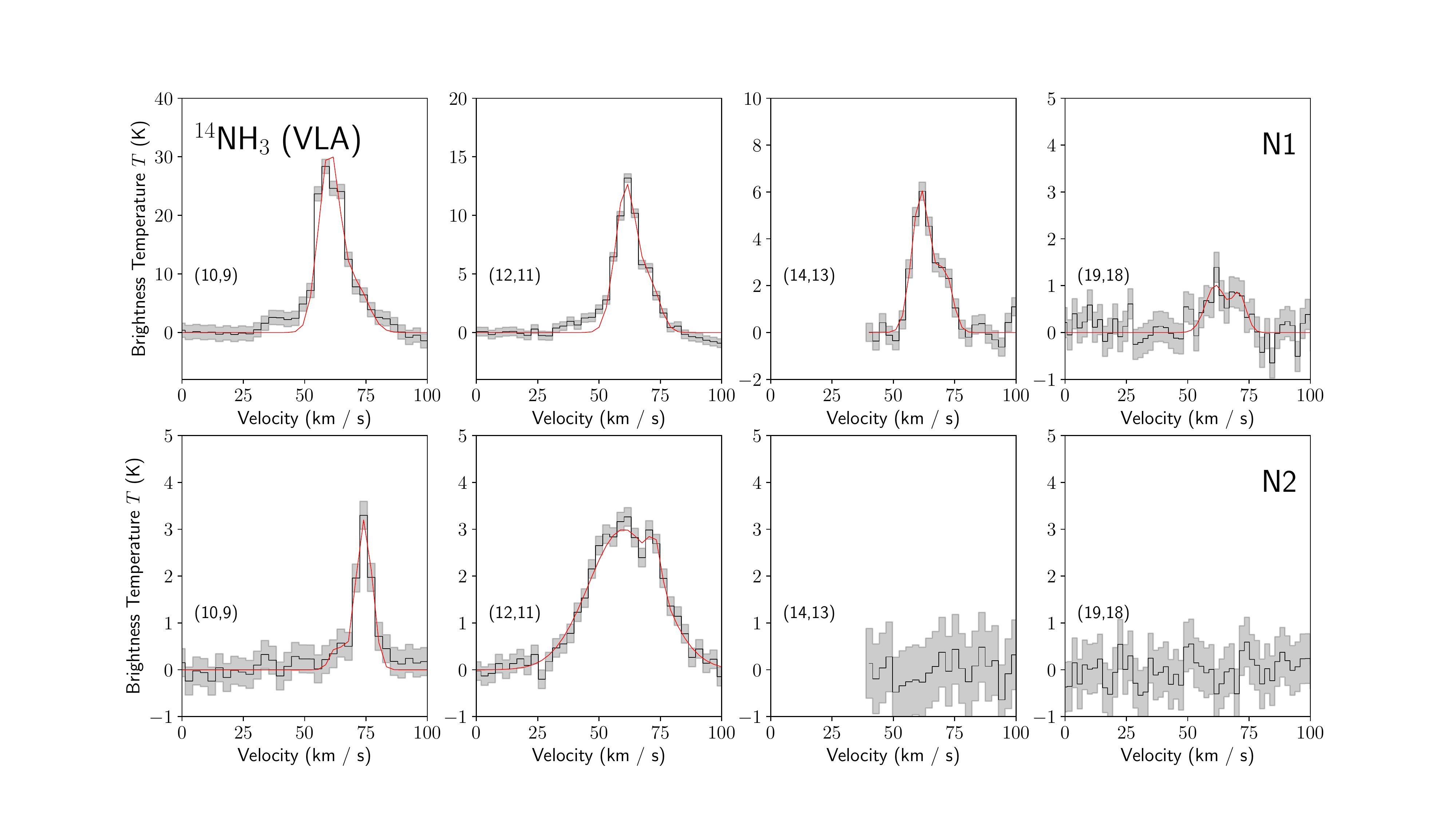}
\caption{Fitted spectra of \ammain\, nonmetastable transitions. For each line, two Gaussian components are fit, for the contribution from the N1 source at 63.5 \kms\, and the N2 source at 73 \kms. The (12,11) line toward N2 is dominated by contamination from the H63$\alpha$ line, tracing ionized gas in the \hii\, regions along the line of sight. }
\label{amnon_fit}
\end{figure*}

\end{document}